\documentclass[a4paper,11pt]{article}
\usepackage{amsmath,amssymb,enumerate,latexsym,amsthm}
\usepackage[USenglish]{babel}
\usepackage{natbib}
\usepackage{setspace,fullpage,graphicx}
\usepackage{xcolor}
\newcommand{\added}[1]{#1}
\newcommand{\deleted}[1]{}

\newcommand{\deletedmath}[1]{}
\usepackage{epstopdf}
\usepackage{caption}
\usepackage{subcaption}
\usepackage{float}
\usepackage{threeparttable}
\usepackage[section]{placeins}
\usepackage{booktabs}
\usepackage{array}
\usepackage{hyperref}
\newcommand{\assign}{:=}
\newcommand{\mathd}{\mathrm{d}}

\newcommand{\nosymbol}{}

\newcommand{\tmem}[1]{{\em #1\/}}
\newcommand{\tmop}[1]{\ensuremath{\operatorname{#1}}}
\newcommand{\tmtextbf}[1]{{\bfseries{#1}}}

\newtheorem{lemma}{Lemma}
\newtheorem{theorem}{Theorem}
\newtheorem{remark}{Remark}
\allowdisplaybreaks

\begin{document}

\title{Testing for Explosiveness in Financial Asset Prices Using High-Frequency Volatility}
\author{H. Peter Boswijk\thanks{Amsterdam School of Economics, University of Amsterdam, PO Box 15867, 1001 NJ Amsterdam, The Netherlands; Email: H.P.Boswijk@uva.nl. }, Jun Yu\thanks{Department of Finance and Business Economics, Faculty of Business Administration, University of Macau, Macao, China; Email: junyu@um.edu.mo.} and Yang Zu\thanks{Department of Economics, University of Macau, Macao, China; Email: yangzu@um.edu.mo.}}
\maketitle

\begin{abstract}
Based on a continuous-time stochastic volatility model with linear drift and jumps,
we develop a test for explosive behavior in financial asset prices at a low
frequency when prices are sampled at a higher frequency. The test exploits
the volatility information in the high-frequency data, while maintaining robustness to jumps using truncation. The method involves first removing jumps and then devolatizing low-frequency logarithmic asset price increments using truncated realized volatility measures. Subsequently, a supremum-type recursive Dickey-Fuller test is performed on the reconstructed sample. The proposed test has a nuisance-parameter-free
asymptotic null distribution and is easy to implement. We study the size and
power properties of the test in Monte Carlo simulations. A real-time
date-stamping strategy based on the reconstructed sample is proposed for the
origination date of the explosive regime. Conditions under which the
real-time date-stamping strategy is consistent are established. The test and
the date-stamping strategy are used to study explosive behavior in recent major cryptocurrency markets and in the S\&P 500 index during the 1990s dotcom bubble period.
\end{abstract}

\noindent {Keywords: Stochastic volatility model; Explosiveness; Asset price bubbles; Cryptocurrency.}

\section{Introduction}

Following the influential paper by {\cite{phillips2011explosive}},
there has been a renewed interest in the empirical identification of explosive behavior in financial asset prices as a means of
detecting economic bubbles. {\cite{phillips2011explosive}} uses a test statistic that is the supremum of recursively implemented Dickey-Fuller ($t$) statistics. The test is a right-tailed test, commonly referred to as the PWY test. This approach
effectively addresses the power deficiency problem commonly encountered in
simple right-tailed unit root tests, a problem extensively discussed in
previous studies such as {\cite{Diba1987,Diba1988}}, {\cite{Flood1980}},
{\cite{Flood1986}} and {\cite{evans1991pitfalls}}. 

Although the explosive behavior of asset prices can be attributed to non-bubble mechanisms (see, for example, \cite{AitSahalia2024}), such behavior in the price series---especially after fundamental
components have been removed---often suggests the presence of a bubble phenomenon. This perspective is consistent with the rational bubble model developed in \cite{blanchard1982bubbles}.
\cite{phillips2011explosive} and \cite{Phillips2013a} extensively discuss how testing for explosiveness can be instrumental in identifying rational bubbles.
When an asset lacks fundamental components, a rapid and explosive increase in its price also carries significant implications for bubble formation.
This is because such a fast price surge is unlikely to be sustainable and is therefore more prone to a crash.
This understanding resonates with Fama's definition of a bubble in his Nobel Lecture (\cite{fama2014two}, p.\ 1475), where he describes a bubble as ``an irrationally strong price increase that implies a predictable strong decline.''

Subsequent literature has expanded the theoretical framework and practical applications
of explosive bubble testing. For example, {\cite{Phillips2011}} studies the
behavior of bubbles during the subprime crisis. {\cite{homm2012testing}}
propose various supremum-type tests for bubbles based on alternative tests
for changes in persistence in time-series models. They also propose to use
the supremum of backwardly implemented recursive Chow-type tests to test
for bubbles. \cite{Phillips2013} considers the problem of testing for multiple
bubbles and proposes a double supremum version of DF $t$-statistics over
all possible subsamples as their test statistic. {\cite{Phillips2013a}} studies the limit theory of bubble date detectors in the context of
multiple bubbles. See also \cite{harvey2015tests},  \cite{harvey2025unit}, {\cite{Breitung2013}}, \cite{astill2023cusum}, \cite{pavlidis2012new} and \cite{shi2015identifying},
among others. These studies collectively contribute to the advancement of our
understanding of bubble detection techniques and provide valuable insights
into the dynamics and implications of explosive behavior in financial
markets.

Existing research on testing asset price bubbles has predominantly been
based on classical discrete-time autoregressive (AR) models related to
the unit-root-testing literature. However, in the finance literature, it is
commonly assumed that asset prices follow a continuous-time model, which can
offer analytical simplicity for pricing derivatives. Furthermore, from an empirical modeling perspective, continuous-time models provide a natural framework for combining data sampled at various frequencies, which is especially relevant to the problem considered in this paper.

In this study, we consider the problem of testing for explosive behavior within the framework of a continuous-time stochastic volatility model featuring a linear drift and a compound Poisson jump component. The stochastic volatility process is specified nonparametrically, so it does not need to satisfy any Markovian or stationarity properties. When volatility is constant and in the absence of jumps, our model corresponds to the well-known Ornstein-Uhlenbeck (OU) process.\footnote{The testing of
unit roots against stationary alternatives in the left tail within the context of the OU process has been explored in previous works
such as {\cite{Phillips1987}}, {\cite{perron1991continuous}}, {\cite{Yu2014}},
{\cite{Zhou2015}} and {\cite{chen2015optimal}}.} To capture the explosive
behavior in the data, we allow the persistence parameter to vary from the
unit root regime to the explosive regime. 

Our procedure involves four steps. First, we utilize the truncated realized variance (TV) estimator of \cite{mancini2009non} to estimate the integrated volatility (IV) over time intervals
corresponding to a lower sampling frequency. Second, we calculate the low-frequency return by aggregating truncated high-frequency returns and then ``devolatize'' the jump-removed lower-frequency returns using the TV estimator. Third, we construct a pseudo-sample by taking the cumulative sum of the devolatized and jump-removed low-frequency returns. Finally, we apply the PWY test to this reconstructed sample to detect explosive behavior, resulting in the TVPWY test.
We show that the TVPWY statistic has the same
asymptotic null distribution as the PWY statistic in the discrete-time model
with constant volatility. As a result, the critical values provided in {\cite{phillips2011explosive}} can be readily used for inference in our TVPWY
test. Because the first three steps can be performed in real-time, the PWY date-stamping procedure can be directly applied to the reconstructed real-time pseudo-sample. We show that the resulting date-stamping strategy can consistently identify the onset time of the explosive regime.

\begin{figure}[h]
    \centering
    \caption{Log-price and pseudo-log-price of Bitcoin, January 1, 2018 to August 31, 2021}
    \includegraphics[width=.9\textwidth]{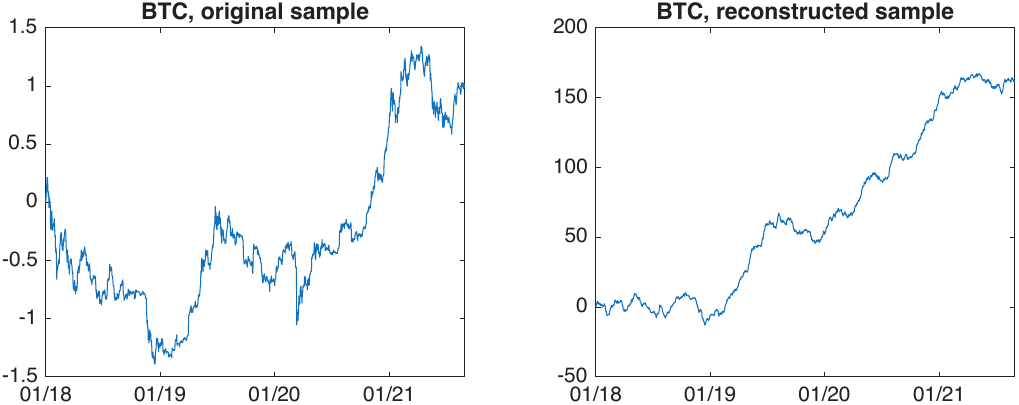}
    \label{fig:motiv}
\end{figure}

Our proposed testing and date-stamping procedure involves applying the PWY test to a reconstructed (specifically, a truncated and devolatized series) time series. This approach has significant implications
for the outcomes of the testing and date-stamping procedure. Figure \ref{fig:motiv} displays the daily Bitcoin log-prices from January 1, 2018 to August 31, 2021 (left panel) and the reconstructed pseudo-sample (right panel). The visual distinction between the two plots is particularly apparent. In the reconstructed sample,
sudden large drops in the Bitcoin price are less apparent. Specifically, the reconstructed sample reduces the apparent magnitude of price drops while largely preserving the relative size of price increases. This is because the original data exhibits higher volatility during price declines. By removing jumps and heteroskedasticity via truncation and devolatization, the reconstructed sample offers a more stable foundation for identifying and date-stamping explosive behavior. Our Monte Carlo simulations and empirical analyses support this intuition.

Despite utilizing high-frequency intraday data, this paper focuses on testing for explosive behavior at the low-frequency level. While explosive behavior can, theoretically, be tested in high-frequency data, the data-generating process at these frequencies is typically more complex than that at lower frequencies. Factors such as diurnal effects and
market microstructure noise need to be taken into account in high-frequency
analyses, which can introduce complications that may invalidate existing
methods. In practice, policymakers and investors often seek to understand
whether an economic time series exhibits explosive behavior at the low-frequency
level, even if they have access to high-frequency data. This preference can
be attributed to various reasons, including the high costs associated with
implementing new policies or rebalancing portfolios. 

This paper builds upon several existing strands of literature. First, it is related to the emerging literature in identifying drift components in the high-frequency data,
such as \cite{COR22}, \cite{LaurentShi2020VolatilityJump}, \cite{LaurentRenoShi2024} and \cite{ShiPhillips2024}.\footnote{Our framework differs from \citet{ShiPhillips2024} in both target and methodology. \citet{ShiPhillips2024} test for \emph{mild drift} in the intraday high-frequency price path using a realized-autocovariance statistic. We test for \emph{explosive} dynamics in the low-frequency price level, using the high-frequency data only to extract the volatility information. Moreover, their test does not differentiate between the direction of deviations from the zero-drift null hypothesis, such as mean-reversion, drift bursts, or explosiveness.
Beyond testing, we also provide a consistent estimator of the origination date of the explosive regime together with its asymptotic theory. Although closely related, our framework neither extends, nor modifies, nor specializes theirs.} Our problem is different in that we are interested in identifying explosive behavior at the lower frequency. Second, it is related to the existing literature on explosive bubble testing accounting for deterministic time-varying volatility (e.g.,  \cite{harvey2015tests}, \cite{harvey2019testing} and \cite{harvey2020sign}). However, these tests do not apply to continuous-time, genuine stochastic volatility models, and especially they cannot deal with jumps considered in this paper. In a recent study, \cite{AitSahaliaJacod2024} test whether volatility is a deterministic function of the price level or a separate stochastic process, and conclude that volatility is stochastic. Third, it is related to \cite{CJ20}, \cite{JK21}, and \cite{JPM22}. Notably, \cite{JK23} uses quadratic variation to identify explosive behavior in asset prices. Furthermore, this research connects to the work of \cite{ATZ23}, where the authors use implied volatility
obtained from options to devolatize the price series, and a CUSUM test to find deviations from the local no-arbitrage condition. Additionally, our work draws on the research of \cite{PY09twostage}, which uses realized volatility to estimate parameters in a two-stage approach.
Relevant parameter estimation techniques are explored by \cite{Corradi2006} and \cite{Todorov2009}.

The structure of the paper is as follows. Section 2 considers a
continuous-time model, defines the truncated realized variance estimator, and
discusses how to construct the pseudo-sample. Section 3 considers the explosiveness testing
problem, defines the TVPWY statistic, and derives the asymptotic null
distribution. Section 4 performs Monte Carlo simulations to study the finite
sample size and power properties of the proposed test, in comparison
with several existing tests. Section 5 considers the real-time
date-stamping problem. Section 6 applies the proposed methodology to two empirical applications. Section 7 concludes the paper. Technical lemmas and mathematical proofs are collected in the
Appendix. 

Throughout the paper, we use $\xrightarrow{d}$ to denote convergence in
distribution and use $\Rightarrow$ to denote weak convergence of a
stochastic process. $\lfloor x \rfloor$
denotes the integer part of a non-negative real number $x$.

\section{The Model, Sampling Scheme, and Pseudo-sample Construction}\label{model}

Consider the following model for the log asset price process $\{z_t \}_{t \geqslant 0}$,
\begin{equation}
    z_t = y_t + J_t, \label{OUJ}
\end{equation}
where $\{ y_t \}$ is the Brownian semimartingale process satisfying
\begin{equation}
    \mathrm{d} y_t = \kappa y_t \mathrm{d} t + \sigma_t \mathrm{d} W_t, \quad t
    \geqslant 0,  \label{OU}
\end{equation}
with $y_0$ a fixed starting value, $\{W_t\}$ a Brownian motion process, $\{\sigma_t\}$ a strictly
positive stochastic volatility process, independent of the $\{W_t\}$ process, and $\kappa$ the persistence parameter. When $\kappa = 0$, the model for $\{y_t\}$ corresponds to the unit root model; when $\kappa > 0$, it
corresponds to an explosive model. $\{ J_t \}$ is a
jump process satisfying
\begin{equation}
\mathd J_t = \xi_t \mathd N_t,  \label{jumps}
\end{equation}
where $\{N_t \}$ is a Poisson process with jump intensity $\lambda$, and
$\xi_t$ is the jump size when it happens at time $t$. This model is an extension of the
continuous-time Ornstein-Uhlenbeck (OU) model considered in {\cite{perron1991continuous}} and {\cite{Zhou2015}} to allow for stochastic
volatility and jumps.

We assume that the log price process $z_t$ is observed at low-frequency $H$ (e.g., one day or one month) and high-frequency $h$ (respectively, 5 minute or one day). To be specific, $z_t$ is observed on a time grid $t_i=iH$, where $i=0, 1, 2, \ldots, n$. Within each low-frequency sampling interval of length $H$, $z_t$ is observed
over a finer time grid with a sampling interval $h$. As explained in Section 1, due to factors such as diurnal effects and local abnormal price movements in high-frequency observations, we utilize high-frequency data to account for volatility and jumps, while testing explosiveness in low-frequency observations. Our asymptotic scheme is such that $H$ is fixed, $h\rightarrow 0$ and $n\rightarrow \infty$. 

\subsection{The Brownian Semimartingale Model and Devolatization}
Our approach consists of two primary elements: devolatization and truncation, which address volatility and jumps in data, respectively. To better motivate the approach, we first consider a simplified case where the jump component is absent from Model (\ref{OUJ}), focusing on the devolatization part of our method.

 Define the difference operator $\Delta$ at the low-frequency $H$ as $\Delta z_{t_i}:=z_{t_{i}}-z_{t_{i-1}}$. When $J_t=0$, one observes $y_t$ directly and $z_t=y_t$. Discretizing the model (\ref{OU}) at
the low observational frequency, we have 
\begin{equation}
\Delta y_{t_i}= \kappa \int_{t_{i-1}}^{t_i} y_u \mathrm{d} u +
\int_{t_{i-1}}^{t_i} \sigma_u \mathrm{d} W_u = \kappa \int_{t_{i-1}}^{t_i}
y_u \mathrm{d} u + \sqrt{\int_{t_{i-1}}^{t_i} \sigma_u^2 \mathrm{d} u}
\times \varepsilon_i,  \label{dOU}
\end{equation}
$i = 1, \ldots, n$, where $\varepsilon_i$ is defined as $\varepsilon_i:=
\int_{t_{i-1}}^{t_i} \sigma_u \mathrm{d} W_u /\sqrt{\int_{t_{i-1}}^{t_i}
\sigma_u^2 \mathrm{d} u}$. Note that this is an exact discretization. Under the independence assumption between $\{\sigma_t\}$ and $\{W_ t\}$, conditional on the $\sigma$ process, $%
\int_{t_{i-1}}^{t_i} \sigma_u \mathrm{d} W_u$ follows a $N(0,%
\int_{t_{i-1}}^{t_i} \sigma_u^2 \mathrm{d} u)$ distribution, and hence $%
\varepsilon_i$'s are independent and identically distributed (i.i.d.) as $N(0,1)$; $\{\varepsilon_i\}$ is also independent
of $\{\sigma_i\}$. 

In this paper, we follow the convention to refer to the quantity $\int_{t_{i-1}}^{t_i} \sigma_u^2\mathrm{d} u$
as the \emph{integrated variance} and its square root as the \emph{integrated volatility}. Let the integrated volatility over the period from $t_{i-1} $ to $t_i$ be denoted as $\omega_i := \sqrt{\int_{t_{i-1}}^{t_i} \sigma_u^2 
\mathrm{d} u}$. When $\kappa = 0$, the model (\ref{dOU}) becomes 
\begin{equation}\label{devol}
    \Delta y_{t_i}  = \omega_i \varepsilon_i, 
\end{equation}
Based on the property that $\varepsilon_i \overset{\mathrm{i.i.d.}}{\sim} N(0,1)$, we construct an infeasible pseudo-sample $x_i^{\ast}$ by setting $x_0^{\ast} = 0$ and 
\begin{equation}\label{cumsumdxstar}
x_i^{\ast} := \sum_{s = 1}^i \frac{\Delta y_{t_s} }{\omega_s} = \sum_{s = 1}^i \varepsilon_s
, \quad i = 1, \ldots, n.
\end{equation}
which is the cumulative sum of i.i.d.\ standard normal variables and follows a random walk model by construction:
\begin{equation}
x_i^{\ast} = x_{i - 1}^{\ast} + \varepsilon_i, \quad i = 1, \ldots, n
,  \label{randomwalk}
\end{equation}
with $x_0^{\ast} = 0$. Note that the variance of $\varepsilon_i$ is unity. Therefore, by testing the existence of a unit root in the pseudo-series $\{x_i^{\ast}\}_{i=0}^{n}$ (e.g., using the classical Dickey-Fuller test), one can potentially test the unit root hypothesis in (\ref{OU}). Compared with the original heteroskedastic sample, where the classical Dickey-Fuller test does not apply directly, the reconstructed sample is devolatized and heteroskedasticity-free. 

\subsection{The Model with Jumps and Truncation}

Consider next the case where $J_t\neq 0$. Under the null hypothesis, the low-frequency return is
\begin{equation}
    \Delta z_{t_i}  = 
    \int_{t_{i - 1}}^{t_i} \sigma_u \mathrm{d} W_u + \int_{t_{i - 1}}^{t_i}
    \xi_u \mathd N_u =\sqrt{\int_{t_{i -
                1}}^{t_i} \sigma_u^2 \mathrm{d} u} \times \varepsilon_i + \sum_{s = N_{t_{i
                - 1}}+1}^{N_{t_i}} J_s,  \label{dOUJ}
\end{equation}
where an additional aggregated jumps component $\sum_{s = N_{t_{i - 1}}+1}^{N_{t_i}} J_s$ appears. Due to this additional component, statistical properties of the pseudo-series $\{x_i^{\ast}\}_{i=0}^{n}$ have completely changed and the pseudo-series no longer follows a random walk process. To overcome this difficulty, we use a truncation approach to address the issue of jumps, which is a common approach used in financial econometrics, and was first proposed by \cite{mancini2009non} for integrated variance estimation.

Truncating the low-frequency return in (\ref{dOUJ}) directly to eliminate the aggregated jump component is not effective, because at the fixed sampling frequency $H$, the magnitude of the diffusion component and jump components are comparable (both bounded in probability), and it is impossible to find a level of truncation to separate the jump components.

Instead, we propose to apply the truncation to high-frequency returns over the sampling interval $h$, which shrinks to 0, and exploit the fact that the magnitude of the diffusion component and the jump component are different when $h\rightarrow 0$. To be specific, for each time interval $[t_{i-1}, t_i]$, assume we also observe $z_t$
over a finer time grid $t_{i-1,j} = t_{i-1}+j h$, with $j=0,1,\ldots,M$ and $h = H/M$,
such that $t_{i-1,0} = t_{i-1}$ coincides with the left end-point and $t_{i-1,M}
= t_{i}=t_{i,0}$ coincides with the right end-point of the interval. Thus, we have a total of $M + 1$
high-frequency observations of $z_t$ with the sampling interval $h$ in the time interval $[t_{i-1}, t_i]$. In a generic low-frequency sampling interval, define the high-frequency-$h$ difference operator $\Delta^h$ as $\Delta^h z_{t_{\boldsymbol{\cdot},j}}:=z_{t_{\boldsymbol{\cdot},j}}-z_{t_{\boldsymbol{\cdot},j-1}}$. We define the jump-robust devolatized pseudo-series by setting $x_0^T = 0$ and
\[ x_i^T \assign \sum_{s = 1}^i \frac{\sum_{j = 1}^M \Delta^h  z_{t_{s - 1, j}} \mathbf{1} \{| \Delta^h z_{t_{s-1,j}} | < \phi_h
    \}}{\omega_s}, \quad i = 1, \ldots, n, \]
where $\phi_h$ represents the truncation level and the superscript $T$ denotes the application of truncation. This approach aggregates the truncated high-frequency returns within the low-frequency sampling interval $H$. 

The intuition of applying truncation to high-frequency returns is as follows: under the null and at high-frequency sampling interval $h$, the diffusion component shrinks and has order $O_p(h^{1/2})$, while the jump component, when present, has order $O_p(1)$ and therefore dominates the diffusion component. Therefore, by an appropriate choice of the truncation level $\phi_h$, we can remove the high-frequency returns that are dominated by jumps. Furthermore, given that the Poisson-type jumps are rare, removing those returns will have a negligible effect on the asymptotic properties of the resulting testing statistic.

The above jump-robust pseudo-series is not feasible for use in practice, because the integrated volatility $\omega_s$ is not observable and must be estimated.
We need a valid integrated variance estimator robust to jumps for this purpose.
We proceed by using \cite{mancini2009non}'s \emph{truncated realized variance} (TV) estimator, given by
\begin{equation}                                                                          
    \hat{\omega}_s^2 \assign\sum_{j = 1}^M (\Delta^h  z_{t_{s - 1, j}})^2 \mathbf{1} \{| \Delta^h z_{t_{s-1,j}} | < \phi_h
    \},
    \label{tvestimator}
\end{equation}
to construct the feasible pseudo-sample. This is defined by setting $x_0 = 0$ and
\begin{equation} \label{pseudosample}
x_i \assign \sum_{s = 1}^i \frac{\sum_{j = 1}^M \Delta^h  z_{t_{s - 1, j}} \mathbf{1} \{| \Delta^h z_{t_{s-1,j}} | < \phi_h
    \}}{\hat{\omega}_s}, \quad i = 1, \ldots, n.
\end{equation} 
The TV estimator $\hat{\omega}_s^2$ in (\ref{tvestimator}) is the sum of those squared high-frequency returns over the
low-frequency interval $[t_{s-1},t_s]$ whose magnitude fall below the truncation threshold $\phi_h$, and it is a jump-robust estimator of the integrated variance $\int_{t_{s-1}}^{t_s}\sigma_u^2\,\mathrm{d}u$
over that interval (equivalently, of $\omega_s^2$).
We use the same truncation level $\phi_h$ for the numerator \added{in (\ref{pseudosample})} and the TV estimator \added{(\ref{tvestimator})}. The feasible pseudo-series $\{x_i\}_{i=0}^n$ no longer follows
an exact random walk model due to the truncation of returns and the estimation errors of the TV estimator. However, we show below that the feasible pseudo-series still behaves like a random walk model asymptotically, in that the normalized feasible series converges weakly to the same Brownian motion limit. For this result, we will need the following assumptions:

\begin{description}
    \item[A1] The volatility process $\{\sigma_t\}$ is continuous, strictly positive,
    with bounded fourth moment  for any $t\geqslant0$, and independent of the $\{W_t\}$
    process.
    
    \item[A2] $H$ is fixed, $n\rightarrow \infty$, $h\rightarrow 0$, and $n h\rightarrow 0$.
    
    \item[A3] The truncation level satisfies $\phi_h = O(h^{1 / 2 - \varepsilon})$, with $0 < \varepsilon
    \leqslant 1 / 4$.
    
    \item[A4] The cumulative distribution function $F_{\xi}(.)$ of the jump sizes is continuous at $0$.
\end{description}

Under A2 and A3, we have $\phi_h \rightarrow 0$ and $\phi_h / h^{1 / 2}
\rightarrow \infty$. Therefore, we set our truncation level between the stochastic orders of the jump and diffusion components, ensuring that the jump component is truncated while the diffusion component remains.
The relative speed condition $n h\rightarrow 0$ in A2 implies that in practice, our method needs the number of high-frequency observations $M$ (per time interval) to be large
relative to the number of low-frequency observations $n$.
Next, we give our first result, which is the weak limit of the normalized pseudo-series.

\begin{theorem}
    \label{thm1}Under Assumptions A1-A4, when $\kappa = 0$,
    
    \begin{equation*}
        \frac{1}{\sqrt{n}} x_{\lfloor n \tau \rfloor} \Rightarrow B_{\tau}, \quad 0 \leqslant \tau \leqslant 1,
    \end{equation*}
    where $B$ is a standard Brownian motion process on $[0,1]$.
\end{theorem}

\begin{remark}
    Our specification of the volatility process permits non-Markovian volatility models. Well-known volatility processes that do not satisfy the Markov property are the fractional Brownian motion and the fractional Ornstein-Uhlenbeck process; see \cite{gatheral2018} and \cite{wang2023}. Increasingly, these non-Markovian models are favored in the recent volatility literature, demonstrating superior performance in forecasting daily realized volatility. Moreover, our volatility process accommodates a wide range of nonstationary behaviors, including unit roots and structural breaks.
\end{remark}

\begin{remark}
    In discrete-time models without jumps, a similar devolatizing approach has been used to address deterministic unconditional heteroskedasticity, see e.g.\ \cite{beare2018} and \cite{harvey2024new}. Even in the absence of jumps, our analysis differs because we focus on a continuous-time stochastic volatility model. When a jump component is present, our proposal, which involves truncation and devolatization to construct a pseudo-sample, is new.

\end{remark}

\begin{remark}
    In time series of equity returns, the leverage effect is often found, 
    whereby volatility shocks are negatively correlated with (lagged) stock returns.
    This would lead to a violation of Assumption A1, and hence of the property that $\{\varepsilon_i\}$
    is an i.i.d.\ $N(0,1)$ sequence. However, there is empirical evidence of i.i.d.\ normality of daily returns that have been standardized
    using daily realized volatility, \emph{despite} the presence of the leverage effect; see, e.g., {\citet*{andersen2001distribution}} and
    \citet[Chapter 3]{christoffersen2012elements}. For many other financial returns, e.g.\ on foreign exchange rates, such leverage effects are not found empirically;
    see, e.g., {\citet*{meyeryu2000}}. Consequently, we maintain the no-leverage assumption in this paper, leaving the relaxation of Assumption A1 for future research.
\end{remark}

\section{Testing Explosiveness}\label{bubbletest}

To consider the explosiveness testing problem in model (\ref{OUJ}), we extend the continuous-time
OU process $y_t$ in (\ref{OU}) to allow for time-varying $\kappa$,
i.e., 
\begin{equation}  \label{tvou}
\mathrm{d} y_t = \kappa_t y_t \mathrm{d} t + \sigma_t \mathrm{d} W_t, \text{ with } \kappa_{t} = \left\{ 
\begin{array}{ll}
0, & t \leqslant t_{i_1} \\ 
\kappa, & t > t_{i_1}
\end{array}
\right. ,
\end{equation}
where we consider a model with a one-time change from the unit root regime to
the explosive regime at $t_{i_1}$ for some integer $i_1<n$. Although the regime change could theoretically occur between high-frequency observations, for ease of exposition, we assume that the change happens only at a low-frequency time point.
The explosiveness testing problem considered is as follows: 
\begin{equation*}
\mathcal{H}_0 : \kappa= 0 \hspace{1em} \mbox{vs.} \hspace{1em} \mathcal{H}_1 : \kappa > 0.
\end{equation*}
That is, we seek to test whether a change of regime happened within the
sample period. The null hypothesis is that no such change occurred.

Assume we have already constructed a feasible pseudo-sample $\{x_1, \ldots,
x_n \}$. For $0 < \tau \leqslant 1$, define
\[ \tmop{TVDF}_{\tau} = \frac{\sum_{s = 2}^{\lfloor \tau n \rfloor} (\Delta
   x_s - \overline{\Delta x_s}) (x_{s - 1} - \overline{x_{s - 1}})}{\left(
   \sum_{s = 2}^{\lfloor \tau n \rfloor} (x_{s - 1} - \overline{x_{s -
   1}})^2 \right)^{1 / 2}}, \]
where $\overline{\Delta x_s} = (1 / (\lfloor \tau n \rfloor -1)) \sum_{s =
2}^{\lfloor \tau n \rfloor} \Delta x_s$ and $\overline{x_{s - 1}} = (1 /
(\lfloor \tau n \rfloor -1)) \sum_{s = 2}^{\lfloor \tau n \rfloor} x_{s - 1}$. Notice that the $\tmop{TVDF}_{\tau}$ defined above is a slightly \emph{modified} classical Dickey-Fuller $t$ statistic
over the subsample $\{x_1, \ldots, x_{\lfloor \tau n \rfloor} \}$. The modification involves imposing the error variance estimator to be equal to one, which is the true error variance under the null hypothesis.
If the null hypothesis is correct, such modification has a negligible effect, as our pseudo-sample $\{x_1, \ldots,x_n \}$ asymptotically follows a random walk model with unit error variance.
Under the alternative hypothesis, it will turn out that such a modification is crucial for proving the consistency of the date-stamping procedure in Section 5.

With this modified Dickey-Fuller statistic
we can define, analogously to PWY, the TVPWY statistic as
\[ \mathrm{TVPWY} \assign \sup_{\tau \in [\tau_0, 1]} \mathrm{TVDF}_{\tau}, \]
where $\tau_0$ is a small fraction to
ensure a reasonable sample size in the smallest subsample $\{x_1, \ldots,
x_{\tau_0} \}$.
\begin{theorem}
\label{thm2}If Assumptions A1 to A4 hold, under the null hypothesis $%
\mathcal{H}_0$, 
\[ \mathrm{TVPWY}\Rightarrow \sup_{\tau\in[\tau_0,1]} \left (\int_0^\tau \widetilde{B}_r\,\mathrm{d}B_r \right) \big/\left(\int_0^\tau \widetilde{B}_r^2\,\mathrm{d}r\right)^{1/2},\]
where $\widetilde{B}_r := B_r - \tau^{-1}\int_{0}^\tau B_u \mathrm{d}u$ denotes a demeaned standard Brownian motion on $[0,\tau]$; this is the same limit as that of the original PWY statistic.
\end{theorem}

It is noteworthy that the asymptotic null distribution of the TVPWY
statistic under a general continuous-time process with nonparametrically
specified stochastic volatility and jumps is the same as the PWY statistic under a
discrete time AR model with homoskedastic errors. Since the
critical values of the PWY test have been tabulated in {\cite{phillips2011explosive}}, the result of Theorem \ref{thm2} implies that
those critical values can be used for the TVPWY test. 

\begin{remark}
By setting our truncation level $\phi_h\rightarrow\infty$, the numerator in the summand of (\ref{pseudosample}) becomes the daily log return, while the TV estimator in the denominator reduces to the classical RV estimator. Based on this pseudo-sample, an analogous RVPWY test can be defined. Clearly, the RVPWY test is not robust to the jumps, but can be useful for datasets where jumps are absent or have been pre-filtered out in advance. In the absence of the jump component, the RVPWY test also has the same asymptotic null distribution as the original PWY statistic. The working paper version of the present paper discusses the RVPWY in more detail.
\end{remark}

\begin{remark}
Theorem \ref{thm2} only describes the behavior of the TVPWY test statistic under the null hypothesis. Under the alternative hypothesis $\kappa >0$, the test statistic will diverge as $n \to \infty$,
such that the test rejecting for large values of TVPWY is consistent. This is a corollary to the asymptotic properties of the date-stamping procedure, analyzed in
Section \ref{sec:RTM}.
\end{remark}

\section{Monte Carlo Simulation}
In this section, we perform Monte Carlo simulations to investigate the size and power properties of the TVPWY test in finite samples and to compare these with those of existing tests. We also include the RVPWY test discussed in Section 3 to examine the impact of jumps. No existing explosive tests permit stochastic volatility and jumps under the null. Nevertheless, we include a few well-known tests for comparison in our simulation experiment. We include the classical PWY test, the recent Heteroskedasticity-Autocorrelation Robust test of \cite{lui2024robust} (LPY test hereafter), and the wild bootstrap PWY test (BTPWY test hereafter) proposed in \cite{harvey2015tests}. The BTPWY test is designed to be robust against unconditional, deterministic heteroskedasticity under the null, but recent developments in the bootstrap literature by \cite{BCGR21} suggest that the wild bootstrap might also provide asymptotically valid inference for the PWY test under stochastic volatility. 

We consider two models as our data-generating processes. We first consider the Heston model with jumps, where we find that the persistence
of the volatility process and the jump intensity play an essential role in the size and power performance of the tests. We then consider our model (\ref{tvou}) with a stationary fractional OU volatility model with jumps, where we find that the performance of our test is essentially unaffected by the roughness of the volatility process, as long as the process is stationary. In both simulations, we use 5,000 Monte Carlo replications, and report empirical size and power at the 5\% level.

\subsection{Heston model}\label{sec:heston}

We first consider the following Heston-type stochastic volatility model with jumps:
\begin{eqnarray*}
    z_t & = & y_t + J_t ,\\
    \mathrm{d} y_t & = & \kappa_t y_t \mathrm{d} t + \sigma_t \mathrm{d}
    W_t^1,\\
    \mathrm{d} \sigma_t^2 & = & a (b - \sigma_t^2) \mathrm{d} t + c
    \sqrt{\sigma_t^2} \mathrm{d} W_t^2,\\
    \mathd J_t & = & \xi_t \mathd N_t
\end{eqnarray*}
where $W_t^1$ and $W_t^2$ are two independent Brownian motions. The volatility process for the Heston model follows a so-called square-root diffusion process, which is non-negative if $\sigma_0^2>0, a>0, b>0$ and $c>0$.
Under these conditions, a stationary solution exists. The parameter $a$ determines the strength of the volatility mean-reversion,
and hence the persistence of shocks to the volatility. The parameter $b$ defines the mean variance, whereas the $c$  parameter defines the volatility of the variance process.
The process has strictly positive sample paths if the Feller condition $2ab > c^2$ is satisfied; in that case, the mean-reversion is strong enough relative to the
volatility-of-volatility $c$ to pull the volatility process back to the level $b$ when it approaches the lower bound $0$. We set the volatility model parameters to $(a,c)=(0.03, 0.10), (0.10,0.10), (0.10, 0.20)$, with $b=0.25$ fixed.
These parameters are taken from \cite{Bollerslev2002} and pertain to a unit time interval corresponding to one day. The parameter configuration $(a,c)=(0.03, 0.10)$
represents a highly persistent, near unit-root volatility process; $(a,c)=(0.10,0.10)$ corresponds to a less persistent variance process; $(a,c)=(0.10, 0.20)$ has a higher variance-of-variance and is close to the non-stationary region. The Feller condition is satisfied in all the three parameter settings.\footnote{In finite sample simulations, negative values of the variance process are still possible.
    In such cases, we take the absolute value of the variance to ensure that the square root operation always results in a real outcome.} We use $\xi_t
    \sim^{i.i.d.} N (- 0.1, 0.1)$ and $N_t$ is a Poisson process with jump
    intensity $\lambda \in \{ 0, 100, 200\}$. 
We simulate one year of daily observations by taking 252 trading days for a year, so $n = 252$. We take daily observations as the low-frequency sampling interval: $H=1$.
The high-frequency sampling interval is taken as $h = 1 / 288$, to simulate 288 five-minute observations within each trading day (assuming 24 trading hours).
Under the null hypothesis, it is assumed that $\kappa_t = 0$ for all $t$, while under the alternative hypothesis, $\kappa_t$ changes from 0 to $\kappa = 0.015$ at time $t$ equal to a fraction $\tau_1=0.5$ of the sample size.
In all the simulations, we set the truncation level locally for each low-frequency interval $s$ as $\phi_{h,s}=c\,\sqrt{\widehat{\mathrm{MedRV}}_s}\,h^{\varpi}$,
where $\widehat{\mathrm{MedRV}}_s$ is the jump-robust median realized variance estimator of \citet{andersen2012jump}: writing $r_{s,j}\equiv\Delta^h z_{t_{s-1,j}}$ for the $j$-th of the $M$ high-frequency returns
in interval $s$,
\[
\widehat{\mathrm{MedRV}}_s=\frac{\pi}{6-4\sqrt{3}+\pi}\,
\frac{M}{M-2}\sum_{j=2}^{M-1}\operatorname{med}\!\big(|r_{s,j-1}|,\,|r_{s,j}|,\,|r_{s,j+1}|\big)^2,
\]
which consistently estimates the local integrated variance $\int_{t_{s-1}}^{t_s}\sigma_u^2\,\mathrm{d} u$ in the presence of jumps.
We use $c=3$ and $\varpi=0.49$. This truncation rule is used throughout all simulation experiments and the empirical analyses.\footnote{Unreported simulations show that the results are fairly robust to variations
in this truncation rule. Results have been obtained for $c=3$ and $c=4$, also in case $\widehat{\mathrm{MedRV}}_s$ is replaced by the full-sample variance of high-frequency returns.
Across all settings, the two rules deliver almost identical size and power results: the largest discrepancy anywhere is about 0.010 (one percentage point). Detailed results are available from the authors upon request.}

\begin{table}[htbp]
    \centering
    \caption{Simulated Sizes of Tests, Heston Volatility}\label{tab:size_heston}
    \begin{tabular}{lccccc}
        \hline
        Parameters & PWY & BTPWY & LPY & RVPWY & TVPWY \\
        \hline
        \hline
        $(a,c)$&\multicolumn{5}{c}{0 Jumps per Year } \\
        \hline
        $(0.03,0.10)$ & 0.0950 & 0.0420 & 0.0516 & 0.0412 & 0.0420 \\
        $(0.10,0.10)$ & 0.0594 & 0.0416 & 0.0458 & 0.0410 & 0.0400 \\
        $(0.10,0.20)$ & 0.1042 & 0.0436 & 0.0476 & 0.0406 & 0.0416 \\
        \hline
        \hline
        $(a,c)$&\multicolumn{5}{c}{100 Jumps per Year } \\
        \hline
        $(0.03,0.10)$ & 0.4364 & 0.3256 & 0.0000 & 0.2608 & 0.0462 \\
        $(0.10,0.10)$ & 0.4372 & 0.3288 & 0.0000 & 0.2580 & 0.0454 \\
        $(0.10,0.20)$ & 0.4400 & 0.3210 & 0.0000 & 0.2612 & 0.0442 \\
        \hline
        \hline
        $(a,c)$&\multicolumn{5}{c}{200 Jumps per Year } \\
        \hline
        $(0.03,0.10)$ & 0.4860 & 0.4030 & 0.0000 & 0.3392 & 0.0440 \\
        $(0.10,0.10)$ & 0.4888 & 0.4034 & 0.0000 & 0.3384 & 0.0440 \\
        $(0.10,0.20)$ & 0.4862 & 0.4014 & 0.0000 & 0.3382 & 0.0436 \\
        \hline
        \hline
\end{tabular}

\end{table}

\begin{table}[htbp]
    \centering
    \caption{Simulated Powers of Tests, Heston Volatility}\label{tab:power_heston}
    \begin{tabular}{lccccc}
        \hline
        Parameters & PWY & BTPWY & LPY & RVPWY & TVPWY \\
        \hline
        \hline
        $(a,c)$&\multicolumn{5}{c}{0 Jumps per Year } \\
        \hline
        $(0.03,0.10)$ & 0.6834 & 0.6362 & 0.5884 & 0.7224 & 0.7206 \\
        $(0.10,0.10)$ & 0.6790 & 0.6564 & 0.5978 & 0.6968 & 0.6896 \\
        $(0.10,0.20)$ & 0.6860 & 0.6210 & 0.5986 & 0.7500 & 0.7512 \\
        \hline
        \hline
        $(a,c)$&\multicolumn{5}{c}{100 Jumps per Year } \\
        \hline
        $(0.03,0.10)$ & 0.4992 & 0.4004 & 0.0002 & 0.6032 & 0.7204 \\
        $(0.10,0.10)$ & 0.4960 & 0.4028 & 0.0004 & 0.5728 & 0.6952 \\
        $(0.10,0.20)$ & 0.5000 & 0.3984 & 0.0004 & 0.6246 & 0.7504 \\
        \hline
        \hline
        $(a,c)$&\multicolumn{5}{c}{200 Jumps per Year } \\
        \hline
        $(0.03,0.10)$ & 0.5102 & 0.4438 & 0.0000 & 0.5292 & 0.7254 \\
        $(0.10,0.10)$ & 0.5152 & 0.4482 & 0.0000 & 0.5038 & 0.6906 \\
        $(0.10,0.20)$ & 0.5120 & 0.4404 & 0.0000 & 0.5416 & 0.7632 \\
        \hline
        \hline
\end{tabular}

\end{table}

Using the finite sample critical values for $n=200$, $r_0=0.137$ in \cite{phillips2011explosive}, we calculate the empirical rejection frequency of the classical PWY test,
the BTPWY test, the LPY test, the RVPWY test, and the TVPWY test at a 5\% level. The empirical size is given in Table \ref{tab:size_heston} under the volatility and jump intensity parameter settings (i.e., $(a,b,c)$ and $\lambda$). In the zero-jump scenario, the classical PWY test is noticeably oversized when the volatility process is persistent (i.e. $(a,c)=(0.03, 0.10)$) and when the variance-of-variance is high (i.e. $(a,c)=(0.10, 0.20)$), but otherwise maintains size close to the nominal level (i.e. when $(a,c)=(0.10, 0.10)$). This shows that the PWY test has a limited level of robustness to stationary stochastic volatility as long as the volatility process is not very persistent and variance-of-variance is not very high. The BTPWY test and the LPY test demonstrate more robustness to stochastic volatility than the original PWY test, and maintain size close to the nominal level for all the volatility parameter settings we consider, even though they are not designed to accommodate stochastic volatility. As expected, RVPWY test and TVPWY test both have size controlled in all the volatility parameter settings as they are designed for this purpose. Situation changes when jumps are present. We see that as the annual jump intensity rises to 100 or 200, the PWY and BTPWY tests exhibit substantial over-rejection, with simulated sizes rising to around $30-50\%$, reflecting a severe inflation in false positives. The LPY test, in contrast, shows severe under-rejection, with a simulated size (virtually) equal to 0 in all cases.
The RVPWY test by design is not robust to jumps, and also shows serious over-rejection, with a simulated size of around $26$--$34\%$.
By contrast, the proposed jump-robust TVPWY test has a much more stable performance in the presence of frequent jumps, consistently producing empirical sizes around and mostly below the nominal level, even when the jump frequency is high. 

 In Table \ref{tab:power_heston}, we present empirical powers of the same tests considered previously, against the alternative with $\kappa_t = 0.015 \times \mathbf{1}\{t/n > 0.5\}$.
 For the zero-jump case, although the original PWY test is more oversized than the TVPWY test, indicating a higher Type I error probability, its power is also lower compared with the TVPWY test. This means that it is inferior to the TVPWY test in terms of both size control (Type I error) and power (Type II error). The BTPWY test and the LPY test also show clearly lower empirical power than the TVPWY test. For the relative performance to the RVPWY test, it is interesting to notice that the RVPWY and TVPWY tests have almost identical empirical power across different volatility specifications, showing that even when there are no jumps, directly using the jump-robust TVPWY test is not a worse choice than the RVPWY test in terms of the rejection power. As the jump frequency increases (to 100 or 200 per year), the PWY, BTPWY, LPY and RVPWY tests all have empirical rejection probability reduced relative to the no-jump case. In particular, the LPY seems to become very conservative in the presence of jumps and essentially shows no testing power, with empirical rejection rate essentially zero. Nevertheless,  the TVPWY test preserves consistently high power (typically above 68\%). This pattern indicates that our proposed TVPWY test can remain robust in size under various realistic jump intensities when no explosive regime is in presence, and effectively detects explosive behavior when it exists without suffering power losses (when compared with the no-jump case), making it the preferred choice among the considered tests when jumps are present.

\subsection{Rough Volatility}
The so-called rough volatility models have gained increasing popularity in the recent literature due to their superior performance in forecasting volatility.
See, e.g., \cite{gatheral2018} and \cite{wang2023}.  Volatility processes in such models do not possess the Markovian property, unlike classical diffusion-type volatility models.
Non-Markovian volatility models are permitted by our assumptions. In this section, we demonstrate the validity of our test for this class of models.
We consider our model (\ref{tvou}) with the log volatility following a fractional Ornstein-Uhlenbeck (fOU) process
\begin{equation} 
    \mathd \ln \sigma_t^2 = a (b - \ln \sigma^2_t) \mathd t + c \mathd B_t^{\mathbb{H}}, \label{fou}
\end{equation}
where $\{ B_t^{\mathbb{H}} \}$ is a fractional Brownian motion (fBm) process with the Hurst index
${\mathbb{H}}$. When ${\mathbb{H}} \in (0, 1 / 2)$, the increments of $B_t^{\mathbb{H}}$ are negatively
correlated; while when ${\mathbb{H}} \in (1 / 2, 1)$, the increments of $B_t^{\mathbb{H}}$ are
positively correlated. When ${\mathbb{H} = 1 / 2}$, 
$\{B_t^{\mathbb{H}} \}$ becomes a standard Brownian motion. Lower values of ${\mathbb{H}}$ correspond to rougher sample path of both $B_t^{\mathbb{H}}$ and $\ln \sigma_t^2$. 

\begin{table}[htbp]
    \centering
    \caption{Simulated Sizes of Tests, Rough Volatility}\label{tab:size_fou}
    \begin{tabular}{lccccc}
        \hline
        Parameters & PWY & BTPWY & LPY & RVPWY & TVPWY \\
        \hline
        \hline
        $(a,b,c,{\mathbb{H}})$&\multicolumn{5}{c}{0 Jumps per Year } \\
        \hline
        $(0.24,2.42,0.70,0.13)$ & 0.0460 & 0.0404 & 0.0440 & 0.0474 & 0.0478 \\
        $(0.02,2.43,0.57,0.83)$ & 0.0498 & 0.0410 & 0.0488 & 0.0474 & 0.0478 \\
        $(2.92,2.52,1.05,0.23)$ & 0.0710 & 0.0428 & 0.0546 & 0.0476 & 0.0450 \\
        \hline
        \hline
        $(a,b,c,{\mathbb{H}})$&\multicolumn{5}{c}{100 Jumps per Year } \\
        \hline
        $(0.24,2.42,0.70,0.13)$ & 0.1000 & 0.0894 & 0.0412 & 0.1178 & 0.0476 \\
        $(0.02,2.43,0.57,0.83)$ & 0.0748 & 0.0648 & 0.0496 & 0.0744 & 0.0504 \\
        $(2.92,2.52,1.05,0.23)$ & 0.2144 & 0.1684 & 0.0260 & 0.2272 & 0.0438 \\
        \hline
        \hline
        $(a,b,c,{\mathbb{H}})$&\multicolumn{5}{c}{200 Jumps per Year } \\
        \hline
        $(0.24,2.42,0.70,0.13)$ & 0.1938 & 0.1786 & 0.0200 & 0.2518 & 0.0646 \\
        $(0.02,2.43,0.57,0.83)$ & 0.1388 & 0.1210 & 0.0428 & 0.1468 & 0.0726 \\
        $(2.92,2.52,1.05,0.23)$ & 0.3536 & 0.2942 & 0.0018 & 0.4008 & 0.0506 \\
        \hline
        \hline
\end{tabular}

\end{table}

\begin{table}[htbp]
    \centering
    \caption{Simulated Powers of Tests, Rough Volatility}\label{tab:power_fou}
    \begin{tabular}{lccccc}
        \hline
        Parameters & PWY & BTPWY & LPY & RVPWY & TVPWY \\
        \hline
        \hline
        $(a,b,c,{\mathbb{H}})$&\multicolumn{5}{c}{0 Jumps per Year } \\
        \hline
        $(0.24,2.42,0.70,0.13)$ & 0.6736 & 0.6720 & 0.5698 & 0.7052 & 0.7080 \\
        $(0.02,2.43,0.57,0.83)$ & 0.6690 & 0.6658 & 0.5760 & 0.6778 & 0.6742 \\
        $(2.92,2.52,1.05,0.23)$ & 0.6784 & 0.6500 & 0.5786 & 0.6856 & 0.6842 \\
        \hline
        \hline
        $(a,b,c,{\mathbb{H}})$&\multicolumn{5}{c}{100 Jumps per Year } \\
        \hline
        $(0.24,2.42,0.70,0.13)$ & 0.6540 & 0.6520 & 0.3988 & 0.6896 & 0.7066 \\
        $(0.02,2.43,0.57,0.83)$ & 0.6676 & 0.6588 & 0.4684 & 0.6736 & 0.6738 \\
        $(2.92,2.52,1.05,0.23)$ & 0.6360 & 0.6082 & 0.2712 & 0.6532 & 0.6886 \\
        \hline
        \hline
        $(a,b,c,{\mathbb{H}})$&\multicolumn{5}{c}{200 Jumps per Year } \\
        \hline
        $(0.24,2.42,0.70,0.13)$ & 0.6284 & 0.6192 & 0.2658 & 0.6782 & 0.7026 \\
        $(0.02,2.43,0.57,0.83)$ & 0.6520 & 0.6404 & 0.3422 & 0.6562 & 0.6694 \\
        $(2.92,2.52,1.05,0.23)$ & 0.6012 & 0.5586 & 0.1242 & 0.6310 & 0.6868 \\
        \hline
        \hline
\end{tabular}

\end{table}

We use three parameter settings $(a,b,c,\mathbb{H})= (0.24,2.42,0.70,0.13)$, $(0.02,2.43,0.57,0.83)$, $(2.92,2.52,1.05,0.23)$. These parameter values are taken from \cite{wang2023}'s estimation results
for real data (i.e.\ \cite{wang2023}, Table 5 for S\&P 500, DJIA, and S\&P 100 series). 
These parameters pertain to a unit time interval corresponding to one year and deliver a stationary volatility process regardless of the Hurst index $\mathbb{H}$ value.
Assuming 6.5 trading hours a day, we simulate one year of 5-minute observations and compute the empirical size of all tests under the null hypothesis of no explosive regime.
We use the same $\kappa$ as in the Heston model simulations for the power simulation. However, since the unit time interval is now one year, the value for $\kappa$ is adjusted to $0.015\times 252$. The jump size, jump intensity specifications and the truncation level used are identical to those of the Heston model case. 

We present the simulated size in Table \ref{tab:size_fou}. We can see that in the zero-jump scenario, all tests remain reasonably accurate, with empirical sizes mostly falling within the $4$--$6\%$ range
across the three roughness settings. The exception is the PWY test, which is slightly oversized at 7.1\% when $(a,b,c,\mathbb{H})=(2.92,2.52,1.05,0.23)$.
However, as the annual jump frequency increases to 100, the PWY, BTPWY, and also RVPWY display size inflation, sometimes exceeding 10\% and rising more sharply under the highest jump intensity (200 jumps per year) case.
By contrast, the proposed TVPWY test generally shows more robust performance, keeping empirical sizes near or below the $5$--$8\%$ range. 

Table \ref{tab:power_fou} presents the power of the same tests against the alternative $\kappa_t = \kappa \times \mathbf{1}\{t/n > \tau_1\}$, where $\tau_1 = 0.5$ as before, but $\kappa$
is now given by $\kappa=0.015\times 252$ to accommodate the unit time interval of one year.
Overall, the TVPWY test exhibits strong detection rates (roughly $0.67$--$0.71$) across the three parameterizations. Increasing the jump intensity does not have an effect on the simulated power. By contrast,  the PWY and BTPWY, although only occasionally valid in size, still show a mild power disadvantage compared with the TVPWY test.  The LPY test is always valid in terms of size, but it suffers considerable power deterioration when the jumps are high. These patterns are in line with what we have observed in the Heston model simulation.

\section{Real-time Monitoring}
\label{sec:RTM}
PWY propose a strategy to date-stamp the start and the end of an explosive regime in real time.
Analogous to PWY, a date-stamping strategy can be constructed based on our TVDF statistic, by comparing the sequence of test statistics $\text{TVDF}_{\tau}$, $\tau\in[\tau_0,1]$,
to the right-tailed critical values of the asymptotic distribution of the standard Dickey-Fuller $t$-statistic.
In particular, letting $\tau_e$ denote the origination date and $\tau_f$ the conclusion date of the explosive period, then analogous to PWY's date-stamping strategy, the following estimators can be defined:
\begin{equation}
    \hat{\tau}_e = \inf_{\tau\geqslant \tau_0}\{\tau: \text{TVDF}_{\tau}>\tmop{cv}_{\beta_n}\}, \quad \hat{\tau}_f = \inf_{\tau\geqslant \hat{\tau}_e+\frac{\log(n)}{n}}\{\tau:  \text{TVDF}_{\tau}<\tmop{cv}_{\beta_n}\},
\end{equation}
where $\tmop{cv}_{\beta_n}$ is the right-side critical value of the DF $t$ distribution corresponding to a significance level of $\beta_n$; $\hat{\tau}_e$ is the first upcrossing time of the monitoring statistic $\text{TVDF}_{\tau}$ over the critical line, and serves as an estimator for the origination date of the explosive regime; while $\hat{\tau}_f$ is the first  downcrossing time of $\text{TVDF}_{\tau}$ with the critical value line after upcrossing, and is used as an estimator for the conclusion time of the explosive regime. In PWY, to achieve consistent estimation for the date stamps $\{\hat{\tau}_e,\hat{\tau}_f\}$,
the significance level $\beta_n$ needs to slowly approach zero asymptotically, and correspondingly $\tmop{cv}_{\beta_n}$ must slowly diverge to infinity.
In practical implementations, PWY recommend setting the significance level in the $1$--$5\%$ range. 

In our analysis, similar to the result in PWY, we find that the upcrossing time $\hat{\tau}_e$ consistently estimates the start date of the explosive regime. However, a downcrossing of the monitoring statistic does not consistently estimate the conclusion time of the explosive regime. Recall that Phillips (2011) considers a model where a bubble collapses back in one period to the level present at the onset of the bubble. In contrast, our truncation mechanism, which is necessary to ensure robustness against jumps, effectively removes such rapid crashes from the constructed pseudo-sample. As a result, \(\hat{\tau}_f\) is less likely to identify the end time of the explosive regime promptly. This is the trade-off inherent in accommodating jumps. 

To study the statistical properties of our date-stamping procedure, we make some further assumptions regarding the persistence parameter $\kappa$ of the time-varying OU process in (\ref{tvou}). The observed log asset price process is still (\ref{OUJ}).
\begin{description}
    \item[A5]  The persistence parameter satisfies $\kappa = c/n^{\alpha}$ with $c > 0$, $0 \leqslant \alpha < 1$.  
\end{description}
The persistence parameter $\kappa$ specifies the magnitude of
deviation from the unit root model of the explosive regime. Assumption A5 includes two cases: the mildly explosive case, where $0<\alpha<1$, and the fixed case, where $\alpha=0$. With this specification of the persistence
parameter, the continuous-time process starts with a unit root regime, then
changes to be {purely explosive or mildly explosive} at time $t_{i_1}$. It could revert to a
unit root regime at a later time, but in this section, our focus is to estimate the onset time of the explosive regime, so we keep this explosive specification until the end of the sample. 

We first show that asymptotically the date stamping estimator $\hat{\tau}_e$ does not signal an origination date if an explosive regime is absent.
\begin{theorem}
    \label{thm3}Under  the null hypothesis of no
    explosive behavior (that is, the continuous-time model given by (\ref{OUJ}), (\ref{jumps}), and (\ref{tvou}) with $\kappa=0$), suppose assumptions A1-A4 hold. Then if
    $\tmop{cv}_{\beta_n} \rightarrow \infty$, the probability of detecting the
    origination of the explosiveness using the TVDF statistic is 0. That is, $P (\hat{\tau}_e \in [\tau_0, 1]) \rightarrow 0$, as $n \rightarrow
    \infty$.
\end{theorem}

Let $\tau_1=i_1/n$ denote the fractional origination date of the explosive regime.
We show in the next theorem that our date stamping estimator $\hat{\tau}_e$ is consistent in estimating the true onset time $\tau_1$, for which purpose we make the following additional assumptions.
\begin{description}
    \item[A6] $n h \log^2(n)    \rightarrow 0$.
    \item[A7] The volatility process is uniformly bounded from above.
\end{description}

\begin{theorem}
    \label{thm4}Under the alternative hypothesis that an explosive regime is
    present (that is, the continuous-time model given by (\ref{OUJ}),
    (\ref{jumps}), and (\ref{tvou})), suppose assumptions A1-A7
    hold. We have $\tmop{TVDF}_{\tau} = O_p (1)$ when $\tau_0 < \tau \leqslant
    \tau_1$; after the inception of the explosive regime, $\tmop{TVDF}_{\tau}$ tends to
    increase, and by the time $\tau_1^{\ast} = \tau_1 + \frac{1}{nH} 
    \left\lfloor \frac{1}{\kappa} \log \left( \frac{\log(n)}{\kappa} \right)
    \right\rfloor$, $\tmop{TVDF}_{\tau}$ diverges. When the explosive magnitude parameter $0 \leqslant \alpha < 1/2$, $\tmop{TVDF}_{\tau}$ diverges at a rate $O_p (n^{(1+\alpha) / 2} \log(n) \wedge n^{2\alpha} \log^2(n))$; while when $1/2\leqslant \alpha < 1$, $\tmop{TVDF}_{\tau}$ diverges at a rate $O_p (n^{1-\alpha/2} )$.

    Consequently, for an unknown $\kappa$, if the
    sequence of critical values $\tmop{cv}_{\beta_n}$ satisfies
    \[ \frac{1}{\tmop{cv}_{\beta_n}} + \frac{\tmop{cv}_{\beta_n}}{ \log^2(n)} \rightarrow 0, \]
    then $\hat{\tau}_e \overset{p}{\rightarrow} \tau_1$ as $n \rightarrow
    \infty$.
\end{theorem}

\begin{remark}
    In PWY, the date-stamping statistic is shown to be diverging after a fixed fraction of time under the specification that $\kappa=c/n^{\alpha}$, with $c>0,\alpha \in (0,1)$. In Theorem \ref{thm4}, we show a similar result for $\alpha \in [0,1)$. We also strengthen PWY's result by showing that the date-stamping statistic diverges after a diminishing fraction of time $\frac{1}{n H} \left\lfloor \frac{1}{\kappa} \log \left(
\frac{\log(n)}{\kappa} \right) \right\rfloor$ (because $\kappa n \rightarrow \infty$). 

It is also interesting to observe that the length of this fraction depends on the strength of the explosiveness signal. When $\kappa$ is larger, i.e. when $\kappa = c/n^{\alpha}$ with $\alpha$ close to 0 in the range $[0,1)$, the diminishing fraction $\frac{1}{n H} \left\lfloor \frac{1}{\kappa} \log \left(
\frac{\log(n)}{\kappa} \right) \right\rfloor$ is smaller. This means that stronger explosive signal requires shorter time to be found. On the other hand, when  $\kappa$ is smaller, e.g. when $\kappa = c/n^{\alpha}$ with $\alpha$ close to 1 in the range $[0,1)$, the fraction is larger, meaning that it takes longer for the detector to find the milder explosiveness.

The slowest rate of divergence of the monitoring statistic is $\log^2(n)$. It happens when $\kappa=c$ is a fixed constant (i.e. $\alpha=0$). Although this may appear counter-intuitive that the speed of monitoring statistic diverges is slowest when the magnitude of explosiveness is strongest, we note that from the discussion in the previous paragraph, this divergence happens after a shortest period of fraction time with order $\log(\log(n))/n$.

Our test statistic and the date-stamping statistic are both based on a modified DF $t$ statistic, where the error variance is not estimated but imposed to be one. This modification plays an instrumental role for proving the consistency of our date-stamping procedure. If the standard DF $t$ statistic is used and the error variance is estimated, the estimated error variance will diverge and make the detecting statistic smaller. Imposing the error variance to be one inflates the statistic such that we can estimate the origination time of the explosive regime at the present diminishing fraction of time. 

\end{remark}

\begin{remark}
    The slightly stronger condition involving the sampling interval in A6 is to ensure the uniform consistency of the TV estimator before time $\tau_1^*$ under the alternative hypothesis. The uniform boundedness assumption of the volatility path is made for technical convenience when analyzing the real-time monitoring procedure under the alternative hypothesis. We acknowledge that it is not satisfied by the DGPs used in the simulation studies. Notice that this assumption is not needed for the validity of the TVPWY test (Theorem 2) and the monitoring procedure (Theorem 3). We conjecture that, given substantial additional technical effort, this condition could be relaxed to certain types of
    moment restrictions. We leave this technical extension for future research.
\end{remark}

\section{Empirical Studies}

\subsection{Explosiveness in Cryptocurrencies}
Cryptocurrencies have emerged as a main innovation in financial markets. As of 2022, the market capitalization of cryptocurrencies was approximately USD 1.8 trillion,
with Bitcoin (BTC) and Ethereum (ETH) making up the majority of that sum, with market capitalizations of approximately USD 750 billion and USD 350 billion, respectively.

Traditionally, cryptocurrencies, especially the decentralized ones, were regarded as financial assets without fundamental value.
In recent years, a strand of literature regarding the economics of using and valuing crypto-tokens has emerged; see \cite*{cong2021a}, \cite*{cong2021b}, \cite{cong2022}, \cite{malinova2023} and \cite{sockin2023}).
For example, \cite{malinova2023} shows that optimally designed token contracts can yield the same payoffs as equity, and, with entrepreneurial moral hazard, tokens can finance a strictly larger set of ventures than equity.
\cite{sockin2023} uses cryptocurrencies as utility tokens to design a decentralized platform to facilitate transactions between users of certain goods or services.
See \cite{biais2023} for a survey of studies published in a special issue of \emph{Management Science} on the topic of cryptocurrencies.

Regardless of the debate over whether cryptocurrencies have fundamental value, it is interesting to know if they exhibit explosive behavior for investment or regulation.
When intra-day cryptocurrency price data is available, it offers the opportunity to derive highly accurate daily volatility measures.
In this empirical application, we use our methodology to detect explosive behavior in cryptocurrencies.
Specifically, we focus on the 24-hour 5-minute data of the two largest cryptocurrencies, BTC and ETH, during the period from January 1, 2018, to August 31, 2021. The data are obtained from Binance. We employ the TVPWY test and the TVDF detector to test and date-stamp potentially explosive behavior in cryptocurrencies.
To facilitate comparison, we also include the results obtained from the PWY test and its associated DF detector.
We use $\tau_0=0.1$ in all our calculations, roughly corresponding to 4.5 months.

\begin{figure}[H]
    \centering
    \caption{Log-Price and Pseudo-Log-Price of Cryptocurrencies}
    \includegraphics[width=.9\textwidth]{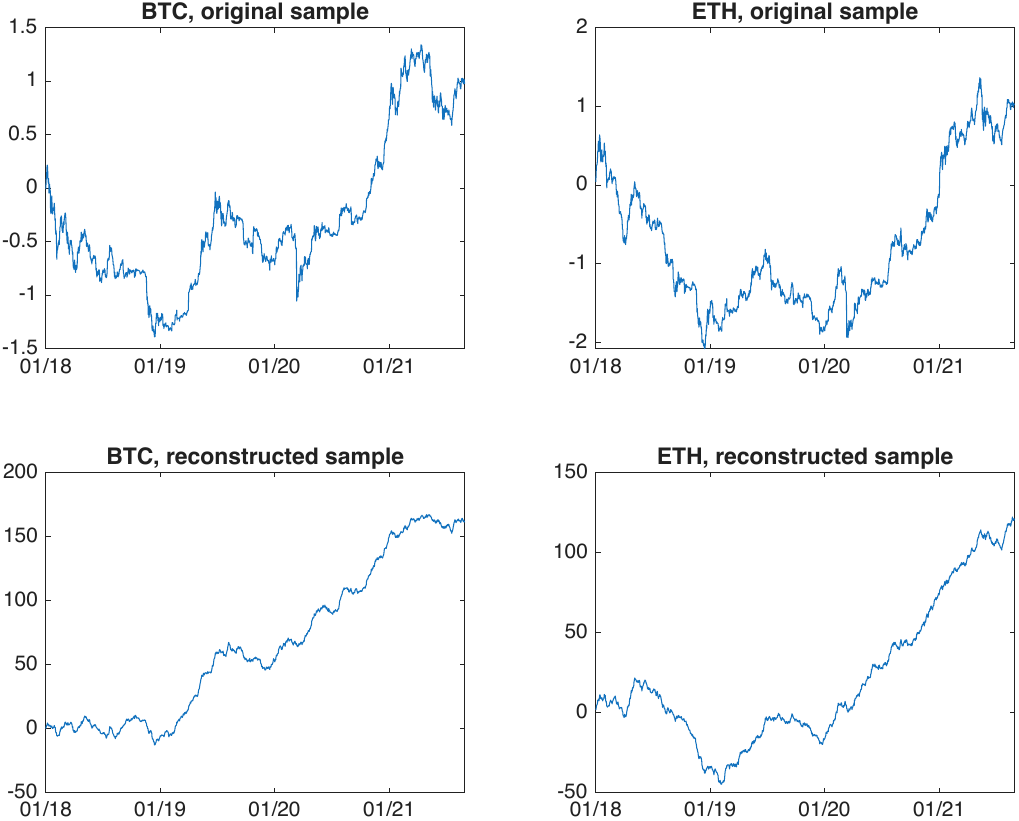}
    \label{fig:lpri}
\end{figure} 

In Figure \ref{fig:lpri}, we plot the log-price series of the two cryptocurrencies. To be transparent about the devolatization and truncation in our proposed methodology, we also plot the pseudo-log-price series for both currencies to facilitate comparison. In Figure \ref{fig:lpri}, the initial points of the original series are normalized to zero to facilitate comparison with the reconstructed samples. 

It can be seen that for the BTC series, the decrease in price at the beginning of the original sample becomes milder in magnitude,
while the price increases observed in the latter part of the original sample are largely preserved.
The plots of the ETH series and its reconstructed sample show a similar pattern.
The difference is due to the higher volatility during the price's decreasing phase. The normalized original series in Figure \ref{fig:lpri} are also used to compute the PWY statistics and the associated DF detectors. Since both statistics are based on a ``with intercept'' version of the DF statistic, they are invariant to the normalization used. Table \ref{empiricaltable} gives the values of the test statistics. The PWY test fails to identify explosive behavior in either price series, while the
TVPWY test finds explosive behavior in both the BTC series and the ETH series at the 1\% level. 

\begin{table}[t]
    \centering
    \begin{threeparttable}
        \caption{Explosiveness Test Statistics and $p$-values}
        \label{empiricaltable}
        \begin{tabular*}{0.7\textwidth}{@{\extracolsep{\fill}} l c c @{}}
            \hline
            & \textbf{PWY} & \textbf{TVPWY}\\
            \hline
            \textbf{BTC}&$0.603$&$3.521^{***}$\\
            \textbf{ETH}&$0.533$&$3.555^{***}$\\
            \hline
\end{tabular*}

        \begin{tablenotes}
            \footnotesize
            \item \emph{Notes}: Critical values are 2.094, 1.468 and 1.184 for the 1\%,
            5\% and 10\% significance levels, respectively, for both tests.
            Rejection at these significance levels is denoted as ***, **, *, respectively.
        \end{tablenotes}
    \end{threeparttable}
\end{table}

\begin{figure}[H]
    \centering
    \caption{DF and TVDF Detectors} \label{dfdate}
    \includegraphics[width=.85\textwidth]{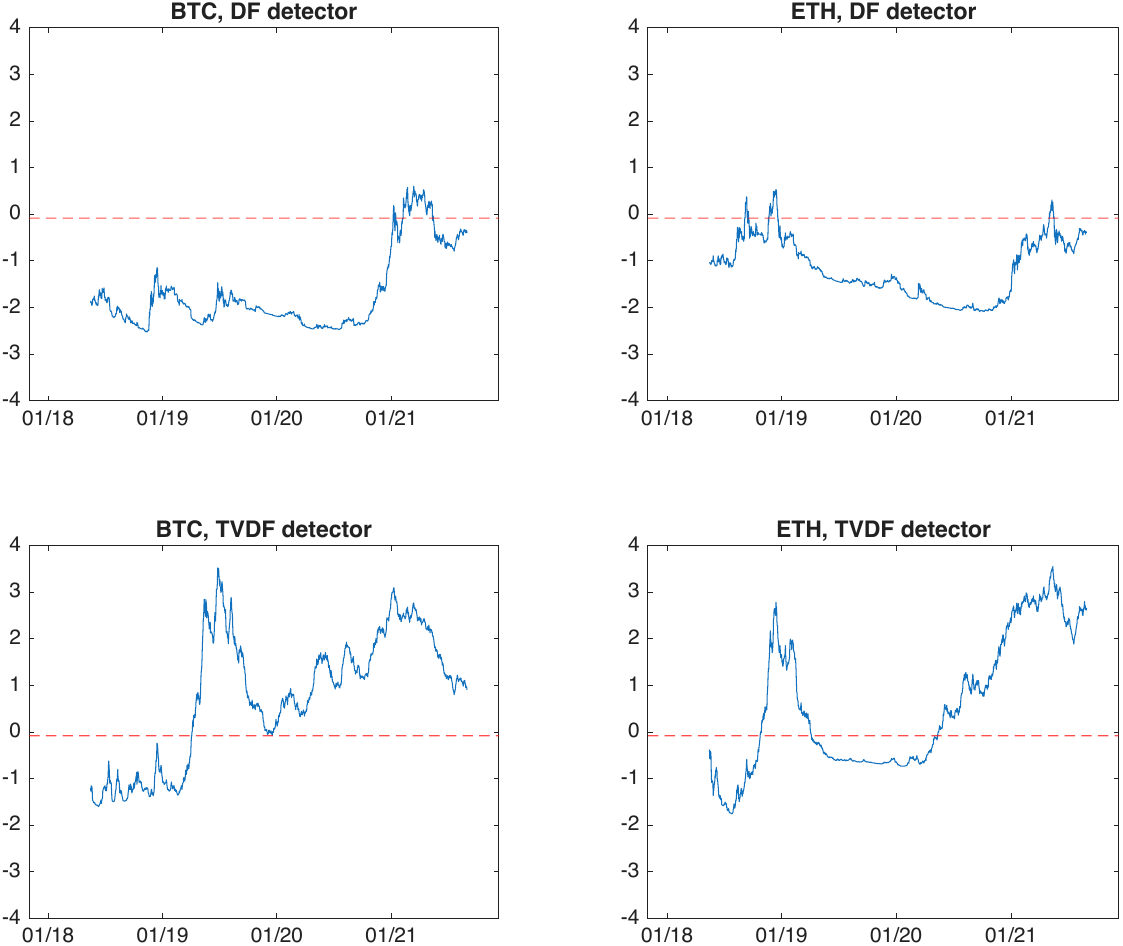}
\end{figure}

We also plot the evolution of the TVDF detector and the DF detector in Figure \ref{dfdate}, together with the critical values of the detector,
which is $-0.08$, the 5\% critical value of the DF distribution. This helps us better understand the difference in the values of the TVDF and TVPWY
statistics and the corresponding DF and PWY statistics. The BTC series underwent two periods of noticeable price increases during the sample: the first period started in early 2019, after a crash at the end of 2018, and lasted for about six months; the second period started after the rapid crash in early 2020 and lasted for about one year. In the reconstructed sample, the fast price decreases preceding both price increase periods become milder in magnitude, such that the two price increase periods become more pronounced. The DF detector, however, only identifies the second fast-increasing period, with a significant delay: the longest period that it stays above the critical line is from February 8, 2021 to May 11, 2021. This brief period is shorter than the 4.5-month minimum window required to calculate the subsample statistics. Furthermore, the maximum of the DF detector for the full sample (0.603) remains below the 10\% critical value. Consequently, we conclude that the DF detector does not find explosive behavior
in this series. In contrast, our TVDF detector crosses the $-0.08$ critical value line since the first period of rapid price increase for the BTC: with a few light touches of the critical line in 2018, it clearly crosses the critical value line from April 4, 2019 and remains above the line until the end of the sample. The maximum of the TVDF statistic, 3.521, is reached around mid 2019, implying a highly significant (1\% level) rejection of the no-explosive hypothesis.

During the same period, the ETH series experienced a noticeable decline from the start of the sample. After a short and mild price recovery starting from late 2018, followed by a decrease, both trends lasted for about six months in the first and second halves of 2019. The ETH price then entered an extended period of increasing prices after a quick crash in the first quarter of 2020. In the reconstructed sample, we observe similar changes to the BTC series, where we find that the decreasing parts of the series become milder and the increasing periods become more pronounced. The PWY and PSY statistics \citep{Phillips2013, Phillips2013a} are known to flag fast downturns as explosive regimes \citep{phillipsshi2020,wangyu2023}. This feature is confirmed when applying the DF detector to the ETH data: the statistic crosses the critical value line several times during the late-2018 price decline. These crossings are all short-lived, however---the longest lasting only about a month---and none is long enough to qualify as an explosive episode. The clearest evidence from our TVDF detector concerns the extended price increase following the first quarter of 2020. The detector also responds to the sharp late-2018 decline, but we do not interpret this as explosive behavior, as it coincides with a crash rather than a price run-up. The DF detector fails to identify the extended increasing period of the ETH price after the first quarter of 2020: it only briefly crosses the $-0.08$ critical line with a long delay in 2021. It stays above the critical line for a few days. Our TVDF detector shows strong evidence of explosiveness for this run-up, it crosses the critical value line on May 12, 2020, with only a short delay, and remains above the critical line towards the end of the sample. The maximum of the TVDF detector reaches 3.555 during this period, ensuring a highly significant rejection of the non-explosiveness hypothesis at the 1\% level.

From the above analysis of the BTC and ETH prices using the classical method and our new method, we find that our test and real-time detector provide an appealing approach to explosiveness identification. 

\subsection{Explosiveness in the monthly S\&P 500 index during the 90s}

The dot-com bubble was a historical economic bubble characterized by excessive speculation during the late 1990s, a time marked by rapid growth in the use and adoption of the Internet by both businesses and consumers. PWY has analyzed the monthly real NASDAQ index and found explosive behavior. In this study, we apply our proposed methodology to investigate potential explosive behavior in the nominal and real S\&P 500 indices during this period. As in previous research, our focus is on the explosive behavior observed in the monthly S\&P 500 indices. By utilizing the methodology presented here, we take advantage of the volatility information contained in the daily S\&P 500 price data, which is widely available but has not been utilized in past analyses.

 The daily level nominal S\&P 500 index is widely available and therefore our methodology can be applied directly. For the real S\&P 500 index, our methodology cannot be used directly because CPI data, which is needed to calculate the real index, is only released at a monthly frequency. We will discuss an empirically meaningful approximation strategy for the monthly volatility of the real S\&P 500 index. The sample period we use is from January 1990 to June 2005. The daily level S\&P 500 index data is downloaded from Yahoo Finance. The monthly level US CPI data is downloaded from the Federal Reserve website. Since we use daily S\&P 500 returns to estimate monthly volatility, accounting for the mean is necessary; therefore, we employ demeaned daily S\&P 500 returns (post-truncation) in calculating the TV estimator.
 
 As with most daily returns, jumps are very rare in the daily nominal and real S\&P 500 indices in our sample. According to our discussion in the last remark of Section 3, we can also use the RVPWY test and the associated real-time detector in this application. We implemented this and found almost identical results as our reported results for TVPWY test and detector below. We omit these results to avoid repetition.  

\subsubsection{Monthly volatility of S\&P500 index}
For the monthly volatility of the real S\&P 500 index, the daily level real S\&P 500 index is not available because of the unavailability of daily level CPI data. Here, we discuss an empirically sound approximation strategy, where the monthly volatility of the real S\&P 500 series can be approximated by the nominal monthly volatility of S\&P 500. 

Denote the nominal S\&P 500 index as $P_t$, and the corresponding consumer price index (CPI) by $\mathit{CPI}_t$. The monthly real S\&P 500 index $RP_t$ is then defined as
\[RP_t=\frac{P_t}{\mathit{CPI}_t}.\]
Then the monthly real S\&P 500 log return is
\[\Delta \log(RP_t)= \Delta \log(P_t) - \Delta \log(\mathit{CPI}_t).\]
Computing the variance on both sides, we have the variance of the monthly S\&P return as 
\[\mathrm{Var}(\Delta \log(RP_t))= \mathrm{Var}(\Delta \log(P_t)) + \mathrm{Var}(\Delta \log(\mathit{CPI}_t))-2\mathrm{Cov}(\Delta \log(P_t),\Delta \log(\mathit{CPI}_t)),\]
which is a sum of the variance of monthly nominal S\&P 500 returns, the variance of monthly CPI inflation, and their covariance. The variance of monthly nominal S\&P 500 returns can be estimated using the daily nominal S\&P 500 prices and the realized variance estimator; however, the CPI data is only available at a monthly frequency, so the other two terms cannot be estimated using the same methodology. Empirical observation confirms that the CPI, which represents the price level for consumer goods and services, exhibits significantly less variability than the stock price index. This can be seen in the variance-covariance matrix of the monthly nominal S\&P 500 return and the monthly CPI inflation, given in Table \ref{tb:cov}. From the table, we can see that at the monthly level, the CPI variance and the CPI-S\&P 500 covariance are approximately 3 orders of magnitude smaller than the S\&P 500 variance. Based on this empirical observation,  we assume  $\mathrm{Var}(\Delta \log(\mathit{CPI}_t))\ll \mathrm{Var}(\Delta \log(P_t))$ and $\mathrm{Cov}(\Delta \log(P_t),\Delta \log(\mathit{CPI}_t)) \ll \mathrm{Var}(\Delta \log(P_t))$, implying that 
\[\mathrm{Var}(\Delta \log(RP_t))\approx \mathrm{Var}(\Delta \log(P_t)).\] 

\begin{table}[H]
    \centering
    \caption{Variance-covariance matrix of monthly nominal S\&P 500 return and monthly CPI return, both expressed as percentages, from January 1990 to June 2005.}
    \label{tb:cov}
    \begin{tabular*}{0.7\textwidth}{@{\extracolsep{\fill}} lcc @{}}
        \toprule
        & \textbf{S\&P 500 Return} & \textbf{CPI Inflation} \\
        \midrule
        \textbf{S\&P 500 Return} & 15.69 & -0.03 \\
        \textbf{CPI Inflation}   & -0.03 & 0.03 \\
        \bottomrule
\end{tabular*}

\end{table}

\begin{figure}[H]
    \centering
    \includegraphics[width=0.9\linewidth]{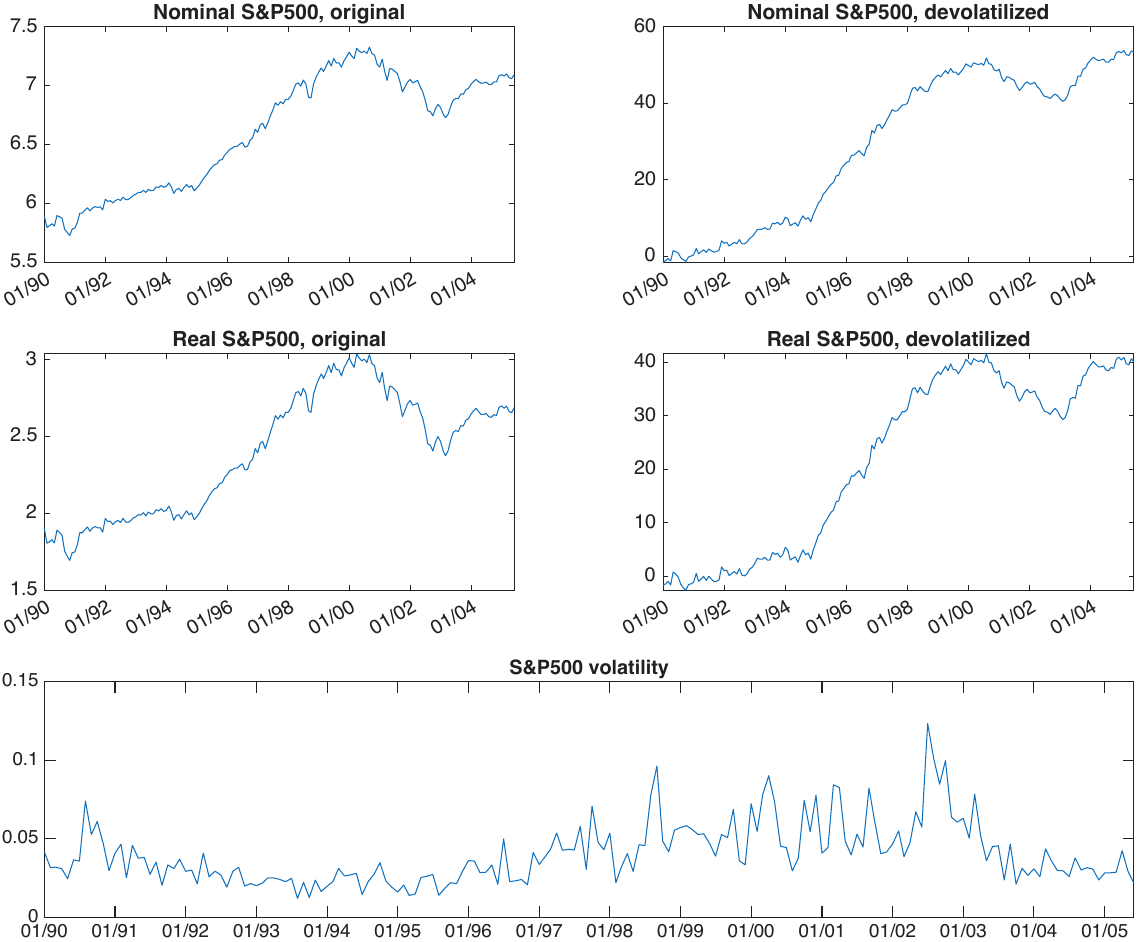}
    \caption{Monthly nominal and real S\&P 500 log-price series, their devolatized (pseudo-log-price) counterparts, and the estimated volatility, 1990--2005.}
    \label{fig:spxseries}
\end{figure}

This observation is useful, as the variance of the real index can be approximated by that of the nominal index, which can in turn be estimated by an estimator for the integrated variance using nominal daily stock index data. Indeed, although CPI will have a nontrivial effect on the level of the real stock index, its impact on the variance is negligible. Since daily stock index data have been widely available since the 1970s, these data can potentially be used to improve our understanding of the volatility structure, even in the real index. 

Applying the TV estimator, we estimate the monthly volatility of the S\&P500 from January 1990 to June 2005. The estimated volatility path, together with the plots of the nominal and real S\&P500 index, is given in Figure \ref{fig:spxseries}. From the volatility estimate, the volatility appears to be generally higher during the build-up of the explosive episode and is obviously higher when the price drops.  

\begin{table}[H]
    \centering
    \caption{Timeline of the explosive episode in the S\&P 500 index.}
    \label{tb:timeline}
    \begin{tabular*}{0.8\textwidth}{@{\extracolsep{\fill}} l | c c | c c @{}}
        \toprule
        & \multicolumn{2}{c|}{Nominal S\&P 500} & \multicolumn{2}{c}{Real S\&P 500}\\
        \cmidrule(lr){2-3} \cmidrule(lr){4-5}
        & PWY & TVPWY & PWY & TVPWY \\
        \midrule
        Test statistic      & 2.0071 & 2.2236 & 2.0016 & 2.2768 \\
        Start of bubble     & Jul 1995 & Jan 1995 & Nov 1995 & Mar 1995 \\
        Downward crossing time & Feb 2001 & Jan 2001 & Feb 2001 & Jan 2001 \\
        Monthly S\&P max  & \multicolumn{2}{c|}{Sep 2000} & \multicolumn{2}{c}{Apr 2000} \\
        \bottomrule
\end{tabular*}

\end{table}

We then apply the testing and date-stamping strategy developed in this paper to both the nominal and real S\&P 500 series, and compare the results with those obtained by PWY's methodology. The results are summarized in Table \ref{tb:timeline}. First, for the test, it seems that the TVPWY test gives slightly higher statistic values than the PWY test for both series, such that the TVPWY test rejects the ``no-explosiveness'' hypothesis at the 1\% level in both nominal and real S\&P 500 series. The PWY test also rejects the hypothesis, but at the 5\% level in both series. 

In the nominal series, the TVDF detector finds the explosive behavior from January 1995, six months earlier than the DF detector by PWY. In the real S\&P 500 series, the TVDF detector finds the explosive behavior from March 1995, which is again earlier than PWY's methodology, by 8 months.
The nominal S\&P500 index reached its maximum in September 2000, and the corresponding real index peaked in April of the same year. From an ex post point of view, it is unlikely that explosive behavior persists after the series reaches its maximum; we therefore take the months when the series reaches its maximum as the end of the explosive regime.
As discussed in Section 5, the downward crossing time of the monitoring statistic cannot estimate the conclusion time of the explosive regime consistently; the downcrossing dates below are reported only as a rough, and possibly delayed, indication of the end of the regime.
In the nominal series, the TVDF detector puts the end of the explosive regime in January 2001, which is a four-month delay relative to the peak month September 2000, and which is one month earlier than the DF detector (February 2001).
In the real series, the DF detector downward crosses the critical line in December 2000, but quickly bounces back to a level above the critical line in the next month, and surely exceeds the critical line from February 2001 onwards, whereas the TVDF detector downward crosses the critical line in January 2001; this represents a 9-month delay relative to the April 2000 peak of the real series, but is one month earlier than the DF detector.
Overall, the TVDF detector is most useful for detecting the \emph{origination} of the explosive regime; the termination date, by contrast, cannot be consistently estimated under our method (as explained in Section~5) and should be interpreted with caution.

\section{Conclusion}
In a new continuous-time framework with stochastic volatility and jumps, by incorporating the volatility information in data sampled at a higher frequency,
we develop a jump-robust and powerful test for the explosive behavior in data sampled at a lower frequency. Monte Carlo simulation results suggest that our test has well-controlled size under stochastic volatility and jumps, suffers no efficiency loss even when used for data without jumps, and maintains high rejection power when jumps are present, strictly dominating the existing tests in the literature. We apply the developed methodology to study explosive behavior in cryptocurrencies and in the nominal and real S\&P 500 index and find interesting results.

\section*{Appendix A: Lemmas}
\renewcommand{\theequation}{A.\arabic{equation}}
\renewcommand{\theHequation}{A.\arabic{equation}}
\setcounter{equation}{0}

In the proofs, $C$ denotes a generic positive constant number, the value of which may change from one equation to the next.

\begin{lemma}
    \label{lemsecmoment}Under the same assumptions as in Theorem 1, letting
    \[ Z_{i-1, j} \assign \left( \int_{t_{i - 1, j - 1}}^{t_{i - 1, j}} \sigma_u
    \mathd W_u \right)^2 - \int_{t_{i - 1, j - 1}}^{t_{i - 1, j}} \sigma_u^2
    \mathd u, \]
    we have
    \[ E \left( \sum_{j = 1}^M Z_{i-1, j} \right)^2 = O (h), \quad \forall i. \]
\end{lemma}

\begin{proof}
    By the property of stochastic integrals, and in particular Ito's isometry,
    $EZ_{i, j} = 0$, and $Z_{i, j}$ is a martingale difference sequence indexed by $j$ for every $i$. Using the Burkholder inequality for the $L_2$ norm (cf.\ \citet{shiryaev1995probability}, p.~499)
    for the martingale difference sequence, we have
    \begin{eqnarray}
        E \left( \sum_{j = 1}^M Z_{i-1, j} \right)^2 & \leqslant & C \times E \left(
        \sum_{j = 1}^M Z_{i-1, j}^2 \right) \nonumber\\
        & = & C \times E \sum_{j = 1}^M \left( \left( \int_{t_{i - 1, j -
                1}}^{t_{i - 1, j}} \sigma_u \mathd W_u \right)^2 - \int_{t_{i - 1, j -
                1}}^{t_{i - 1, j}} \sigma_u^2 \mathd u \right)^2 \nonumber\\
        & \leqslant & 2 C \times E \sum_{j = 1}^M \left( \left( \int_{t_{i - 1, j
                - 1}}^{t_{i - 1, j}} \sigma_u \mathd W_u \right)^4 + \left( \int_{t_{i -
                1, j - 1}}^{t_{i - 1, j}} \sigma_u^2 \mathd u \right)^2 \right), 
        \label{secondmoment}
    \end{eqnarray}
    where in the
    last step we use the generic inequality $(a + b)^2 \leqslant 2 a^2 + 2 b^2$
    for any real numbers $a$ and $b$. Now, by the Burkholder-Davis-Gundy inequality, for all $i, j$,
    \begin{eqnarray*}
        E \left( \int_{t_{i - 1, j - 1}}^{t_{i - 1, j}} \sigma_u \mathd W_u
        \right)^4 & \leqslant & CE \left( \int_{t_{i - 1, j - 1}}^{t_{i - 1, j}}
        \sigma_u^2 \mathd u \right)^2 = O (h^2),
    \end{eqnarray*}
    by the boundedness of the fourth moment of the volatility process in Assumption A1. For the same reason,
    the second term in the brackets in (\ref{secondmoment}) also satisfies that,
    for all $i, j$,
    \[ E \left( \int_{t_{i - 1, j - 1}}^{t_{i - 1, j}} \sigma_u^2 \mathd u
    \right)^2 = O (h^2) . \]
    Continuing on (\ref{secondmoment}), we have
    \[ E \left( \sum_{j = 1}^M Z_{i-1, j} \right)^2 = O (h^2 M) = O (h),\quad \forall i, \]
    which completes the proof.
\end{proof}

\begin{lemma}
    \label{unicons}Under the same assumptions as in Theorem 1, for $i = 1, 2,
    \ldots, n$,
    \[ E (\hat{\omega}_i^2 - \omega_i^2)^2 = O (h) . \]
\end{lemma}

\begin{proof}
    Under the null, the high-frequency return has the following decomposition
    \[ \Delta^h z_{t_{i, j}} = \int_{t_{i, j - 1}}^{t_{i, j}} \sigma_s \mathd
    W_s + \int_{t_{i, j - 1}}^{t_{i, j}} \xi_s \mathd N_s = : \Delta^h C_{i,
        j} + \Delta^h J_{i, j}, \]
    where $C_{i, j}$ and $J_{i, j}$ are defined implicitly to denote the
    continuous and jump parts of the process, respectively. Using these
    notations, for $i = 0, \ldots, n - 1$,
    \begin{eqnarray*}
        \hat{\omega}_{i + 1}^2 - \omega_{i + 1}^2 & = & \sum_{j = 1}^M (\Delta^h
        C_{i, j} + \Delta^h J_{i, j})^2  \text{\tmtextbf{1}} \{| \Delta^h z_{t_{i,
                j}} | < \phi_h \} - \omega_{i+1}^2\\
        & = & \left( \sum_{j = 1}^M (\Delta^h C_{i, j})^2 - \omega_{i+1}^2 \right) +
        \sum_{j = 1}^M (\Delta^h C_{i, j})^2  \text{\tmtextbf{1}} \{| \Delta^h
        z_{t_{i, j}} | \geqslant \phi_h \}\\
        &  & + \sum_{j = 1}^M (\Delta^h J_{i, j})^2  \text{\tmtextbf{1}} \{|
        \Delta^h z_{t_{i, j}} | < \phi_h \} + 2 \sum_{j = 1}^M (\Delta^h C_{i, j})
        (\Delta^h J_{i, j})  \text{\tmtextbf{1}} \{| \Delta^h z_{t_{i, j}} | <
        \phi_h \}\\
        & = : & A + B + C + D.
    \end{eqnarray*}
    By the inequality $(A + B + C + D)^2 \leqslant 4 (A^2 + B^2 + C^2 + D^2)$,
    we proceed by evaluating the second moments of $A, B, C$ and $D$ terms
    individually.
    
    First, applying the result of Lemma \ref{lemsecmoment}, we have $EA^2 = O
    (h)$.
    
    For the second term $B$, we have
    \begin{eqnarray*}
        B & = & \sum_{j = 1}^M (\Delta^h C_{i, j})^2 \text{\tmtextbf{1}} \{|
        \Delta^h z_{t_{i, j}} | \geqslant \phi_h, \Delta^h J_{i, j} = 0\} +
        \sum_{j = 1}^M (\Delta^h C_{i, j})^2 \text{\tmtextbf{1}} \{| \Delta^h
        z_{t_{i, j}} | \geqslant \phi_h, \Delta^h J_{i, j} \neq 0\}\\
        & = & B 1 + B 2.
    \end{eqnarray*}
    
    We look at $B 1$ first. Using the inequality that $\left( \sum_{i = 1}^n a_i
    \right)^2 \leqslant n \left( \sum_{i = 1}^n a_i^2 \right)$ for any real
    numbers $a_1, \ldots, a_n$, 
    \begin{eqnarray*}
        E |B1|^2 & \leqslant & M \sum_{j = 1}^M E (| \Delta^h C_{i, j} |^4
        \text{\tmtextbf{1}} \{| \Delta^h C_{i, j} | \geqslant \phi_h, \Delta^h
        J_{i, j} = 0\})\\
        & \leqslant & M \sum_{j = 1}^M (E (\Delta^h C_{i, j})^8)^{1 / 2}  (P (|
        \Delta^h C_{i, j} | \geqslant \phi_h))^{1 / 2},
    \end{eqnarray*}
    by the Cauchy-Schwarz inequality for the expectation. By definition $E
    (\Delta^h C_{i, j})^8 = O (h^4)$. Using the fact that $\Delta^h C_{i,
            j} / \sqrt{\int_{t_{i, j - 1}}^{t_{i, j}} \sigma_s^2 \mathd s} \sim N (0, 1)$ and letting $\Phi (\cdot)$ be the cumulative distribution function of a standard
    normal distribution, we have
    \[ P (| \Delta^h C_{i, j} | \geqslant \phi_h) = P \left( \frac{\Delta^h
        C_{i, j}}{\sqrt{\int_{t_{i, j - 1}}^{t_{i, j}} \sigma_s^2 \mathd s}}
    \geqslant \frac{\phi_h}{\sqrt{\int_{t_{i, j - 1}}^{t_{i, j}} \sigma_s^2
            \mathd s}} \right) = 1 - \Phi (b_h), \]
    where $b_h = \phi_h / \sqrt{\int_{t_{i, j - 1}}^{t_{i, j}} \sigma_s^2 \mathd
        s}$. Since $\int_{t_{i, j - 1}}^{t_{i, j}} \sigma_s^2 \mathd s = O (h)$, by
    Assumption A3, $\phi_h = O (h^{1 / 2 - \varepsilon})$ for $\varepsilon > 0$,
    we have $b_h = O (h^{- \varepsilon}) \rightarrow \infty$. Letting $\varphi (\cdot)$
    be the probability density function of the standard normal distribution, and using
    the standard result that $(1 - \Phi (x)) / \varphi (x) \sim 1 / x$ as $x
    \rightarrow \infty$ we have
    \[ P (| \Delta^h C_{i, j} | \geqslant \phi_h) = 1 - \Phi (b_h) \sim
    \frac{\varphi (b_h)}{b_h} = O \left( \frac{1}{b_h \exp (b_h^2 / 2)}
    \right), \]
    and hence we have $E |B1|^2 = O \left( M^2 h^2  \sqrt{1/(b_h \exp (b_h^2 / 2))}
    \right) = O \left( \sqrt{1/(b_h \exp (b_h^2 / 2))} \right) = o (h)$,
    as this converges to 0 faster than any polynomial rate of $h$ due to the
    exponential function applied to $b_h = O (h^{- \varepsilon})$. 
    
    For the term
    $B 2$, notice that conditional on the event that there are $N$ jumps happened on the
    $i$th low-frequency interval, there are $N$ nonzero terms in the above sum.
    Using the law of iterated expectations,
    \begin{eqnarray*}
        E |B2|^2 & = & E \left( E \left( \left( \sum_{1 \leqslant j \leqslant M,
            \Delta^h J_{i, j} \neq 0} (\Delta^h C_{i, j})^2 \text{\tmtextbf{1}} \{|
        \Delta^h z_{t_{i, j}} | \geqslant \phi_h, \Delta^h J_{i, j} \neq 0\}
        \right)^2 |N_i = N \right) \right)\\
        & = & E (N_i^2 h^2),
    \end{eqnarray*}
    where we use a similar argument in evaluating the conditional expectation as
    B1. The difference here is that there are only $N_i$ non-zero terms in the
    sum. Because $EN_i^2 = \lambda H + (\lambda H)^2 = O (1)$, $E |B2|^2 =
    O (h^2) = o (h)$.
    
    The term $C$ is the sum of untruncated squared jump increments in the high
    frequency. Notice that it is very unlikely that a jump is not truncated,
    because $\Delta^h J_{i, j} = O_p (1)$ when they actually happen, while
    $\phi_h \rightarrow 0$. Also, for the same reason as discussed for $B 2$,
    there is only at most a finite number of such increments that are non-zero
    over an interval $H$. Both factors reduce the magnitude of $C$. In
    particular, we have
    \begin{eqnarray*}
        C & = & \sum_{1 \leqslant j \leqslant M, \Delta^h J_{i, j} \neq 0}
        (\Delta^h J_{i, j})^2 \text{\tmtextbf{1}} \{| \Delta^h z_{t_{i, j}} | <
        \phi_h, \Delta^h J_{i, j} \neq 0\}\\
        & \leqslant & \sum_{1 \leqslant j \leqslant M, \Delta^h J_{i, j} \neq 0}
        (\Delta^h J_{i, j})^2 \text{\tmtextbf{1}} \{| \Delta^h J_{i, j} | < \phi_h
        + | \Delta^h C_{i, j} |, \Delta^h J_{i, j} \neq 0\}\\
        & \leqslant & \sum_{1 \leqslant j \leqslant M, \Delta^h J_{i, j} \neq 0}
        (\phi_h + | \Delta^h C_{i, j} |)^2  \text{\tmtextbf{1}} \{| \Delta^h J_{i,
            j} | < \phi_h + | \Delta^h C_{i, j} |\},
    \end{eqnarray*}
    where in the second step we use a result that if $|X + Y| < Z$ for positive
    $Z$, then $|X| < Z + |Y|$. Now notice that because $\phi_h + | \Delta^h
    C_{i, j} | = \phi_h  (1 + o_p (1)) \rightarrow 0$,
    \[ P (| \Delta^h J_{i, j} | < \phi_h + | \Delta^h C_{i, j} |) = (F_{\xi}
    (\phi_h) - F_{\xi} (- \phi_h))  (1 + o (1)) = O (\phi_h), \]
    by the continuity assumption of the cumulative distribution function of the
    jump size $\xi_t$. Then use the same argument for $B 2$, we have $EC^2 = E
    (N_i^2 \phi_h^4  \sqrt{\phi_h}) = O (\phi_h^{9 / 2}) = O (h^{(1 / 2 -
        \varepsilon) 9 / 2})$. Since $\varepsilon \leqslant 1 / 4$ by Assumption A3, it
    follows that $(1 / 2 - \varepsilon) 9 / 2 = 9 / 8 > 1$ and $EC^2 = o (h)$.
    
    The term $D$ can be analyzed in a similar way to $C$.
    \begin{eqnarray*}
        D & \leqslant & \sum_{1 \leqslant j \leqslant M, \Delta^h J_{i, j} \neq 0}
        | (\Delta^h C_{i, j}) (\Delta^h J_{i, j}) | \text{\tmtextbf{1}} \{|
        \Delta^h z_{t_{i, j}} | < \phi_h, \Delta^h J_{i, j} \neq 0\}\\
        & \leqslant & \sum_{1 \leqslant j \leqslant M, \Delta^h J_{i, j} \neq 0}
        | (\Delta^h C_{i, j}) (\Delta^h J_{i, j}) | \text{\tmtextbf{1}} \{|
        \Delta^h J_{i, j} | < \phi_h + | \Delta^h C_{i, j} |, \Delta^h J_{i, j}
        \neq 0\}\\
        & \leqslant & \sum_{1 \leqslant j \leqslant M, \Delta^h J_{i, j} \neq 0}
        | \Delta^h C_{i, j} |  (\phi_h + | \Delta^h C_{i, j} |) 
        \text{\tmtextbf{1}} \{| \Delta^h J_{i, j} | < \phi_h + | \Delta^h C_{i, j}
        |, \Delta^h J_{i, j} \neq 0\}.
    \end{eqnarray*}
    Notice
    \begin{eqnarray*}
        &  & E | \Delta^h C_{i, j} |^2  (\phi_h + | \Delta^h C_{i, j} |)^2 
        \text{\tmtextbf{1}} \{| \Delta^h J_{i, j} | < \phi_h + | \Delta^h C_{i, j}
        |, \Delta^h J_{i, j} \neq 0\}\\
        & \leqslant & (E| \Delta^h C_{i, j} |^8)^{1 / 4}  (E (\phi_h + | \Delta^h
        C_{i, j} |)^8)^{1 / 4}  (P (| \Delta^h J_{i, j} | < \phi_h + | \Delta^h
        C_{i, j} |))^{1 / 2}\\
        & = & O (h) O (\phi_h^2) O (\phi_h^{1 / 2}) = O (h \phi_h^{5 / 2}) .
    \end{eqnarray*}
    Applying the same argument as before, we have
    \[ E (D^2) = E (N_i^2 h \phi_h^{5 / 2}) = O (h \phi_h^{5 / 2}) = o (h), \]
    as $\phi_h \rightarrow 0$. In total, the claim of the lemma is proved.
\end{proof}

\begin{lemma}
    \label{estimatorbound}Under the same assumptions as in Theorem 1, $\max_{1 \leqslant i \leqslant n} |
    \omega_i^{- 1} | = O_p (1)$ and $\max_{1 \leqslant i \leqslant n} |
    \hat{\omega}_i^{- 1} | = O_p (1)$.
\end{lemma}

\begin{proof}
It is clear that $\max_{1 \leqslant i \leqslant n} | \omega_i^{- 1} | = O_p (1)$ is a
    consequence of the strict positiveness of $\sigma$ in Assumption A1. For the
    realized variance $\hat{\omega}_i^2$, first notice that
    \[ \min_{1 \leqslant i \leqslant n} | \omega_i^2 | \leqslant \min_{1
        \leqslant i \leqslant n} | \hat{\omega}_i^2 | + \min_{1 \leqslant i
        \leqslant n} | \omega_i^2 - \hat{\omega}_i^2 | \leqslant \min_{1
        \leqslant i \leqslant n} | \hat{\omega}_i^2 | + \max_{1 \leqslant i
        \leqslant n} | \omega_i^2 - \hat{\omega}_i^2 |, \]
    which implies that
    \begin{eqnarray*}
        \min_{1 \leqslant i \leqslant n} | \hat{\omega}_i^2 | & \geqslant &
        \min_{1 \leqslant i \leqslant n} | \omega_i^2 | -_{\nosymbol} \max_{1
            \leqslant i \leqslant n} | \omega_i^2 - \hat{\omega}_i^2 | .
    \end{eqnarray*}
    Again, by the strict positivity of the volatility process, $\min_{1 \leqslant i \leqslant n} \omega_i^2 > c$;
    together with the result of Lemma \ref{unicons} we have %
    $\min_{1 \leqslant i \leqslant n} | \hat{\omega}_i^2 | \geqslant c + o_p (1)$
    for a positive constant $c$, and therefore we have $\max_{1 \leqslant
        i \leqslant n} | \hat{\omega}_i^{- 2} | = O_p (1)$, and the second claim of the lemma is proved.
\end{proof}

\begin{lemma}
    \label{lemsol}If the time-varying mean-reversion parameter $\kappa_t$
    satisfies
    \[ \kappa_t = \left\{ \begin{array}{ll}
        0, & t \leqslant t_{\lfloor \tau_1 n \rfloor},\\
        \kappa, & t_{\lfloor \tau_1 n \rfloor} < t \leqslant n .
    \end{array} \right. \]
    The solution of the process $y_t$ is
    \[ y_t = \left\{ \begin{array}{ll}
        y_0+\int_0^t \sigma_u \mathd W_u & t \leqslant t_{\lfloor \tau_1 n
            \rfloor},\\
        y_{t_{\lfloor \tau_1 n \rfloor}} e^{\kappa (t - t_{\lfloor \tau_1 n
                \rfloor})} + e^{\kappa t}  \int_{t_{\lfloor \tau_1 n \rfloor}}^t
        \sigma_u e^{- \kappa u} \mathd W_u & t_{\lfloor \tau_1 n \rfloor} < t
        \leqslant n.
    \end{array} \right. \]
\end{lemma}

\begin{proof}
    First, notice that for the {\tmem{constant}} mean-reversion parameter model
    given by
    \[ \mathd y_t = \kappa y_t \mathd t + \sigma_t \mathd W_t, \]
    the solution is:
    \[ y_t = y_0 e^{\kappa t} + e^{\kappa t}  \int_0^t \sigma_u e^{- \kappa u}
    \mathd W_u . \]
    When $t \leqslant t_{\lfloor \tau_1 n \rfloor}$, the process has a constant
    time-varying parameter $\kappa = 0$, and the process starts from $y_0$.
    So the claimed solution in this regime follows easily.
    
    When $t_{\lfloor \tau_1 n \rfloor} < t \leqslant n$, the process has a
    constant time-varying parameter $\kappa > 0$, and the process starts from
    the last point of the previous unit regime $y_{\lfloor \tau_1 n \rfloor} =
    \int_0^{\lfloor \tau_1 n \rfloor} \sigma_u \mathd W_u$. It is
    straightforward to check that the solution in this regime is
    \[ y_t = y_{t_{\lfloor \tau_1 n \rfloor}} e^{\kappa (t - t_{\lfloor \tau_1 n
            \rfloor})} + e^{\kappa t}  \int_{t_{\lfloor \tau_1 n \rfloor}}^t \sigma_u
    e^{- \kappa u} \mathd W_u . \]
    The proof of the lemma is then complete.
\end{proof}

\begin{remark}
    From the result of the above lemma, we can see that in the unit root regime,
    the order of the process is $n^{1 / 2}$. In the explosive regime,
 the process is dominated by the initial value's effect (i.e., the term $y_{t_{\lfloor \tau_1 n \rfloor}} e^{\kappa (t - t_{\lfloor \tau_1 n
            \rfloor})}$)
    of the regime. This is because the solution of this
    regime takes the form $y_{t_{\lfloor \tau_1 n \rfloor}} e^{\kappa (t - t_{\lfloor \tau_1 n
            \rfloor})} + e^{\kappa (t - t_{\lfloor \tau_1 n \rfloor})} \int_{t_{\lfloor
            \tau_1 n \rfloor}}^t \sigma_u e^{- \kappa (u - t_{\lfloor \tau_1 n \rfloor})}
    \mathd W_u$. In the first term, $y_{t_{\lfloor \tau_1 n \rfloor}}$ has order
    $n^{1 / 2}$; while in the second term, the order of the $\int_{t_{\lfloor
            \tau_1 n \rfloor}}^t \sigma_u e^{- \kappa (u - t_{\lfloor \tau_1 n \rfloor})}
    \mathd W_u$ term is smaller as by Ito's
    isometry,
    \begin{eqnarray*}
        E \left( \int_{t_{\lfloor \tau_1 n \rfloor}}^t \sigma_u e^{- \kappa (u -
            t_{\lfloor \tau_1 n \rfloor})} \mathd W_u \right)^2 & = & \int_{t_{\lfloor
                \tau_1 n \rfloor}}^t E \sigma_u^2 e^{- 2 \kappa (u - t_{\lfloor \tau_1 n
                \rfloor})} \mathd u\\
        & \leqslant & C \int_{t_{\lfloor \tau_1 n \rfloor}}^t e^{- 2 \kappa (u -
            t_{\lfloor \tau_1 n \rfloor})} \mathd u\\
        & = & - C \frac{1}{2 \kappa} \left( e^{- 2 \kappa (t - t_{\lfloor \tau_1 n
                \rfloor})} - 1 \right) .
    \end{eqnarray*}
    When $t - t_{\lfloor \tau_1 n \rfloor}$ is finite, $\kappa (t - t_{\lfloor \tau_1 n \rfloor})$
    has order $\kappa$ as $\kappa \rightarrow 0$, and the above can be expanded as
    \[ - C \frac{1}{2 \kappa} \left ( - 2 \kappa (t - t_{\lfloor \tau_1 n \rfloor}) + \frac{1}{2}(-
    2 \kappa (t - t_{\lfloor \tau_1 n \rfloor}))^2 + \cdots \right ) = O (1) . \]
    When $t = t_{\lfloor \tau n \rfloor}$ for some $\tau > \tau_1$, $t -
    t_{\lfloor \tau_1 n \rfloor} = (\lfloor \tau n \rfloor - \lfloor \tau_1 n
    \rfloor) H = O (n)$ and it follows that $- 2 \kappa (t - t_{\lfloor \tau_1 n
        \rfloor}) \sim - 2 n^{1 - \alpha} \rightarrow - \infty$ and therefore $e^{- 2
        \kappa (t - t_{\lfloor \tau_1 n \rfloor})} - 1 \rightarrow - 1$. We then have
    \[ - C \frac{1}{2 \kappa} \left( e^{- 2 \kappa (t - t_{\lfloor \tau_1 n
            \rfloor})} - 1 \right) \sim \frac{1}{2 \kappa} = O (n^{\alpha}) = o (n), \]
    and it follows that $\int_{t_{\lfloor \tau_1 n \rfloor}}^t \sigma_u e^{- \kappa (u -
        t_{\lfloor \tau_1 n \rfloor})} \mathd W_u = o_p (n^{1 / 2})$. In either case,
    the Ito integral is dominated by the $y_{t_{\lfloor \tau_1 n \rfloor}}$ term, which is $O_p (n^{1 / 2})$. 
\end{remark}

\begin{lemma}
    \label{unicons1}Under the same assumptions as in Theorem 4, with
    $i_1^{\ast} = \lfloor \tau_1^{\ast} n \rfloor$ ($\tau_1^{\ast}$ is also
    defined in Theorem 4), we have
    \[ \max_{1 \leqslant i \leqslant i_1^{\ast}} | \hat{\omega}_i^2 - \omega_i^2
    | = O_p \left( \sqrt{nh} \vee n h \log^2 (n) \right) . \]
\end{lemma}

\begin{proof}
    Under the alternative hypothesis and in the unit root regime, the result of
    Lemma \ref{unicons} imply that $\max_{1 \leqslant i \leqslant i_1} |
    \hat{\omega}_i^2 - \omega_i^2 | = O_p \left( \sqrt{nh} \right)$, by noticing
    that
    \[P \left( \max_{1 \leqslant i \leqslant i_1} | \hat{\omega}_i^2 -
        \omega_i^2 | \geqslant \eta \sqrt{hn} \right)  \leqslant  \sum_{i =
            1}^{i_1} P \left( | \hat{\omega}_i^2 - \omega_i^2 | \geqslant \eta
        \sqrt{hn} \right)\leqslant \sum_{i = 1}^{i_1} E | \hat{\omega}_i^2 - \omega_i^2 |^2 /
        (\eta^2 hn)=O (1).\]
    When the process gets to the explosive regime, the high-frequency return has
    the following decomposition:
    \[ \Delta^h z_{t_{i, j}} = \kappa \int_{t_{i, j - 1}}^{t_{i, j}} y_u \mathd
    u + \int_{t_{i, j - 1}}^{t_{i, j}} \sigma_s \mathd W_s + \int_{t_{i, j -
            1}}^{t_{i, j}} \xi_s \mathd N_s = : \Delta^h D_{i, j} + \Delta^h C_{i, j}
    + \Delta^h J_{i, j}, \]
    where $C_{i, j}$ and $J_{i, j}$ are defined as in the proof of Lemma
    \ref{unicons}. $D_{i, j}$ is the drift part of the process. Using these
    notations, for $i_1 \leqslant i \leqslant i_1^{\ast} - 1$,
    \begin{eqnarray*}
        \hat{\omega}_{i + 1}^2 - \omega_{i + 1}^2 & = & \sum_{j = 1}^M (\Delta^h
        D_{i, j} + \Delta^h C_{i, j} + \Delta^h J_{i, j})^2  \text{\tmtextbf{1}}
        \{| \Delta^h z_{t_{i, j}} | < \phi_h \} - \omega_{i + 1}^2\\
        & = & \left( \sum_{j = 1}^M (\Delta^h C_{i, j})^2 - \omega_{i + 1}^2
        \right) + \sum_{j = 1}^M (\Delta^h C_{i, j})^2  \text{\tmtextbf{1}} \{|
        \Delta^h z_{t_{i, j}} | \geqslant \phi_h \}\\
        &  & + \sum_{j = 1}^M (\Delta^h J_{i, j})^2  \text{\tmtextbf{1}} \{|
        \Delta^h z_{t_{i, j}} | < \phi_h \} + 2 \sum_{j = 1}^M (\Delta^h C_{i, j})
        (\Delta^h J_{i, j})  \text{\tmtextbf{1}} \{| \Delta^h z_{t_{i, j}} | <
        \phi_h \}\\
        &  & + \sum_{j = 1}^M (\Delta^h D_{i, j})^2  \text{\tmtextbf{1}} \{|
        \Delta^h z_{t_{i, j}} | < \phi_h \} + 2 \sum_{j = 1}^M (\Delta^h D_{i, j})
        (\Delta^h J_{i, j})  \text{\tmtextbf{1}} \{| \Delta^h z_{t_{i, j}} | <
        \phi_h \}\\
        &  & + 2 \sum_{j = 1}^M (\Delta^h D_{i, j})  (\Delta^h C_{i, j}) 
        \text{\tmtextbf{1}} \{| \Delta^h z_{t_{i, j}} | < \phi_h \}\\
        & = : & A + B + C + D + E + F + G,
    \end{eqnarray*}
    where the term $A$ is exactly the same as in the proof of Lemma
    \ref{unicons}, and we have
    \[ \max_{i_1 \leqslant i \leqslant i_1^{\ast} - 1} |A| = O_p \left( \sqrt{nh} \right) . \]
    
    Terms $B, C, D$ are closely related to the corresponding terms in the proof of
    Lemma \ref{unicons}, with the difference that the high-frequency return
    $\Delta^h z_{t_{i, j}}$ (in the indicator function) now has an extra drift increment
    term. The order of the drift increment term can be analyzed as follows: using the solution of the continuous part, the drift
    increments satisfy
    \begin{eqnarray*}
        \Delta^h D_{i, j} & = & \kappa \int_{t_{i, j - 1}}^{t_{i, j}}
        y_{t_{\lfloor \tau_1 n \rfloor}} e^{\kappa (u - t_{\lfloor \tau_1 n
                \rfloor})} \mathd u (1 + o_p (1))\\
        & = & \kappa e^{- \kappa t_{\lfloor \tau_1 n \rfloor}} y_{t_{\lfloor
                \tau_1 n \rfloor}}  \int_{t_{i, j - 1}}^{t_{i, j}} e^{\kappa u} \mathd u
        (1 + o_p (1))\\
        & = & e^{- \kappa t_{\lfloor \tau_1 n \rfloor}} y_{t_{\lfloor \tau_1 n
                \rfloor}}  (e^{\kappa (t_{i, j})} - e^{\kappa (t_{i, j - 1})})  (1 + o_p
        (1))\\
        & = & e^{\kappa (t_{i, j} - t_{\lfloor \tau_1 n \rfloor})} y_{t_{\lfloor
                \tau_1 n \rfloor}}  (1 - e^{- \kappa h})  (1 + o_p (1))\\
        & = & \kappa he^{\kappa (t_{i, j} - t_{\lfloor \tau_1 n \rfloor})}
        y_{t_{\lfloor \tau_1 n \rfloor}}  (1 + o_p (1)) .
    \end{eqnarray*}
    For $t_{i_1} \leqslant t_i \leqslant t_{i_1^{\ast} - 1}$, we have $t_i -
    t_{i_1} \leqslant \left\lfloor \kappa^{-1} \log \left(\kappa^{-1} \log(n) \right) \right\rfloor$, the above has order at most $O_p (hn^{1 / 2}\log(n))$,
    which is $o_p (1)$ under the assumption A6. Compared
    with the diffusion part, which has order $O_p (h^{1 / 2})$, the drift
    increment is dominated because $hn^{1 / 2}\log(n) / h^{1 / 2} \linebreak = (nh)^{1 / 2}\log(n)
    \rightarrow 0$ under A6. Therefore, when $i_1 \leqslant i \leqslant i_1^{\ast} - 1$, these drift increments are dominated by the corresponding diffusion increments, implying that the truncation conditions are asymptotically the same as the ``no drift" case and thus the derived order for the pointwise second moments ($o(h)$) of the corresponding $B, C, D$ terms in Lemma \ref{unicons} still hold here, and these pointwise moment orders imply the uniform order $\max_{i_1 \leqslant i \leqslant
        i_1^{\ast} - 1} (\{B, C, D\}) = o_p \left( \sqrt{nh} \right)$.
    
    We now look at the terms $E, F, G$, which are related to the drift term. For term $E$, make a further
    decomposition
    \[ E = \sum_{j = 1}^M (\Delta^h D_{i, j})^2 + \sum_{j = 1}^M (\Delta^h D_{i,
        j})^2  \text{\tmtextbf{1}} \{| \Delta^h z_{t_{i, j}} | \geqslant \phi_h
    \} = E 1 + E 2. \]
    using the solution in Lemma \ref{lemsol},
    \begin{eqnarray*}
        \max_{i_1 \leqslant i \leqslant i_1^{\ast} - 1} |E1| & = & \max_{i_1
            \leqslant i \leqslant i_1^{\ast} - 1} \kappa^2  \sum_{j = 1}^M \left(
        \int_{t_{i, j - 1}}^{t_{i, j}} y_u \mathd u \right)^2\\
        & = & \max_{i_1 \leqslant i \leqslant i_1^{\ast} - 1} \kappa^2  \sum_{j =
            1}^M \left( \int_{t_{i, j - 1}}^{t_{i, j}} y_{t_{\lfloor \tau_1 n
                \rfloor}} e^{\kappa (u - t_{\lfloor \tau_1 n \rfloor})} \mathd u (1 + o_p
        (1)) \right)^2\\
        & = & \max_{i_1 \leqslant i \leqslant i_1^{\ast} - 1}  \sum_{j = 1}^M
        \left( y_{t_{\lfloor \tau_1 n \rfloor}} e^{\kappa (t_{i, j - 1} -
            t_{\lfloor \tau_1 n \rfloor})} (e^{\kappa (t_{i, j} - t_{i, j - 1})} - 1)
        (1 + o_p (1)) \right)^2\\
        & = & y_{t_{\lfloor \tau_1 n \rfloor}}^2  (e^{\kappa h} - 1)^2 e^{- 2
            \kappa t_{\lfloor \tau_1 n \rfloor}} \max_{i_1 \leqslant i \leqslant
            i_1^{\ast} - 1}  \sum_{j = 1}^M e^{2 \kappa t_{i, j - 1}}  (1 + o_p (1))\\
        & = & y_{t_{\lfloor \tau_1 n \rfloor}}^2  (e^{\kappa h} - 1)^2 e^{- 2
            \kappa t_{\lfloor \tau_1 n \rfloor}} \max_{i_1 \leqslant i \leqslant
            i_1^{\ast} - 1} e^{2 \kappa t_i}  \frac{1 - e^{- 2 \kappa
            H}}{1 - e^{- 2 \kappa
            h}}\\
        & = & y_{t_{\lfloor \tau_1 n \rfloor}}^2  (e^{\kappa h} - 1)^2 e^{- 2
            \kappa t_{\lfloor \tau_1 n \rfloor}}  (2 \kappa h)^{- 1} e^{2 \kappa
            \lfloor \tau_1^{\ast} n \rfloor}  (1 - e^{- 2 \kappa H})\\
        & = & O_p \left( n (\kappa h)^2 (2 \kappa h)^{- 1} e^{2 \kappa
            \frac{1}{\kappa} \log \left( \frac{\log(n)}{\kappa} \right)} (\kappa) \right)\\
        & = & O_p (nh\log^2(n))
    \end{eqnarray*}
    The $E 2$ term can be analyzed in the same way as the term $B$ in the proof
    of Lemma \ref{unicons}, by looking at the sum when the jump is zero and when the
    jump is not zero. Since the second moment of the drift term $E (\Delta^h
    D_{i, j})^2$ is dominated by the second moment of the diffusion term $E
    (\Delta^h C_{i, j})^2$ at the high frequency, the derivation in Lemma
    \ref{unicons} goes through in the same way, and it can be shown the same
    second moment bound $o(h)$ holds for $E 2$, which implies that $\max |E2| = o_p \left(
    \sqrt{nh} \right)$. $F$ and $G$ are cross-product terms, and by the
    Cauchy-Schwarz inequality, they cannot be larger than the largest of
    the term $\sum_{j = 1}^M (\Delta^h C_{i, j})^2  \text{\tmtextbf{1}} \{|
    \Delta^h z_{t_{i, j}} | < \phi_h \}$ (which is $A + B$), the term $C$ and
    the term $E$. Therefore, we have
    \[ \max_{i_1 \leqslant i \leqslant i_1^{\ast} - 1} | \hat{\omega}_{i + 1}^2 -
    \omega_{i + 1}^2 | = O_p (n h \log^2 (n)) . \]
    Hence,
    \[ \max_{1 \leqslant i \leqslant i_1^{\ast}} | \hat{\omega}_i^2 - \omega_i^2 |
    = O_p (\sqrt{nh} \vee n h \log^2 (n)), \]
    which is $o_p (1)$ under Assumption A6.
\end{proof}

\begin{lemma}
    \label{leminc}Assume the same assumptions as in Theorem 4.
    
    When $0\leqslant \alpha < 1/2$ (i.e. when $\kappa n^{1 / 2} \rightarrow \infty$), \ then for $i_1 \leqslant i
    \leqslant i_1^{\ast}$,
    \[ \Delta x_i = O_p \left( \kappa n^{1 / 2} e^{\kappa (t_i - t_{i_1})}
    \right) , \]
    which is diverging.
    
    When $1/2\leqslant \alpha < 1$ (i.e. when $\kappa n^{1 / 2} \rightarrow 0$ or a constant), $\Delta x_i$ remains bounded in probability and does not immediately diverge when the explosive regime just starts. However, it diverges after the data build up for a certain period of time.
    Let $l^{\ast} = i_1 + \left\lfloor \kappa^{-1} \log \left( \kappa^{-1} n^{-1/2} \log (n) \right) \right\rfloor$, for $l^{\ast} \leqslant
    i \leqslant i_1^{\ast}$, $\Delta x_i = O_p \left( \kappa n^{1 / 2} e^{\kappa (t_i - t_{i_1})}
    \right)$, which diverges.
\end{lemma}

\begin{proof}
    Using the notations defined in Lemma \ref{unicons1},
    \begin{eqnarray*}
        \Delta x_i & = & \frac{\sum_{j = 1}^M (\Delta^h D_{i - 1, j} + \Delta^h
            C_{i - 1, j} + \Delta^h J_{i - 1, j})  \text{\tmtextbf{1}} \{| \Delta^h
            z_{t_{i - 1, j}} | < \phi_h \}}{\hat{\omega}_i}\\
        & = & \frac{\sum_{j = 1}^M (\Delta^h D_{i - 1, j} + \Delta^h C_{i - 1,
                j})}{\hat{\omega}_i} - \frac{\sum_{j = 1}^M (\Delta^h D_{i - 1, j}) 
            \text{\tmtextbf{1}} \{| \Delta^h z_{t_{i - 1, j}} | \geqslant \phi_h
            \}}{\hat{\omega}_i}\\
        &  & - \frac{\sum_{j = 1}^M (\Delta^h C_{i - 1, j})  \text{\tmtextbf{1}}
            \{| \Delta^h z_{t_{i - 1, j}} | \geqslant \phi_h \}}{\hat{\omega}_i} +
        \frac{\sum_{j = 1}^M (\Delta^h J_{i - 1, j})  \text{\tmtextbf{1}} \{|
            \Delta^h z_{t_{i - 1, j}} | < \phi_h \}}{\hat{\omega}_i}\\
        & = & A - B - C + D.
    \end{eqnarray*}
    We first look at $A$. Notice that $\Delta^h D_{i - 1, j} + \Delta^h C_{i -
        1, j} = \Delta^h y_{t_{i - 1, j}}$, which is the high-frequency return of
    the $\{ y \}$ process. Notice that the numerator of $A$ is the
    aggregated high-frequency return within the low-frequency interval, we therefore have:
    \begin{eqnarray*}
        A & = & \frac{\sum_{j = 1}^M \Delta^h y_{t_{i - 1, j}}}{\hat{\omega}_i}
         =  \frac{\Delta y_{t_i}}{\hat{\omega}_i}
         =  \frac{\kappa \int_{t_{i - 1}}^{t_i} y_u \mathd u}{\hat{\omega}_i} +
        \frac{\omega_i}{\hat{\omega}_i} \varepsilon_i\\
        & = & \frac{\kappa \int_{t_{i - 1}}^{t_i} y_u \mathd u}{\omega_i} +
        \kappa \left( \frac{\int_{t_{i - 1}}^{t_i} y_u \mathd u}{\hat{\omega}_i} -
        \frac{\int_{t_{i - 1}}^{t_i} y_u \mathd u}{\omega_i} \right) +
        \frac{\omega_i}{\hat{\omega}_i} \varepsilon_i\\
        & = & A 1 + A 2 + A 3.
    \end{eqnarray*}
    Using the dominating part of the solution of the continuous-time process in
    the explosive regime as derived in Lemma \ref{lemsol}, we have
    \begin{eqnarray*}
        A 1 & = & \frac{\kappa \int_{t_{i - 1}}^{t_i} \left( y_{t_{\lfloor \tau_1
                    n \rfloor}} e^{\kappa (u - t_{\lfloor \tau_1 n \rfloor})} \right) \mathd u
            (1 + o_p (1))}{\omega_i}\\
        & = & \frac{\kappa y_{t_{\lfloor \tau_1 n \rfloor}} e^{- \kappa
                t_{\lfloor \tau_1 n \rfloor}}}{\omega_i}  \int_{t_{i - 1}}^{t_i} e^{\kappa
            u} \mathd u (1 + o_p (1))\\
        & = & \frac{\kappa y_{t_{\lfloor \tau_1 n \rfloor}} e^{- \kappa
                t_{\lfloor \tau_1 n \rfloor}}}{\omega_i \kappa} e^{\kappa t_i}  (1 - e^{-
            \kappa H})  (1 + o_p (1))\\
        & = & O_p \left( \kappa n^{1 / 2} e^{\kappa (t_i - t_{i_1})} \right) .
    \end{eqnarray*}
    Next,
    \[ A 2 \leqslant \kappa \left| \int_{t_{i - 1}}^{t_i} y_u \mathd u \right|
    \left| \frac{1}{\hat{\omega}_i \omega_i} \right|  | \hat{\omega}_i -
    \omega_i | = o_p \left( \left| \frac{\kappa \int_{t_{i - 1}}^{t_i} y_u
        \mathd u}{\omega_i} \right| \right), \]
   where use Lemma \ref{unicons1} and that the results of Lemma \ref{estimatorbound} also hold under the assumptions of Theorem 4. Finally, $A 3 = O_p (1)$ follows also from Lemma \ref{unicons1}.
    
    Looking at these three terms, when $\kappa n^{1 / 2} \rightarrow \infty$, $A 1$ dominates $A 2$ and $A 3$
    as soon as the explosive regime starts, and it follows that $A = O_p \left(
    \kappa n^{1 / 2} e^{\kappa (t_i - t_{i_1})} \right)$ when $i_1 \leqslant i
    \leqslant i_1^{\ast}$. When $\kappa n^{1 / 2} \rightarrow 0$ or a constant, $A 1$ is no larger
    than $A 3$ when the explosive regime just starts, and it will take some time
    for the $e^{\kappa (t_i - t_{i_1})}$ term to build up, such that $A 1$
    becomes dominant. When $i \geqslant l^{\ast}$, $t_i - t_{i_1}$ is at least
    $\left\lfloor \kappa^{-1} \log \left(\kappa^{-1} n^{-1/2} \log (n) \right) \right\rfloor$ and it follows that $A 1$ is at least
    $O_p (\log (n))$ and becomes dominant and we have $A = O_p \left( \kappa
    n^{1 / 2} e^{\kappa (t_i - t_{i_1})} \right)$.
    
    By similar calculations used in the proof of Lemma \ref{unicons} and Lemma \ref{unicons1}, it can
    be shown that $B, C, D = o_p (1)$ under the assumptions of Theorem 4. We
    omit the details to avoid repetition. In total, the claim of the Lemma is
    proved.
\end{proof}

\begin{lemma}
    \label{lemincsum}Let the assumptions of Theorem 4 hold.\\
    When $0\leqslant \alpha < 1/2$ (i.e. when $\kappa n^{1 / 2} \rightarrow \infty$),
    \[ \sum_{i = i_1 + 1}^{i_1^{\ast}} \Delta x_i = O_p \left( n^{1 / 2}
    e^{\kappa (t_{i_1^{\ast}} - t_{i_1})} \right) . \]
    When $1/2\leqslant \alpha < 1$ (i.e. when $\kappa n^{1 / 2} \rightarrow 0$ or a constant),
    \[ \sum_{i = i_1 + 1}^{i_1^{\ast}} \Delta x_i = O_p \left( n^{1 / 2}
    e^{\kappa (t_{i_1^{\ast}} - t_{l^{\ast} + 1})} \right), \]
    where $l^{\ast}$ is defined in Lemma \ref{leminc}.
    
    \ 
\end{lemma}

\begin{proof}
    When $0\leqslant \alpha < 1/2$ (i.e. when $\kappa n^{1 / 2} \rightarrow \infty$), by definition, we have
    \begin{eqnarray*}
        \sum_{i = i_1 + 1}^{i_1^{\ast}} \Delta x_i & = & \sum_{i = i_1 +
            1}^{i_1^{\ast}} \frac{\kappa \int_{t_{i - 1}}^{t_i} y_u \mathd
            u}{\hat{\omega}_i} + \sum_{i = i_1 + 1}^{i_1^{\ast}}
        \frac{\omega_i}{\hat{\omega}_i} \varepsilon_i\\
        & = & \sum_{i = i_1 + 1}^{i_1^{\ast}} \frac{\kappa \int_{t_{i - 1}}^{t_i}
            y_u \mathd u}{\omega_i} + \sum_{i = i_1 + 1}^{i_1^{\ast}} \kappa \left(
        \frac{\int_{t_{i - 1}}^{t_i} y_u \mathd u}{\hat{\omega}_i} -
        \frac{\int_{t_{i - 1}}^{t_i} y_u \mathd u}{\omega_i} \right) + \sum_{i =
            i_1 + 1}^{i_1^{\ast}} \frac{\omega_i}{\hat{\omega}_i} \varepsilon_i.
    \end{eqnarray*}
    First, using the dominating part of the solution of the process in the
    explosive regime as derived in Lemma \ref{lemsol}, we have
    \begin{eqnarray*}
        \sum_{i = i_1 + 1}^{i_1^{\ast}} \frac{\int_{t_{i - 1}}^{t_i} y_u \mathd
            u}{\omega_i} & = & \sum_{i = i_1 + 1}^{i_1^{\ast}} \frac{\int_{t_{i -
                    1}}^{t_i} \left( y_{t_{\lfloor \tau_1 n \rfloor}} e^{\kappa (u -
                t_{\lfloor \tau_1 n \rfloor})} \right) \mathd u}{\omega_i}(1+o_p(1))\\
        & = & \sum_{i = i_1 + 1}^{i_1^{\ast}} \frac{y_{t_{\lfloor \tau_1 n
                    \rfloor}} e^{- \kappa t_{\lfloor \tau_1 n \rfloor}}}{\omega_i \kappa} e^{\kappa
            t_{i - 1}}  (e^{\kappa (t_i - t_{i - 1})} - 1)(1+o_p(1))\\
        & = & y_{t_{\lfloor \tau_1 n \rfloor}}  \sum_{i = i_1 + 1}^{i_1^{\ast}}
        \frac{1}{\omega_i} e^{\kappa (t_{i - 1} - t_{\lfloor \tau_1 n \rfloor})}(1+o_p(1))\\
        & = & y_{t_{\lfloor \tau_1 n \rfloor}} O \left( \sum_{i = i_1 +
            1}^{i_1^{\ast}} e^{\kappa (t_{i - 1} - t_{\lfloor \tau_1 n \rfloor})}
        \right)(1+o_p(1))\\
        & = & y_{t_{\lfloor \tau_1 n \rfloor}} O \left( \frac{1 - e^{\kappa (t_{i_1^*} -
                t_{i_1})}}{1 - e^{\kappa H}} \right)(1+o_p(1))\\
        & = & O_p (\kappa^{- 1} n^{1 / 2} e^{\kappa (t_{i_1^*} - t_{i_1})}) ,
    \end{eqnarray*}
where in the second equality we use $t_i-t_{i-1}=H$ and $e^{\kappa H}= 1+ \kappa H + o(\kappa)$. Next,
\[
        \left| \sum_{i = i_1 + 1}^{i_1^{\ast}} \left( \frac{\int_{t_{i - 1}}^{t_i}
            y_u \mathd u}{\hat{\omega}_i} - \frac{\int_{t_{i - 1}}^{t_i} y_u \mathd
            u}{\omega_i} \right) \right|  \leqslant  \sum_{i = i_1 + 1}^{i_1^{\ast}}
        \left| \int_{t_{i - 1}}^{t_i} y_u \mathd u \right| \frac{1}{|
            \hat{\omega}_i \omega_i |}  | \hat{\omega}_i - \omega_i |
         =  o_p \left( \sum_{i = i_1 + 1}^{i_1^{\ast}} \left| \frac{\int_{t_{i
                    - 1}}^{t_i} y_u \mathd u}{\omega_i} \right| \right),
\]
    using the uniform consistency of the realized variance estimator in Lemma
    \ref{unicons1}. Finally, \linebreak $\sum_{i = i_1 + 1}^{i_1^{\ast}} (\omega_i/\hat{\omega}_i)
    \varepsilon_i = O_p (\kappa^{- 1} \log (\kappa^{- 1} \log (n)))$ because
    $i_1^{\ast} - i_1 = O (\kappa^{- 1} \log (\kappa^{- 1} \log (n)))$. This is
    $o_p (n^{1 / 2})$ because $\kappa = O (n^{- \alpha})$ with $\alpha < 1 / 2$.
    The proof for the case $\kappa n^{1 / 2} \rightarrow \infty$ is then finished.

    When $1/2\leqslant \alpha < 1$ (i.e. when $\kappa n^{1 / 2} \rightarrow 0$ or a constant), the proof is similar,
    but we have to decompose the summation before and after the time $l^{\ast}$:
    \[ \sum_{i = i_1 + 1}^{i_1^{\ast}} \Delta x_i = \sum_{i = i_1 +
        1}^{l^{\ast}} \Delta x_i + \sum_{i = l^{\ast} + 1}^{i_1^{\ast}} \Delta
    x_i . \]
    First look at the second term in the above decomposition.
    \begin{eqnarray*}
        \sum_{i = l^{\ast} + 1}^{i_1^{\ast}} \Delta x_i & = & \sum_{i = l^{\ast} +
            1}^{i_1^{\ast}} \frac{\kappa \int_{t_{i - 1}}^{t_i} y_u \mathd
            u}{\hat{\omega}_i} + \sum_{i = l^{\ast} + 1}^{i_1^{\ast}}
        \frac{\omega_i}{\hat{\omega}_i} \varepsilon_i\\
        & = & \sum_{i = l^{\ast} + 1}^{i_1^{\ast}} \frac{\kappa \int_{t_{i -
                    1}}^{t_i} y_u \mathd u}{\omega_i} + \sum_{i = l^{\ast} + 1}^{i_1^{\ast}}
        \kappa \left( \frac{\int_{t_{i - 1}}^{t_i} y_u \mathd u}{\hat{\omega}_i} -
        \frac{\int_{t_{i - 1}}^{t_i} y_u \mathd u}{\omega_i} \right) + \sum_{i =
            l^{\ast} + 1}^{i_1^{\ast}} \frac{\omega_i}{\hat{\omega}_i} \varepsilon_i.
    \end{eqnarray*}
    For the first term, using the dominating part of the solution of the process
    in the explosive regime as derived in Lemma \ref{lemsol}, we have
    \begin{eqnarray*}
        \sum_{i = l^{\ast} + 1}^{i_1^{\ast}} \frac{\int_{t_{i - 1}}^{t_i} y_u
            \mathd u}{\omega_i} & = & \sum_{i = l^{\ast} + 1}^{i_1^{\ast}}
        \frac{\int_{t_{i - 1}}^{t_i} \left( y_{t_{\lfloor \tau_1 n \rfloor}}
            e^{\kappa (u - t_{\lfloor \tau_1 n \rfloor})} \right) \mathd
            u}{\omega_i}\\
        & = & \sum_{i = l^{\ast} + 1}^{i_1^{\ast}} \frac{y_{t_{\lfloor \tau_1 n
                    \rfloor}} e^{- \kappa t_{\lfloor \tau_1 n \rfloor}}}{\omega_i \kappa} e^{\kappa
            t_i}  (1 - e^{\kappa H})\\
        & = & y_{t_{\lfloor \tau_1 n \rfloor}}  \sum_{i = l^{\ast} +
            1}^{i_1^{\ast}} \frac{1}{\omega_i} e^{\kappa (t_i - t_{\lfloor \tau_1 n
                \rfloor})}\\
        & = & y_{t_{\lfloor \tau_1 n \rfloor}} O \left( \sum_{i = l^{\ast} +
            1}^{i_1^{\ast}} e^{\kappa (t_i - t_{\lfloor \tau_1 n \rfloor})} \right)\\
        & = & y_{t_{\lfloor \tau_1 n \rfloor}} O \left( \frac{1 - e^{\kappa
                (t_{i_1^{\ast}} - t_{l^{\ast} + 1})}}{1 - e^{\kappa H}} \right)\\
        & = & O_p (\kappa^{- 1} n^{1 / 2} e^{\kappa (t_{i_1^{\ast}} - t_{l^{\ast}
                + 1})}) .
    \end{eqnarray*}
    Next,
    \begin{eqnarray*}
        \left| \sum_{i = l^{\ast} + 1}^{i_1^{\ast}} \left( \frac{\int_{t_{i -
                    1}}^{t_i} y_u \mathd u}{\hat{\omega}_i} - \frac{\int_{t_{i - 1}}^{t_i} y_u
            \mathd u}{\omega_i} \right) \right| & \leqslant & \sum_{i = l^{\ast} +
            1}^{i_1^{\ast}} \left| \int_{t_{i - 1}}^{t_i} y_u \mathd u \right|
        \frac{1}{| \hat{\omega}_i \omega_i |}  | \hat{\omega}_i - \omega_i |\\
        & = & o_p \left( \sum_{i = l^{\ast} + 1}^{i_1^{\ast}} \left|
        \frac{\int_{t_{i - 1}}^{t_i} y_u \mathd u}{\omega_i} \right| \right),
    \end{eqnarray*}
    using the uniform consistency of the realized variance estimator in Lemma
    \ref{unicons1}. Finally, \linebreak $\sum_{i = l^{\ast} + 1}^{i_1^{\ast}}
    (\omega_i/\hat{\omega}_i) \varepsilon_i = O_p (\kappa^{- 1} \log (\sqrt{n}))$ because $(i_1^{\ast} - l^{\ast}) = O (\kappa^{- 1} \log (\sqrt{n}))$, and in total we have that the first term is dominant and
    \[ \sum_{i = l^{\ast} + 1}^{i_1^{\ast}} \Delta x_i = O_p (n^{1 / 2} e^{\kappa
        (t_{i_1^{\ast}} - t_{l^{\ast} + 1})}) = O_p (n), \]
   using $t_{i_1^{\ast}} - t_{l^{\ast} + 1} = \kappa^{-1} \log \left(
    \sqrt{n} \right)$.
    
Now we look at the first term $\sum_{i = i_1 + 1}^{l^{\ast}} \Delta x_i$,
where $l^{\ast} = i_1 + \left\lfloor \kappa^{-1} \log \left(
\kappa^{-1} n^{-1/2} \log (n) \right) \right\rfloor$. We make a
further decomposition of it,
\[ \sum_{i = i_1 + 1}^{i_1 + \left\lfloor \frac{1}{\kappa} \log \left(
    \frac{1}{\kappa}  \frac{\log (n)}{\sqrt{n}} \right) \right\rfloor} \Delta
x_i = \sum_{i = i_1 + 1}^{i_1 + \left\lfloor \frac{1}{\kappa} \log \left(
    \frac{1}{\kappa}  \frac{1}{\sqrt{n}} \right) \right\rfloor} \Delta x_i +
\sum_{i = i_1 + \left\lfloor \frac{1}{\kappa} \log \left( \frac{1}{\kappa} 
    \frac{1}{\sqrt{n}} \right) \right\rfloor + 1}^{i_1 + \left\lfloor
    \frac{1}{\kappa} \log \left( \frac{1}{\kappa}  \frac{\log (n)}{\sqrt{n}}
    \right) \right\rfloor} \Delta x_i = A 1 + A 2 . \]

From the proof of Lemma \ref{leminc}, $\Delta x_i = O_p \left( \kappa n^{1 / 2}
e^{\kappa (t_i - t_{i_1})} \right)$. It follows that the $\Delta x_i$'s in $A
1$ are all $O_p (1)$, and thus the first summation \
\[ A 1 = O_p \left( \frac{1}{\kappa} \log \left( \frac{1}{\kappa} 
\frac{1}{\sqrt{n}} \right) \right), \]
which is dominated by $O_p (n)$ under our assumptions. The $\Delta x_i$'s in
the $A 2$ may be diverging, but the magnitude of the individual terms cannot
be larger than the terms when $i > l^{\ast}$, and there are in total
\[ \frac{1}{\kappa} \log \left( \frac{1}{\kappa}  \frac{\log (n)}{\sqrt{n}}
\right) - \frac{1}{\kappa} \log \left( \frac{1}{\kappa}  \frac{1}{\sqrt{n}}
\right) = \frac{1}{\kappa} \log (\log (n)) \]
terms in $A 2$, which has a lower order than the number of items summed up $O
\left( \kappa^{- 1} \log \left( \sqrt{n} \right) \right)$ in $\sum_{i =
    l^{\ast} + 1}^{i_1^{\ast}} \Delta x_i$. Therefore, $A 2$ sums up a lower
number of smaller terms than the sum $\sum_{i = l^{\ast} + 1}^{i_1^{\ast}}
\Delta x_i$, so it is dominated by $\sum_{i = l^{\ast} + 1}^{i_1^{\ast}}
\Delta x_i$. In summary, the result for the $1 / 2 \leqslant \alpha < 1$ case
is also proved.
    
\end{proof}

\section*{Appendix B: Proof of Theorems}
\renewcommand{\theequation}{B.\arabic{equation}}
\renewcommand{\theHequation}{B.\arabic{equation}}
\setcounter{equation}{0}

\begin{proof}
    (of Theorem \ref{thm1}) Using the decomposition and the $\Delta^h C_{i, j}$
    and $\Delta^h J_{i, j}$ notations defined in Lemma \ref{unicons}, we have
    \begin{eqnarray*}
        \frac{1}{\sqrt{n}} x_{\lfloor n \tau \rfloor} & = & \frac{1}{\sqrt{n}}  \sum_{i = 1}^{\lfloor n
            \tau \rfloor} \frac{\sum_{j = 1}^M (\Delta^h z_{t_{i - 1, j}} 
            \text{\tmtextbf{1}} \{| \Delta^h z_{t_{i - 1, j}} | < \phi_h
            \})}{\hat{\omega}_i}\\
        & = & \frac{1}{\sqrt{n}}  \sum_{i = 1}^{\lfloor n \tau \rfloor} \frac{\sum_{j = 1}^M
            (\Delta^h C_{i - 1, j} + \Delta^h J_{i - 1, j})  \text{\tmtextbf{1}} \{|
            \Delta^h z_{t_{i - 1, j}} | < \phi_h \}}{\hat{\omega}_i}\\
        & = & \frac{1}{\sqrt{n}}  \sum_{i = 1}^{\lfloor n \tau \rfloor} \frac{\sum_{j = 1}^M
            \Delta^h C_{i - 1, j}}{\hat{\omega}_i} - \frac{1}{\sqrt{n}}  \sum_{i =
            1}^{\lfloor n \tau \rfloor} \frac{\sum_{j = 1}^M \Delta^h C_{i - 1, j} 
            \text{\tmtextbf{1}} \{| \Delta^h z_{t_{i - 1, j}} | \geqslant \phi_h
            \}}{\hat{\omega}_i}\\
        &  & + \frac{1}{\sqrt{n}}  \sum_{i = 1}^{\lfloor n \tau \rfloor} \frac{\sum_{j = 1}^M
            \Delta^h J_{i - 1, j}  \text{\tmtextbf{1}} \{| \Delta^h z_{t_{i - 1, j}} |
            < \phi_h \}}{\hat{\omega}_i}\\
        & = & A - B + C.
    \end{eqnarray*}
    First, notice that $\sum_{j = 1}^M \Delta^h C_{i - 1, j} = \int_{t_{i -
            1}}^{t_i} \sigma_s \mathd W_s$; it follows that
    \begin{eqnarray*}
        A & = & \frac{1}{\sqrt{n}}  \sum_{i = 1}^{\lfloor n \tau \rfloor} \frac{\omega_i
            \varepsilon_i}{\hat{\omega}_i}\\
        & = & \frac{1}{\sqrt{n}}  \sum_{i = 1}^{\lfloor n \tau \rfloor} \varepsilon_i +
        \frac{1}{\sqrt{n}}  \sum_{i = 1}^{\lfloor n \tau \rfloor} \left(
        \frac{\omega_i}{\hat{\omega}_i} - 1 \right) \varepsilon_i\\
        & = : & A 1 + A 2.
    \end{eqnarray*}
    The first term $A 1 \Rightarrow B_{\tau}$. The second term
    \begin{eqnarray}
        A 2 & = & \left| \frac{1}{\sqrt{n}}  \sum_{i = 1}^{\lfloor n \tau \rfloor} \left(
        \frac{\omega_i}{\hat{\omega}_i} - 1 \right) \varepsilon_i \right|
        \nonumber\\
        & = & \left| \frac{1}{\sqrt{n}}  \sum_{i = 1}^{\lfloor n \tau \rfloor} (\omega_i -
        \hat{\omega}_i) \varepsilon_i \omega_i^{- 1} \right| \nonumber\\
        & = & \left| \frac{1}{\sqrt{n}}  \sum_{i = 1}^{\lfloor n \tau \rfloor} (\omega_i^2 -
        \hat{\omega}_i^2) \varepsilon_i \omega_i^{- 1} (\omega_i +
        \hat{\omega}_i)^{- 1} \right| \nonumber\\
        & \leqslant & \max_{1 \leqslant i \leqslant n} (\omega_i^{- 1} (\omega_i
        + \hat{\omega}_i)^{- 1})  \left| \frac{1}{\sqrt{n}}  \sum_{i = 1}^{\lfloor n
            \tau \rfloor } \varepsilon_i (\omega_i^2 - \hat{\omega}_i^2) \right| \nonumber\\
        & \leqslant & \max_{1 \leqslant i \leqslant n} (\omega_i^{- 1} (\omega_i
        + \hat{\omega}_i)^{- 1})  \left( \frac{1}{n}  \sum_{i = 1}^{\lfloor n \tau \rfloor}
        \varepsilon_i^2 \right)^{1 / 2} \left( \sum_{i = 1}^{\lfloor n \tau \rfloor} (\omega_i^2
        - \hat{\omega}_i^2)^2 \right)^{1 / 2},  \label{ineqweakcon}
    \end{eqnarray}
    using the Cauchy-Schwarz inequality in the last step. Using the results of
    Lemma \ref{unicons}, Lemma \ref{estimatorbound} and the fact that
    $\frac{1}{n}  \sum_{t = 1}^{\lfloor n \tau \rfloor} \varepsilon_t^2 = O_p (1)$, we have
    the above is of order $O_p (nh)$. Under the imposed assumption $nh
    \rightarrow 0$ we have thus shown that $A 2 = o_p (1)$.
    
    For the second term $B$,
    \begin{eqnarray*}
        B & = & \frac{1}{\sqrt{n}}  \sum_{i = 1}^{\lfloor n \tau \rfloor} \frac{\sum_{j = 1}^M
            (\Delta^h C_{i - 1, j}) \text{\tmtextbf{1}} \{| \Delta^h z_{t_{i - 1, j}}
            | \geqslant \phi_h, \Delta^h J_{i - 1, j} = 0\}}{\hat{\omega}_i}\\
        &  & + \frac{1}{\sqrt{n}}  \sum_{i = 1}^{\lfloor n \tau \rfloor} \frac{\sum_{j = 1}^M
            \Delta^h C_{i - 1, j} \text{\tmtextbf{1}} \{| \Delta^h z_{t_{i - 1, j}} |
            \geqslant \phi_h, \Delta^h J_{i - 1, j} \neq 0\}}{\hat{\omega}_i}\\
        & = & B 1 + B 2.
    \end{eqnarray*}
    
    We look at $B 1$. Notice that
    \[ |B1| \leqslant \max_i | \hat{\omega}_i^{- 1} | \frac{1}{\sqrt{n}} 
    \sum_{i = 1}^{\lfloor n \tau \rfloor} \sum_{j = 1}^M | \Delta^h C_{i - 1, j} |
    \text{\tmtextbf{1}} \{| \Delta^h C_{i - 1, j} | \geqslant \phi_h,
    \Delta^h J_{i - 1, j} = 0\} \]
    Now,
    \begin{eqnarray*}
        &  & E \left( \frac{1}{\sqrt{n}}  \sum_{i = 1}^{\lfloor n \tau \rfloor} \sum_{j = 1}^M
        | \Delta^h C_{i - 1, j} | \text{\tmtextbf{1}} \{| \Delta^h C_{i - 1, j} |
        \geqslant \phi_h, \Delta^h J_{i - 1, j} = 0\} \right)\\
        & \leqslant & \frac{1}{\sqrt{n}}  \sum_{i = 1}^{\lfloor n \tau \rfloor} \sum_{j = 1}^M
        (E (\Delta^h C_{i - 1, j})^2)^{1 / 2}  (P (| \Delta^h C_{i - 1, j} |
        \geqslant \phi_h))^{1 / 2} .
    \end{eqnarray*}
    $E (\Delta^h C_{i - 1, j})^2 = O (h)$. Again, using the notation $b_h$ and
    the intermediate result derived in Lemma \ref{unicons},
    \[ P (| \Delta^h C_{i - 1, j} | \geqslant \phi_h) = \frac{1}{b_h \exp (b_h^2
        / 2)}, \]
    we have
    \[ E |B1| = O \left( n^{- 1 / 2} nMh^{1 / 2}  \sqrt{\frac{1}{b_h \exp
            (b_h^2 / 2)}} \right) = O \left( \sqrt{\frac{n}{hb_h \exp (b_h^2 / 2)}}
    \right) = o (1), \]
    by Assumption A3. It follows that $|B1| = o_p (1)$. $B 2$ satisfies
    \begin{eqnarray*}
        |B2| & \leqslant & \max_i | \hat{\omega}_i^{- 1} | \frac{1}{\sqrt{n}} 
        \sum_{i = 1}^{\lfloor n \tau \rfloor} \sum_{j = 1}^M | \Delta^h C_{i - 1, j} |
        \text{\tmtextbf{1}} \{| \Delta^h z_{t_{i - 1, j}} | \geqslant \phi_h,
        \Delta^h J_{i - 1, j} \neq 0\}.
    \end{eqnarray*}
    Conditional on there being $N$ jumps happening at time $T = nH$, there are $N$
    nonzero terms in the above double sum. Using the conditioning argument as in
    the proof of Lemma \ref{unicons},
    \begin{eqnarray*}
        E \sum_{i = 1}^{\lfloor n \tau \rfloor} \sum_{j = 1}^M | \Delta^h C_{i - 1, j} |
        \text{\tmtextbf{1}} \{| \Delta^h z_{t_{i - 1, j}} | \geqslant \phi_h,
        \Delta^h J_{i - 1, j} \neq 0\} & \leqslant & E (E (N \cdot | \Delta^h C_{i
            - 1, j} ||N_T = N))\\
        & = & h^{1 / 2} E (N_T) = O (h^{1 / 2} n),
    \end{eqnarray*}
    because $E (N_T) = O (n)$. Now using $\max_i | \hat{\omega}_i^{- 1} | = O_p
    (1)$, we have $|B2| = O_p (n^{- 1 / 2} h^{1 / 2} n) = O_p ((nh)^{1 / 2}) =
    o_p (1)$ as $nh \rightarrow 0$. In total, $B = o_p (1)$.
    
    For the term $C$,
    \begin{eqnarray*}
        |C| & = & \left| \frac{1}{\sqrt{n}}  \sum_{i = 1}^{\lfloor n \tau \rfloor} \frac{\sum_{j
                = 1}^M \Delta^h J_{i - 1, j}  \text{\tmtextbf{1}} \{| \Delta^h z_{t_{i -
                    1, j}} | < \phi_h \}}{\hat{\omega}_i} \right|\\
        & = & \left| \frac{1}{\sqrt{n}}  \sum_{i = 1}^{\lfloor n \tau \rfloor} \frac{\sum_{j =
                1}^M \Delta^h J_{i - 1, j} \text{\tmtextbf{1}} \{| \Delta^h z_{t_{i - 1,
                    j}} | < \phi_h, \Delta^h J_{i - 1, j} \neq 0\}}{\hat{\omega}_i} \right|\\
        & \leqslant & \frac{\max_i | \hat{\omega}_i^{- 1} |}{\sqrt{n}}  \sum_{i =
            1}^{\lfloor n \tau \rfloor} \sum_{j = 1}^M | \Delta^h J_{i - 1, j} | \text{\tmtextbf{1}}
        \{| \Delta^h z_{t_{i - 1, j}} | < \phi_h, \Delta^h J_{i - 1, j} \neq 0\}\\
        & \leqslant & \frac{\max_i | \hat{\omega}_i^{- 1} |}{\sqrt{n}}  \sum_{i =
            1}^{\lfloor n \tau \rfloor} \sum_{j = 1}^M | \Delta^h J_{i - 1, j} | \text{\tmtextbf{1}}
        \{| \Delta^h J_{i - 1, j} | < \phi_h + | \Delta^h C_{i - 1, j} |, \Delta^h
        J_{i - 1, j} \neq 0\}\\
        & \leqslant & \frac{\max_i | \hat{\omega}_i^{- 1} |}{\sqrt{n}}  \sum_{i =
            1}^{\lfloor n \tau \rfloor} \sum_{j = 1}^M (\phi_h + | \Delta^h C_{i - 1, j} |) 
        \text{\tmtextbf{1}} \{| \Delta^h J_{i - 1, j} | < \phi_h + | \Delta^h C_{i
            - 1, j} |, \Delta^h J_{i - 1, j} \neq 0\}\\
        & = & \frac{\max_i | \hat{\omega}_i^{- 1} |}{\sqrt{n}} \phi_h  (1 + o_p
        (1))  \sum_{1 \leqslant i \leqslant \lfloor n \tau \rfloor, 1 \leqslant j \leqslant M,
            \Delta^h J_{i - 1, j} \neq 0} \text{\tmtextbf{1}} \{| \Delta^h J_{i - 1,
            j} | < \phi_h + | \Delta^h C_{i - 1, j} |\}.
    \end{eqnarray*}
    Since $P (| \Delta^h J_{i - 1, j} | < \phi_h + | \Delta^h C_{i - 1, j} |) =
    O (\phi_h)$ as in the proof of Lemma \ref{unicons}, using again the
    conditioning argument on the number of jumps above,
    \[ E \left( \sum_{1 \leqslant i \leqslant \lfloor n \tau \rfloor, 1 \leqslant j \leqslant
        M, \Delta^h J_{i - 1, j} \neq 0} \text{\tmtextbf{1}} \{| \Delta^h J_{i -
        1, j} | < \phi_h + | \Delta^h C_{i - 1, j} |\} \right) = E (N_T) \phi_h =
    O (n \phi_h) . \]
    It follows that $|C| = O (n^{- 1 / 2} \phi_h n \phi_h) = O ((n \phi_h^4)^{1
        / 2})$. By Assumption A3, $n \phi_h^4 = nh^{2 - 4 \varepsilon}$, and that
    $\varepsilon \leqslant 1 / 4$, it follows that $2 - 4 \varepsilon \geqslant1$, so $n
    \phi_h^4 \rightarrow 0$ as $h \rightarrow 0$ and $nh \rightarrow 0$.
    Therefore, $|C| = o_p (1)$. In total, the claim of the theorem is proved.
\end{proof}

\begin{proof}
    (of Theorem \ref{thm2}) In view of the proof of Theorem \ref{thm1},
    we have
    \[ \tmop{TVDF}_{\tau} \Rightarrow \frac{\int_0^{\tau} \widetilde{B}_r \mathd B_r}{\left(
        \int_0^{\tau} \widetilde{B}_r^2 \mathd r \right)^{1 / 2}}, \]
    where $\widetilde{B}_r := B_r - \tau^{-1}\int_{0}^\tau B_u \mathd u$ denotes demeaned standard Brownian motion on $[0,\tau]$ and the weak convergence holds for the process $\tmop{TVDF}_{\tau}$
    indexed by $\tau$. Applying the argument in \cite{phillips2011explosive} for the supremum taken over the index $\tau$, the claimed result of
    this theorem follows.
\end{proof}

\begin{proof}
    (of Theorem \ref{thm3}) This is an easy consequence of Theorem \ref{thm2} so we omit the details.
\end{proof}

\begin{proof}
    (of Theorem \ref{thm4}) Recall the TVDF statistic for the sample $\{x_s
    \}_{s = 0}^{\lfloor \tau n \rfloor}$ is defined as
    \[ \tmop{TVDF}_{\tau} = \frac{\sum_{s = 2}^{\lfloor \tau n \rfloor} (\Delta
        x_s - \overline{\Delta x_s})  (x_{s - 1} - \overline{x_{s - 1}})}{\left(
        \sum_{s = 2}^{\lfloor \tau n \rfloor} (x_{s - 1} - \overline{x_{s -
                1}})^2 \right)^{1 / 2}}, \]
    where
    \begin{equation}
        \label{nume} \sum_{s = 2}^{\lfloor \tau n \rfloor} (\Delta x_s -
        \overline{\Delta x_s})  (x_{s - 1} - \overline{x_{s - 1}}) = \sum_{s =
            2}^{\lfloor \tau n \rfloor} \Delta x_s x_{s - 1} - \frac{1}{\lfloor \tau n
            \rfloor - 1}  (x_{\lfloor \tau n \rfloor} - x_1) \left( \sum_{s =
            2}^{\lfloor \tau n \rfloor} x_{s - 1} \right),
    \end{equation}
    and 
    \begin{equation}
        \label{deno} \sum_{s = 2}^{\lfloor \tau n \rfloor} (x_{s - 1} -
        \overline{x_{s - 1}})^2 = \sum_{s = 2}^{\lfloor \tau n \rfloor} (x_{s -
            1})^2 - \frac{1}{\lfloor \tau n \rfloor - 1} \left( \sum_{s = 2}^{\lfloor
            \tau n \rfloor} x_{s - 1} \right)^2.
    \end{equation}
    In the following, we will show
    \[ \tmop{TVDF}_{\tau} = \left\{\begin{array}{ll}
        O_p (1), & \tau_0 < \tau \leqslant \tau_1 ,\\
        O_p (n^{(1 + \alpha) / 2} \log (n) \wedge n^{2 \alpha} \log^2 (n))
        \rightarrow \infty, & \tau = \tau_1^{\ast}, 0 \leqslant \alpha < 1 /
        2,\\
        O_p (n^{1 - \alpha / 2}) \rightarrow \infty, & \tau = \tau_1^{\ast}, 1
        / 2 \leqslant \alpha < 1,
    \end{array}\right. \]
    where $\tau_1^{\ast} = \tau_1 + \frac{1}{nH}  \left\lfloor \kappa^{-1}
    \log \left( \kappa^{-1} \log (n) \right) \right\rfloor$. Applying the
    arguments in {\cite{PY09}} and {\cite{PS18}}, the claimed consistency result
    for the estimator of the start date of the explosive regime follows. We now derive the
    above order results under the alternative hypothesis.
    
    \subsubsection*{1) $\tmop{TVDF}_{\tau} = O_p (1)$ for $\tau_0 < \tau
        \leqslant \tau_1$.}
    
    This can be shown in the same way as the proof of Theorem 2.
    
    \subsubsection*{2a) $\tmop{TVDF}_{\tau_1^{\ast}} = O_p (n^{(1 + \alpha) / 2}
        \log (n) \wedge n^{2 \alpha} \log^2 (n)) \rightarrow \infty$ when $0
        \leqslant \alpha < 1 / 2$.}
    
    Notice that by the definition of $\tau_1^{\ast}$, we have $H (\lfloor
    \tau_1^{\ast} n \rfloor - \lfloor \tau_1 n \rfloor) = \left\lfloor
    \kappa^{-1} \log \left( \kappa^{-1} \log (n) \right) \right\rfloor$.
    We first examine the behavior of the numerator $\sum_{s = 2}^{\lfloor
        \tau_1^{\ast} n \rfloor} (\Delta x_s - \overline{\Delta x_s})  (x_{s - 1} -
    \overline{x_{s - 1}})$. We examine each part of the expansion (\ref{nume})
    separately. For the term $\sum_{s = 2}^{\lfloor \tau_1^{\ast} n \rfloor}
    \Delta x_s x_{s - 1}$, notice that
    \begin{eqnarray*}
        \sum_{s = 2}^{\lfloor \tau_1^{\ast} n \rfloor} \Delta x_s x_{s - 1} & = &
        \sum_{s = \lfloor \tau_1 n \rfloor + 1}^{\lfloor \tau_1^{\ast} n \rfloor}
        \Delta x_s x_{s - 1} + \sum_{s = 2}^{\lfloor \tau_1 n \rfloor} \Delta x_s
        x_{s - 1}\\
        & = & \left( \sum_{s = \lfloor \tau_1 n \rfloor + 1}^{\lfloor
            \tau_1^{\ast} n \rfloor} \Delta x_s x_{s - 1} \right) + O_p (n) .
    \end{eqnarray*}
    Now, using the strategy in dealing with $\hat{\omega}_s$'s in the
    denominator as in the proof of Lemma \ref{leminc} and Lemma \ref{lemincsum}, we
    have
    \begin{eqnarray}
        &  & \sum_{s = \lfloor \tau_1 n \rfloor + 1}^{\lfloor \tau_1^{\ast} n
            \rfloor} \Delta x_s x_{s - 1} \nonumber\\
        & = & \kappa^2 y_{t_{\lfloor \tau_1 n \rfloor}}^2  \sum_{s = \lfloor
            \tau_1 n \rfloor + 1}^{\lfloor \tau_1^{\ast} n \rfloor} e^{\kappa H (s - 1
            - \lfloor \tau_1 n \rfloor)}  (e^{\kappa H (s - 2 - \lfloor \tau_1 n
            \rfloor)} + \cdots + 1) (1 + o_p (1)) \nonumber\\
        & = & \kappa^2 y_{t_{\lfloor \tau_1 n \rfloor}}^2  \sum_{s = \lfloor
            \tau_1 n \rfloor + 1}^{\lfloor \tau_1^{\ast} n \rfloor} e^{\kappa H (s - 1
            - \lfloor \tau_1 n \rfloor)}  \frac{e^{\kappa H (s - 1 - \lfloor \tau_1 n
                \rfloor)} - 1}{e^{\kappa H} - 1} (1 + o_p (1)) \nonumber\\
        & = & \kappa^2 y_{t_{\lfloor \tau_1 n \rfloor}}^2 e^{2 \kappa H (\lfloor
            \tau n \rfloor - 1 - \lfloor \tau_1 n \rfloor)}  \sum_{s = \lfloor \tau_1
            n \rfloor + 1}^{\lfloor \tau_1^{\ast} n \rfloor} \frac{e^{2 \kappa H (s -
                1 - \lfloor \tau n \rfloor)} - e^{\kappa H (s - 1 - \lfloor \tau n
                \rfloor)}}{e^{\kappa H} - 1} (1 + o_p (1)) \nonumber\\
        & = & ne^{2 \kappa H (\lfloor \tau_1^{\ast} n \rfloor - 1 - \lfloor
            \tau_1 n \rfloor)} (1 + o_p (1)), \nonumber\\
        & = & O_p \left( \frac{n \log^2 (n)}{\kappa^2} \right), \nonumber
    \end{eqnarray}
    after straightforward calculations involving power series, and we use $H
    (\lfloor \tau_1^{\ast} n \rfloor - 1 - \lfloor \tau_1 n \rfloor) = O \left(
    \kappa ^{-1} \log \left(\kappa^{-1} \log (n) \right) \right)$.
    Therefore, in total, we have
    \begin{equation}
        \sum_{s = 2}^{\lfloor \tau_1^{\ast} n \rfloor} \Delta x_s x_{s - 1} = O_p
        \left( \frac{n \log^2 (n)}{\kappa^2} \right) + O_p (n) = O_p \left(
        \frac{n \log^2 (n)}{\kappa^2} \right) . \label{nume1}
    \end{equation}
    In the following proof, we will repeatedly use the results of Lemma
    \ref{leminc} and Lemma \ref{lemincsum}, along with the strategy for handling
    the $\hat{\omega}_s$ terms in the denominator, as in the proofs of these
    lemmas. For brevity, we will omit explicit references to these lemmas
    throughout.
    
    For the $(x_{\lfloor \tau_1^{\ast} n \rfloor} - x_1)$ term, we have
    \begin{eqnarray}
        &  & x_{\lfloor \tau_1^{\ast} n \rfloor} - x_1 \nonumber\\
        & = & (x_{\lfloor \tau_1^{\ast} n \rfloor} - x_{\lfloor \tau_1 n
            \rfloor}) + (x_{\lfloor \tau_1 n \rfloor} - x_1) \nonumber\\
        & = & \kappa y_{\lfloor \tau_1 n \rfloor}  (e^{\kappa H (\lfloor
            \tau_1^{\ast} n \rfloor - 1 - \lfloor \tau_1 n \rfloor)} + e^{\kappa H
            (\lfloor \tau_1^{\ast} n \rfloor - 2 - \lfloor \tau_1 n \rfloor)} + \cdots
        + 1) (1 + o_p (1)) + O_p (n^{1 / 2}) \nonumber\\
        & = & O_p (n^{1 / 2} e^{\kappa H (\lfloor \tau_1^{\ast} n \rfloor -
            \lfloor \tau_1 n \rfloor)}) (1 + o_p (1)) + O_p (n^{1 / 2}) \nonumber\\
        & = & O_p \left( n^{1 / 2}  \frac{\log (n)}{\kappa} \right) + O_p (n^{1 /
            2}) \nonumber\\
        & = & O_p \left( n^{1 / 2}  \frac{\log (n)}{\kappa} \right) \label{nume2}       
    \end{eqnarray}
    For the $\sum_{s = 2}^{\lfloor \tau_1^{\ast} n \rfloor} x_{s - 1}$ term,
    again using similar calculations as before,
    \begin{eqnarray}
        \sum_{s = 2}^{\lfloor \tau_1^{\ast} n \rfloor} x_{s - 1} & = & \sum_{s =
            \lfloor \tau_1 n \rfloor + 1}^{\lfloor \tau_1^{\ast} n \rfloor} x_{s - 1}
        + \sum_{s = 2}^{\lfloor \tau_1 n \rfloor} x_{s - 1} \nonumber\\
        & = & \kappa y_{\lfloor \tau_1 n \rfloor}  \sum_{s = \lfloor \tau_1 n
            \rfloor + 1}^{\lfloor \tau_1^{\ast} n \rfloor} (e^{\kappa H (s - 2 -
            \lfloor \tau_1 n \rfloor)} + e^{\kappa H (s - 3 - \lfloor \tau_1 n
            \rfloor)} + \cdots + 1) (1 + o_p (1)) + O_p (n^{3 / 2}) \nonumber\\
        & = & O_p \left( \kappa n^{1 / 2}  \frac{1}{\kappa^2} e^{\kappa H
            (\lfloor \tau_1^{\ast} n \rfloor - \lfloor \tau_1 n \rfloor)} \right) +
        O_p (n^{3 / 2}) \nonumber\\
        & = & O_p \left( \frac{n^{1 / 2} \log (n)}{\kappa^2} \right) + O_p (n^{3
            / 2}) \nonumber\\
        & = & O_p (n^{3 / 2}), \label{nume3} 
    \end{eqnarray}
    by noticing that $\alpha < 1 / 2$ so the $n^{3 / 2}$ rate is higher.
    
    Collecting the results in (\ref{nume1}), (\ref{nume2}) and (\ref{nume3}), we
    have that the numerator has order
    \begin{eqnarray}
        \sum_{s = 2}^{\lfloor \tau_1^{\ast} n \rfloor} (\Delta x_s -
        \overline{\Delta x_s})  (x_{s - 1} - \overline{x_{s - 1}}) & = & O_p
        \left( \frac{n \log^2 (n)}{\kappa^2} \right) + \frac{1}{n} O_p \left( n^{1
            / 2}  \frac{\log (n)}{\kappa} \right) O_p (n^{3 / 2}) \nonumber\\
        & = & O_p \left( \frac{n \log^2 (n)}{\kappa^2} \right) . \label{numeorder} 
    \end{eqnarray}
    We then look at the behavior of $\sum_{s = 2}^{\lfloor \tau_1^{\ast} n
        \rfloor} (x_{s - 1} - \overline{x_{s - 1}})^2$ in the denominator. Using the
    expansion (\ref{deno}), we again examine its terms separately. For the
    $\sum_{s = 2}^{\lfloor \tau_1^{\ast} n \rfloor} (x_{s - 1})^2$ term, by the
    similar calculation as before,
    
    \begin{eqnarray}
        \sum_{s = 2}^{\lfloor \tau_1^{\ast} n \rfloor} (x_{s - 1})^2& = & \sum_{s = \lfloor \tau_1 n \rfloor + 1}^{\lfloor \tau_1^{\ast} n
            \rfloor} (x_{s - 1})^2 + \sum_{s = 2}^{\lfloor \tau_1 n \rfloor} (x_{s -
            1})^2 \nonumber\\
        & = & \left( \sum_{s = \lfloor \tau_1 n \rfloor + 1}^{\lfloor \tau_1^{\ast}
            n \rfloor} (x_{s - 1})^2 \right) + O_p (n^2) \nonumber\\
        & = & \left( \sum_{s = \lfloor \tau_1 n \rfloor + 1}^{\lfloor \tau_1^{\ast}
            n \rfloor} (y_{\lfloor \tau_1 n \rfloor}^2 e^{2 \kappa H (s - 1 - \lfloor
            \tau_1 n \rfloor)}) \right) (1 + o_p (1)) + O_p (n^2) \nonumber\\
        & = & O_p (\kappa^{- 1} ne^{2 \kappa H (\lfloor \tau_1^{\ast} n \rfloor - 1
            - \lfloor \tau_1 n \rfloor)}) + O_p (n^2) \nonumber\\
        & = & O_p (\kappa^{- 3} n \log^2 (n) \vee n^2) . \nonumber
    \end{eqnarray}
    The order of the $\sum_{s = 2}^{\lfloor \tau_1^{\ast} n \rfloor} x_{s - 1}$
    has already been derived in (\ref{nume3}). Therefore,
    \begin{equation}
        \sum_{s = 2}^{\lfloor \tau_1^{\ast} n \rfloor} (x_{s - 1} - \overline{x_{s
                - 1}})^2 = O_p (\kappa^{- 3} n \log^2 (n) \vee n^2) + n^{- 1} O_p (n^3) =
        O_p (\kappa^{- 3} n \log^2 (n) \vee n^2) . \label{denoorder}
    \end{equation}
    Summarizing the result of (\ref{numeorder}) and (\ref{denoorder}),
    \[ \tmop{TVDF}_{\tau} = O_p \left( \frac{n \log^2 (n) /
        \kappa^2}{\sqrt{(\kappa^{- 3} n \log^2 (n) \vee n^2)}} \right) = O_p ((n
    / \kappa)^{1 / 2} \log (n) \wedge \kappa^{- 2} \log^2 (n)), \]
    which is $O_p (n^{(1 + \alpha) / 2} \log (n) \wedge n^{2 \alpha} \log^2 (n))
    \rightarrow \infty$, for $0 \leqslant \alpha < 1 / 2$.
    
    \subsubsection*{2b) $\tmop{TVDF}_{\tau_1^{\ast}} = O_p (n^{1 - \alpha / 2})
        \rightarrow \infty$ when $1 / 2 \leqslant \alpha < 1$.}
    
    As in the proof for 2a), we first examine the behavior of the numerator
    $\sum_{s = 2}^{\lfloor \tau_1^{\ast} n \rfloor} (\Delta x_s -
    \overline{\Delta x_s})  (x_{s - 1} - \overline{x_{s - 1}})$ and examine each
    part of the expansion (\ref{nume}) separately. For $\sum_{s = 2}^{\lfloor
        \tau_1^{\ast} n \rfloor} \Delta x_s x_{s - 1}$, we make the following
    decomposition
    \begin{eqnarray*}
        \sum_{s = 2}^{\lfloor \tau_1^{\ast} n \rfloor} \Delta x_s x_{s - 1} & = &
        \sum_{s = 2}^{\lfloor \tau_1 n \rfloor} \Delta x_s x_{s - 1} + \sum_{s =
            \lfloor \tau_1 n \rfloor + 1}^{l^{\ast}} \Delta x_s x_{s - 1} + \sum_{s =
            l^{\ast} + 1}^{\lfloor \tau_1^{\ast} n \rfloor} \Delta x_s x_{s - 1} .
    \end{eqnarray*}
    The first term has order $O_p (n)$. Using the same strategy as in the proof
    for 2a) case, the third term has order $\sum_{s = l^{\ast} + 1}^{\lfloor
        \tau_1^{\ast} n \rfloor} \Delta x_s x_{s - 1} = O_p (ne^{2 \kappa H (\lfloor
        \tau_1^{\ast} n \rfloor - 1 - l^{\ast})})$. Because
    \begin{equation}
        H (\lfloor \tau_1^{\ast} n \rfloor - 1 - l^{\ast}) = \frac{1}{\kappa} \log
        \left( \frac{\log (n)}{\kappa} \right) - \frac{1}{\kappa} \log \left(
        \frac{1}{\kappa}  \frac{\log (n)}{\sqrt{n}} \right) = \frac{1}{\kappa}
        \log \left( \sqrt{n} \right), \label{ltotau1star}
    \end{equation}
    the third term becomes $\sum_{s = l^{\ast} + 1}^{\lfloor \tau_1^{\ast} n
        \rfloor} \Delta x_s x_{s - 1} = O_p (n^2)$. Using the same argument as in
    the proof of Lemma \ref{lemincsum}, the second term $\sum_{s = \lfloor
        \tau_1 n \rfloor + 1}^{l^{\ast}} \Delta x_s x_{s - 1}$ cannot be larger than
    the third term. In total, we have
    \begin{equation}
        \sum_{s = 2}^{\lfloor \tau_1^{\ast} n \rfloor} \Delta x_s x_{s - 1} = O_p
        (n^2) . \label{sixteen1}
    \end{equation}
    For the $(x_{\lfloor \tau_1^{\ast} n \rfloor} - x_1)$ term, we make the
    decomposition
    \begin{eqnarray*}
        x_{\lfloor \tau_1^{\ast} n \rfloor} - x_1 & = & (x_{\lfloor \tau_1^{\ast}
            n \rfloor} - x_{l^{\ast}}) + (x_{l^{\ast}} - x_{\lfloor \tau_1 n \rfloor})
        + (x_{\lfloor \tau_1 n \rfloor} - x_1) .
    \end{eqnarray*}
    Using similar calculations as before, we can show that the term $(x_{\lfloor
        \tau_1^{\ast} n \rfloor} - x_{l^{\ast}})$ is dominant, and
    \begin{eqnarray}
      &&  x_{\lfloor \tau_1^{\ast} n \rfloor} - x_1 \nonumber \\ & = & \kappa y_{\lfloor \tau_1 n
            \rfloor}  (e^{\kappa H (\lfloor \tau_1^{\ast} n \rfloor - 1 - \lfloor \tau_1 n
            \rfloor)} + e^{\kappa H (\lfloor \tau_1^{\ast} n \rfloor - 2 - \lfloor \tau_1 n
            \rfloor)} + \cdots + e^{\kappa H (l^{\ast} - 1 - \lfloor \tau_1 n
            \rfloor)} )  (1 + o_p (1)) \nonumber\\
        & = & O_p (n^{1 / 2} e^{\kappa H (\lfloor \tau_1^{\ast} n \rfloor -
            l^{\ast})}) . \nonumber
    \end{eqnarray}
    Using (\ref{ltotau1star}), the above becomes
    \begin{equation}
        x_{\lfloor \tau_1^{\ast} n \rfloor} - x_1 = O_p (n) . \label{nume21}
    \end{equation}
    For the $\sum_{s = 2}^{\lfloor \tau_1^{\ast} n \rfloor} x_{s - 1}$ term.
    Again using similar calculations as before
    \begin{eqnarray}
        &  & \sum_{s = 2}^{\lfloor \tau_1^{\ast} n \rfloor} x_{s - 1} \nonumber\\
        & = & \kappa y_{\lfloor \tau_1 n \rfloor}  \sum_{s = l^{\ast} +
            1}^{\lfloor \tau_1^{\ast} n \rfloor} (e^{\kappa H (s - 2 - \lfloor \tau_1
            n \rfloor)} + e^{\kappa H (s - 3 - \lfloor \tau_1 n \rfloor)} + \cdots +
        e^{\kappa H (l^{\ast} - \lfloor \tau_1 n \rfloor)} )  (1 + o_p (1)) + O_p (n^{3 / 2}) \nonumber\\
        & = & O_p \left( \kappa n^{1 / 2}  \frac{1}{\kappa^2} e^{\kappa H
            (\lfloor \tau_1^{\ast} n \rfloor - l^{\ast})} \right) + O_p (n^{3 / 2})
        \nonumber\\
        & = & O_p \left( \frac{n}{\kappa} \right) + O_p (n^{3 / 2}) = O_p \left(
        \frac{n}{\kappa} \right), \label{nume31} 
    \end{eqnarray}
    where we again use (\ref{ltotau1star}), and we notice that since now $\alpha
    > 1 / 2$, $n / \kappa = n^{1 + \alpha}$, which has power higher than 3/2.
    
    Collecting the results in (\ref{sixteen1}), (\ref{nume21}) and
    (\ref{nume31}), we have that the numerator
    \begin{eqnarray}
        \sum_{s = 2}^{\lfloor \tau_1^{\ast} n \rfloor} (\Delta x_s -
        \overline{\Delta x_s})  (x_{s - 1} - \overline{x_{s - 1}}) & = & O_p (n^2)
        + \frac{1}{n} O_p (n) O_p \left( \frac{n}{\kappa} \right) \nonumber\\
        & = & O_p (n^2) .  \label{numeorder1}
    \end{eqnarray}
    We then look at the order of $\sum_{s = 2}^{\lfloor \tau_1^{\ast} n \rfloor}
    (x_{s - 1} - \overline{x_{s - 1}})^2$ in the denominator. Using the
    expansion (\ref{deno}), we again examine its terms separately. For the
    $\sum_{s = 2}^{\lfloor \tau_1^{\ast} n \rfloor} (x_{s - 1})^2$ term, by
    similar calculations as before
    \begin{eqnarray}
        \sum_{s = 2}^{\lfloor \tau_1^{\ast} n \rfloor} (x_{s - 1})^2 & = & \sum_{s
            = 2}^{\lfloor \tau_1 n \rfloor} (x_{s - 1})^2 + \sum_{s = \lfloor \tau_1 n
            \rfloor + 1}^{l^{\ast}} (x_{s - 1})^2 + \sum_{s = l^{\ast} + 1}^{\lfloor
            \tau_1^{\ast} n \rfloor} (x_{s - 1})^2 \nonumber\\
        & = & \left( \sum_{s = l^{\ast} + 1}^{\lfloor \tau_1^{\ast} n \rfloor}
        (x_{s - 1})^2 \right)  (1 + o_p (1)) + O_p (n^2) \nonumber\\
        & = & O_p (\kappa^{- 1} ne^{2 \kappa H (\lfloor \tau_1^{\ast} n \rfloor -
            1 - l^{\ast})}) + O_p (n^2) \nonumber\\
        & = & O_p (\kappa^{- 1} n^2) + O_p (n^2) = O_p (\kappa^{- 1} n^2) .
     \nonumber
    \end{eqnarray}
    The order of the $\sum_{s = 2}^{\lfloor \tau_1^{\ast} n \rfloor} x_{s - 1}$
    has already been derived in (\ref{nume31}), therefore
    \begin{equation}
        \sum_{s = 2}^{\lfloor \tau_1^{\ast} n \rfloor} (x_{s - 1} - \overline{x_{s
                - 1}})^2 = O_p (\kappa^{- 1} n^2) + n^{- 1} O_p \left(
        \frac{n^2}{\kappa^2} \right) = O_p (\kappa^{- 1} n^2) .
        \label{denoorder111}
    \end{equation}
    Summarizing the result of (\ref{numeorder1}) and (\ref{denoorder111}),
    \[ \tmop{TVDF}_{\tau} = O_p \left( \frac{n^2}{\sqrt{\kappa^{- 1} n^2}}
    \right) = O_p (n \kappa^{1 / 2}) = O_p (n^{1 - \alpha / 2}) \rightarrow
    \infty, \]
    because $1 / 2 \leqslant \alpha < 1$.
\end{proof}

\bibliographystyle{chicago}
\bibliography{bubblebib}

\begin{thebibliography}{}

\bibitem[\protect\citeauthoryear{A{\"i}t-Sahalia, Beissner, Cheridito, and
  Matthys}{A{\"i}t-Sahalia et~al.}{2024}]{AitSahalia2024}
A{\"i}t-Sahalia, Y., P.~Beissner, P.~Cheridito, and F.~Matthys (2024).
\newblock Asset pricing in an economy with changing sentiment and price
  feedback.
\newblock {\em Manuscript, Princeton University\/}.

\bibitem[\protect\citeauthoryear{A{\"i}t-Sahalia and Jacod}{A{\"i}t-Sahalia and
  Jacod}{2024}]{AitSahaliaJacod2024}
A{\"i}t-Sahalia, Y. and J.~Jacod (2024).
\newblock Test whether volatility is stochastic.
\newblock {\em Manuscript, Princeton University\/}.

\bibitem[\protect\citeauthoryear{Andersen, Bollerslev, Diebold, and
  Ebens}{Andersen et~al.}{2001}]{andersen2001distribution}
Andersen, T.~G., T.~Bollerslev, F.~X. Diebold, and H.~Ebens (2001).
\newblock The distribution of realized stock return volatility.
\newblock {\em Journal of Financial Economics\/}~{\em 61\/}(1), 43--76.

\bibitem[\protect\citeauthoryear{Andersen, Dobrev, and Schaumburg}{Andersen
  et~al.}{2012}]{andersen2012jump}
Andersen, T.~G., D.~Dobrev, and E.~Schaumburg (2012).
\newblock Jump-robust volatility estimation using nearest neighbor truncation.
\newblock {\em Journal of Econometrics\/}~{\em 169\/}(1), 75--93.

\bibitem[\protect\citeauthoryear{Andersen, Todorov, and Zhou}{Andersen
  et~al.}{2025}]{ATZ23}
Andersen, T.~G., V.~Todorov, and B.~Zhou (2025).
\newblock Real-time detection of local no-arbitrage violations.
\newblock {\em Quantitative Economics\/}~{\em 16}, 459--495.

\bibitem[\protect\citeauthoryear{Astill, Harvey, Leybourne, Taylor, and
  Zu}{Astill et~al.}{2023}]{astill2023cusum}
Astill, S., D.~I. Harvey, S.~J. Leybourne, A.~R. Taylor, and Y.~Zu (2023).
\newblock Cusum-based monitoring for explosive episodes in financial data in
  the presence of time-varying volatility.
\newblock {\em Journal of Financial Econometrics\/}~{\em 21\/}(1), 187--227.

\bibitem[\protect\citeauthoryear{Beare}{Beare}{2018}]{beare2018}
Beare, B.~K. (2018).
\newblock Unit root testing with unstable volatility.
\newblock {\em Journal of Time Series Analysis\/}~{\em 39\/}(6), 816--835.

\bibitem[\protect\citeauthoryear{Biais, Capponi, Cong, Gaur, and
  Giesecke}{Biais et~al.}{2023}]{biais2023}
Biais, B., A.~Capponi, L.~Cong, V.~Gaur, and K.~Giesecke (2023).
\newblock Advances in blockchain and crypo economics.
\newblock {\em Management Science\/}~{\em 69\/}(11), 6417--6426.

\bibitem[\protect\citeauthoryear{Blanchard and Watson}{Blanchard and
  Watson}{1982}]{blanchard1982bubbles}
Blanchard, O.~J. and M.~W. Watson (1982).
\newblock Bubbles, rational expectations and financial markets.
\newblock In P.~Wachtel (Ed.), {\em Crises in the Economic and Financial
  Structure}, pp.\  295--316. Lexington, MA: D.C. Heath and Company.

\bibitem[\protect\citeauthoryear{Bollerslev and Zhou}{Bollerslev and
  Zhou}{2002}]{Bollerslev2002}
Bollerslev, T. and H.~Zhou (2002).
\newblock Estimating stochastic volatility diffusion using conditional moments
  of integrated volatility.
\newblock {\em Journal of Econometrics\/}~{\em 109\/}(1), 33--65.

\bibitem[\protect\citeauthoryear{Boswijk, Cavaliere, Georgiev, and
  Rahbek}{Boswijk et~al.}{2021}]{BCGR21}
Boswijk, H.~P., G.~Cavaliere, I.~Georgiev, and A.~Rahbek (2021).
\newblock Bootstrapping non-stationary stochastic volatility.
\newblock {\em Journal of Econometrics\/}~{\em 224\/}(1), 161--180.

\bibitem[\protect\citeauthoryear{Breitung and Kruse}{Breitung and
  Kruse}{2013}]{Breitung2013}
Breitung, J. and R.~Kruse (2013).
\newblock When bubbles burst: Econometric tests based on structural breaks.
\newblock {\em Statistical Papers\/}~{\em 54\/}(4), 911--930.

\bibitem[\protect\citeauthoryear{Chen and Yu}{Chen and
  Yu}{2015}]{chen2015optimal}
Chen, Y. and J.~Yu (2015).
\newblock Optimal jackknife for unit root models.
\newblock {\em Statistics \& Probability Letters\/}~{\em 99}, 135--142.

\bibitem[\protect\citeauthoryear{Choi and Jarrow}{Choi and Jarrow}{2020}]{CJ20}
Choi, S.~H. and R.~Jarrow (2020).
\newblock Testing the local martingale theory of bubbles using
  cryptocurrencies.
\newblock {\em Available at SSRN 3701960\/}.

\bibitem[\protect\citeauthoryear{Christensen, Oomen, and Ren{\`o}}{Christensen
  et~al.}{2022}]{COR22}
Christensen, K., R.~Oomen, and R.~Ren{\`o} (2022).
\newblock The drift burst hypothesis.
\newblock {\em Journal of Econometrics\/}~{\em 227\/}(2), 461--497.

\bibitem[\protect\citeauthoryear{Christoffersen}{Christoffersen}{2012}]{christoffersen2012elements}
Christoffersen, P.~F. (2012).
\newblock {\em Elements of Financial Risk Management}.
\newblock Academic Press.

\bibitem[\protect\citeauthoryear{Cong, Li, and Wang}{Cong
  et~al.}{2022}]{cong2022}
Cong, L., Y.~Li, and N.~Wang (2022).
\newblock Token-based platform finance.
\newblock {\em Journal of Financial Economics\/}~{\em 144\/}(3), 971--991.

\bibitem[\protect\citeauthoryear{Cong, He, and Li}{Cong
  et~al.}{2021}]{cong2021b}
Cong, L.~W., Z.~He, and J.~Li (2021).
\newblock Decentralized mining in centralized pools.
\newblock {\em The Review of Financial Studies\/}~{\em 34\/}(3), 1191--1235.

\bibitem[\protect\citeauthoryear{Cong, Li, and Wang}{Cong
  et~al.}{2021}]{cong2021a}
Cong, L.~W., Y.~Li, and N.~Wang (2021).
\newblock Tokenomics: Dynamic adoption and valuation.
\newblock {\em The Review of Financial Studies\/}~{\em 34\/}(3), 1105--1155.

\bibitem[\protect\citeauthoryear{Corradi and Distaso}{Corradi and
  Distaso}{2006}]{Corradi2006}
Corradi, V. and W.~Distaso (2006).
\newblock Semi-parametric comparison of stochastic volatility models using
  realized measures.
\newblock {\em The Review of Economic Studies\/}~{\em 73\/}(3), 635--667.

\bibitem[\protect\citeauthoryear{Diba and Grossman}{Diba and
  Grossman}{1987}]{Diba1987}
Diba, B.~T. and H.~I. Grossman (1987).
\newblock On the inception of rational bubbles.
\newblock {\em The Quarterly Journal of Economics\/}, 697--700.

\bibitem[\protect\citeauthoryear{Diba and Grossman}{Diba and
  Grossman}{1988}]{Diba1988}
Diba, B.~T. and H.~I. Grossman (1988).
\newblock Explosive rational bubbles in stock prices?
\newblock {\em The American Economic Review\/}, 520--530.

\bibitem[\protect\citeauthoryear{Evans}{Evans}{1991}]{evans1991pitfalls}
Evans, G.~W. (1991).
\newblock Pitfalls in testing for explosive bubbles in asset prices.
\newblock {\em The American Economic Review\/}~{\em 81\/}(4), 922--930.

\bibitem[\protect\citeauthoryear{Fama}{Fama}{2014}]{fama2014two}
Fama, E.~F. (2014).
\newblock Two pillars of asset pricing.
\newblock {\em American Economic Review\/}~{\em 104\/}(6), 1467--1485.

\bibitem[\protect\citeauthoryear{Flood and Garber}{Flood and
  Garber}{1980}]{Flood1980}
Flood, R.~P. and P.~M. Garber (1980).
\newblock Market fundamentals versus price-level bubbles: {T}he first tests.
\newblock {\em The Journal of Political Economy\/}~{\em 88\/}(4), 745--770.

\bibitem[\protect\citeauthoryear{Flood and Hodrick}{Flood and
  Hodrick}{1986}]{Flood1986}
Flood, R.~P. and R.~J. Hodrick (1986).
\newblock Asset price volatility, bubbles, and process switching.
\newblock {\em The Journal of Finance\/}~{\em 41\/}(4), 831--842.

\bibitem[\protect\citeauthoryear{Gatheral, Jaisson, and Rosenbaum}{Gatheral
  et~al.}{2018}]{gatheral2018}
Gatheral, J., T.~Jaisson, and M.~Rosenbaum (2018).
\newblock Volatility is rough.
\newblock {\em Quantitative Finance\/}~{\em 18\/}(6), 933--949.

\bibitem[\protect\citeauthoryear{Harvey, Leybourne, Sollis, and Taylor}{Harvey
  et~al.}{2016}]{harvey2015tests}
Harvey, D.~I., S.~J. Leybourne, R.~Sollis, and A.~R. Taylor (2016).
\newblock Tests for explosive financial bubbles in the presence of
  non-stationary volatility.
\newblock {\em Journal of Empirical Finance\/}~{\em 38}, 548--574.

\bibitem[\protect\citeauthoryear{Harvey, Leybourne, Tatlow, and Zu}{Harvey
  et~al.}{2025}]{harvey2025unit}
Harvey, D.~I., S.~J. Leybourne, B.~S. Tatlow, and Y.~Zu (2025).
\newblock Unit root tests for explosive financial bubbles in the presence of
  deterministic level shifts.
\newblock {\em Oxford Bulletin of Economics and Statistics\/}~{\em 87\/}(5),
  1204--1235.

\bibitem[\protect\citeauthoryear{Harvey, Leybourne, Taylor, and Zu}{Harvey
  et~al.}{2024}]{harvey2024new}
Harvey, D.~I., S.~J. Leybourne, A.~M.~R. Taylor, and Y.~Zu (2024).
\newblock A new heteroskedasticity-robust test for explosive bubbles.
\newblock {\em Journal of Time Series Analysis\/}~{\em 46\/}(5), 846--866.

\bibitem[\protect\citeauthoryear{Harvey, Leybourne, and Zu}{Harvey
  et~al.}{2019}]{harvey2019testing}
Harvey, D.~I., S.~J. Leybourne, and Y.~Zu (2019).
\newblock Testing explosive bubbles with time-varying volatility.
\newblock {\em Econometric Reviews\/}~{\em 38}, 1131--1151.

\bibitem[\protect\citeauthoryear{Harvey, Leybourne, and Zu}{Harvey
  et~al.}{2020}]{harvey2020sign}
Harvey, D.~I., S.~J. Leybourne, and Y.~Zu (2020).
\newblock Sign-based unit root tests for explosive financial bubbles in the
  presence of deterministically time-varying volatility.
\newblock {\em Econometric Theory\/}~{\em 36\/}(1), 122--169.

\bibitem[\protect\citeauthoryear{Homm and Breitung}{Homm and
  Breitung}{2012}]{homm2012testing}
Homm, U. and J.~Breitung (2012).
\newblock Testing for speculative bubbles in stock markets: a comparison of
  alternative methods.
\newblock {\em Journal of Financial Econometrics\/}~{\em 10\/}(1), 198--231.

\bibitem[\protect\citeauthoryear{Jarrow, Protter, and Martin}{Jarrow
  et~al.}{2022}]{JPM22}
Jarrow, R., P.~Protter, and J.~S. Martin (2022).
\newblock Asset price bubbles: Invariance theorems.
\newblock {\em Frontiers of Mathematical Finance\/}~{\em 1\/}(2), 161--188.

\bibitem[\protect\citeauthoryear{Jarrow and Kwok}{Jarrow and Kwok}{2021}]{JK21}
Jarrow, R.~A. and S.~S. Kwok (2021).
\newblock Inferring financial bubbles from option data.
\newblock {\em Journal of Applied Econometrics\/}~{\em 36\/}(7), 1013--1046.

\bibitem[\protect\citeauthoryear{Jarrow and Kwok}{Jarrow and Kwok}{2023}]{JK23}
Jarrow, R.~A. and S.~S. Kwok (2023).
\newblock A study on asset price bubble dynamics: Explosive trend or quadratic
  variation?
\newblock {\em Available at SSRN 4356169\/}.

\bibitem[\protect\citeauthoryear{Laurent, Renò, and Shi}{Laurent
  et~al.}{2024}]{LaurentRenoShi2024}
Laurent, S., R.~Renò, and S.~Shi (2024).
\newblock Realized drift.
\newblock {\em Journal of Econometrics\/}.
\newblock In press.

\bibitem[\protect\citeauthoryear{Laurent and Shi}{Laurent and
  Shi}{2020}]{LaurentShi2020VolatilityJump}
Laurent, S. and S.~Shi (2020).
\newblock Volatility estimation and jump detection for drift--diffusion
  processes.
\newblock {\em Journal of Econometrics\/}~{\em 217\/}(2), 259--290.

\bibitem[\protect\citeauthoryear{Lui, Phillips, and Yu}{Lui
  et~al.}{2024}]{lui2024robust}
Lui, Y.~L., P.~C. Phillips, and J.~Yu (2024).
\newblock Robust testing for explosive behavior with strongly dependent errors.
\newblock {\em Journal of Econometrics\/}~{\em 238\/}(2), 105626.

\bibitem[\protect\citeauthoryear{Malinova and Park}{Malinova and
  Park}{2023}]{malinova2023}
Malinova, K. and A.~Park (2023).
\newblock Tokenomics: {W}hen tokens beat equity.
\newblock {\em Management Science\/}~{\em 69\/}(11), 6568--6583.

\bibitem[\protect\citeauthoryear{Mancini}{Mancini}{2009}]{mancini2009non}
Mancini, C. (2009).
\newblock Non-parametric threshold estimation for models with stochastic
  diffusion coefficient and jumps.
\newblock {\em Scandinavian Journal of Statistics\/}~{\em 36\/}(2), 270--296.

\bibitem[\protect\citeauthoryear{Meyer and Yu}{Meyer and
  Yu}{2000}]{meyeryu2000}
Meyer, R. and J.~Yu (2000).
\newblock {BUGS} for a bayesian analysis of stochastic volatility models.
\newblock {\em The Econometrics Journal\/}~{\em 3\/}(2), 198–--215.

\bibitem[\protect\citeauthoryear{Pavlidis, Paya, and Peel}{Pavlidis
  et~al.}{2017}]{pavlidis2012new}
Pavlidis, E.~G., I.~Paya, and D.~A. Peel (2017).
\newblock Testing for speculative bubbles using spot and forward prices.
\newblock {\em International Economic Review\/}~{\em 58\/}(4), 1191--1226.

\bibitem[\protect\citeauthoryear{Perron}{Perron}{1991}]{perron1991continuous}
Perron, P. (1991).
\newblock A continuous time approximation to the unstable first-order
  autoregressive process: The case without an intercept.
\newblock {\em Econometrica: Journal of the Econometric Society\/}, 211--236.

\bibitem[\protect\citeauthoryear{Phillips}{Phillips}{1987}]{Phillips1987}
Phillips, P. C.~B. (1987).
\newblock Toward a unified asymptotic theory for autoregression.
\newblock {\em Biometrika\/}~{\em 74}, 533--547.

\bibitem[\protect\citeauthoryear{Phillips and Shi}{Phillips and
  Shi}{2019}]{phillipsshi2020}
Phillips, P. C.~B. and S.~Shi (2019).
\newblock Detecting financial collapse and ballooning sovereign risk.
\newblock {\em Oxford Bulletin of Economics and Statistics\/}~{\em 81\/}(6),
  1336--1361.

\bibitem[\protect\citeauthoryear{Phillips, Shi, and Yu}{Phillips
  et~al.}{2015a}]{Phillips2013}
Phillips, P. C.~B., S.~Shi, and J.~Yu (2015a).
\newblock Testing for multiple bubbles: Historical episodes of exuberance and
  collapse in the {S\&P} 500.
\newblock {\em International Economic Review\/}~{\em 56\/}(4), 1043--1078.

\bibitem[\protect\citeauthoryear{Phillips, Shi, and Yu}{Phillips
  et~al.}{2015b}]{Phillips2013a}
Phillips, P. C.~B., S.~Shi, and J.~Yu (2015b).
\newblock Testing for multiple bubbles: Limit theory of real-time detectors.
\newblock {\em International Economic Review\/}~{\em 56\/}(4), 1079--1134.

\bibitem[\protect\citeauthoryear{Phillips and Shi}{Phillips and
  Shi}{2018}]{PS18}
Phillips, P. C.~B. and S.-P. Shi (2018).
\newblock Financial bubble implosion and reverse regression.
\newblock {\em Econometric Theory\/}~{\em 34\/}(4), 705--753.

\bibitem[\protect\citeauthoryear{Phillips, Wu, and Yu}{Phillips
  et~al.}{2011}]{phillips2011explosive}
Phillips, P. C.~B., Y.~Wu, and J.~Yu (2011).
\newblock Explosive behavior in the 1990s {NASDAQ}: When did exuberance
  escalate asset values?
\newblock {\em International Economic Review\/}~{\em 52\/}(1), 201--226.

\bibitem[\protect\citeauthoryear{Phillips and Yu}{Phillips and
  Yu}{2009a}]{PY09}
Phillips, P. C.~B. and J.~Yu (2009a).
\newblock Limit theory for dating the origination and collapse of mildly
  explosive periods in time series data.
\newblock {\em Unpublished Manuscript, Singapore Management University\/}.

\bibitem[\protect\citeauthoryear{Phillips and Yu}{Phillips and
  Yu}{2009b}]{PY09twostage}
Phillips, P. C.~B. and J.~Yu (2009b).
\newblock A two-stage realized volatility approach to estimation of diffusion
  processes with discrete data.
\newblock {\em Journal of Econometrics\/}~{\em 150}, 139--150.

\bibitem[\protect\citeauthoryear{Phillips and Yu}{Phillips and
  Yu}{2011}]{Phillips2011}
Phillips, P. C.~B. and J.~Yu (2011).
\newblock Dating the timeline of financial bubbles during the subprime crisis.
\newblock {\em Quantitative Economics\/}~{\em 2\/}(3), 455--491.

\bibitem[\protect\citeauthoryear{Shi and Phillips}{Shi and
  Phillips}{2026}]{ShiPhillips2024}
Shi, S. and P.~C.~B. Phillips (2026).
\newblock Uncovering mild drift in asset prices with intraday high-frequency
  data.
\newblock {\em Journal of Econometrics\/}~{\em 253}, 1--29.

\bibitem[\protect\citeauthoryear{Shi and Song}{Shi and
  Song}{2015}]{shi2015identifying}
Shi, S. and Y.~Song (2015).
\newblock Identifying speculative bubbles using an infinite hidden {M}arkov
  model.
\newblock {\em Journal of Financial Econometrics\/}~{\em 14\/}(1), 159--184.

\bibitem[\protect\citeauthoryear{Shiryaev}{Shiryaev}{1995}]{shiryaev1995probability}
Shiryaev, A. (1995).
\newblock {\em Probability (2nd edn)}.
\newblock Springer: New York, NY.

\bibitem[\protect\citeauthoryear{Sockin and Xiong}{Sockin and
  Xiong}{2023}]{sockin2023}
Sockin, M. and W.~Xiong (2023).
\newblock A model of crypocurrencies.
\newblock {\em Management Science\/}~{\em 69\/}(11), 6625--6640.

\bibitem[\protect\citeauthoryear{Todorov}{Todorov}{2009}]{Todorov2009}
Todorov, V. (2009).
\newblock Estimation of continuous-time stochastic volatility models with jumps
  using high-frequency data.
\newblock {\em Journal of Econometrics\/}~{\em 148\/}(2), 131--148.

\bibitem[\protect\citeauthoryear{Wang, Xiao, and Yu}{Wang
  et~al.}{2023}]{wang2023}
Wang, X., W.~Xiao, and J.~Yu (2023).
\newblock Modeling and forecasting realized volatility with the fractional
  {O}rnstein-{U}hlenbeck process.
\newblock {\em Journal of Econometrics\/}~{\em 232\/}(2), 389--415.

\bibitem[\protect\citeauthoryear{Wang and Yu}{Wang and Yu}{2022}]{wangyu2023}
Wang, X. and J.~Yu (2022).
\newblock Bubble testing under polynomial trends.
\newblock {\em The Econometrics Journal\/}~{\em 26\/}(1), 25--44.

\bibitem[\protect\citeauthoryear{Yu}{Yu}{2014}]{Yu2014}
Yu, J. (2014).
\newblock Econometric analysis of continuous time models: A survey of {P}eter
  {P}hillips's work and some new results.
\newblock {\em Econometric Theory\/}~{\em 30\/}(04), 737--774.

\bibitem[\protect\citeauthoryear{Zhou and Yu}{Zhou and Yu}{2015}]{Zhou2015}
Zhou, Q. and J.~Yu (2015).
\newblock Asymptotic theory for linear diffusions under alternative sampling
  schemes.
\newblock {\em Economics Letters\/}~{\em 128}, 1--5.

\end{thebibliography}

\end{document}